\documentclass[preprint,12pt]{elsarticle}
\pdfoutput=1
\biboptions{sort&compress}

\usepackage[usenames,dvipsnames]{xcolor}
\usepackage[colorlinks=true, linkcolor=Black, citecolor=Black, urlcolor=Black, filecolor=Black]{hyperref}
\usepackage{longtable}
\usepackage{soul}
\usepackage{epsfig}
\usepackage{amssymb}
\usepackage{amsmath}
\usepackage{graphicx}
\usepackage{verbatim}
\usepackage{xspace}
\usepackage{hyperref}
\usepackage{float}
\DeclareMathOperator{\Tr}{Tr} 

\newcommand{\nc}{\newcommand}
\nc{\be}{\begin{equation}}
\nc{\ee}{\end{equation}}
\nc{\bea}{\begin{eqnarray}}
\nc{\eea}{\end{eqnarray}}

\newcommand{\dg}{\dagger}

\begin{document}

\renewcommand{\arraystretch}{1.5}

\begin{frontmatter}

\title{
Quark masses and mixings in minimally parameterized 
UV completions of the Standard Model}

\author{\quad Reinhard Alkofer\footnote{reinhard.alkofer@uni-graz.at}$^{\mathrm{a}}$, Astrid Eichhorn\footnote{eichhorn@cp3.sdu.dk}$^{\mathrm{b,c}}$, Aaron Held\footnote{a.held@imperial.ac.uk}$^{\mathrm{d,c}}$, \\
Carlos M. Nieto\footnote{cnieto@sissa.it}$^{\mathrm{e,f,g,h}}$, Roberto Percacci\footnote{percacci@sissa.it}$^{\mathrm{e,f}}$, Markus Schr\"ofl\footnote{markus.schroefl@edu.uni-graz.at}$^{\mathrm{a}}$ \\[0.4cm]}
\address{$^{\mathrm{a}}$Institute of Physics, NAWI Graz, University of Graz, Universit\"atsplatz 5, 8010 Graz, Austria \\
$^{\mathrm{b}}$CP3-Origins, University of Southern Denmark, Campusvej 55, DK-5230 Odense M, Denmark \\
$^{\mathrm{c}}$Institut f\"ur Theoretische Physik, Universit\"at Heidelberg, Philosophenweg 16, 69120 Heidelberg, Germany \\
$^{\mathrm{d}}$Blackett Laboratory, Imperial College, London SW7 2AZ, United Kingdom \\
$^{\mathrm{e}}$International School for Advanced Studies, via Bonomea 265, I-34136 Trieste, Italy \\
$^{\mathrm{f}}$INFN, Sezione di Trieste, Italy \\
$^{\mathrm{g}}$International Centre for Theoretical Physics, Strada Costiera 11, 34151 Trieste, Italy \\
$^{\mathrm{h}}$Universidad Industrial de Santander, Cra 27 Calle 9, Bucaramanga, Colombia.
}

\begin{abstract}
We explore a simple parameterization of new physics that results in an ultraviolet complete gauge-quark sector of the Standard Model. Specifically, we add an antiscreening contribution to the beta functions of the gauge couplings and a flavor-independent, antiscreening contribution to the beta functions of the Yukawa couplings. These two free parameters give rise to an intricate web of Renormalization Group fixed points. Their predictive power extends to the flavor structure and mixing patterns, which we investigate to demonstrate that some of the free parameters of the Standard Model could be determined by the Renormalization Group flow.
\end{abstract}

\end{frontmatter}



\section{Introduction}\label{sec:intro}

The spectrum of masses of elementary particles is an intriguing mystery of the Standard Model (SM),
generally referred to as the ``flavor puzzle''. Why is the top quark so much heavier than the other quarks?
More generally, what generates the complicated mixing pattern of the quarks and the huge differences between quark masses? Many ideas have been put forward to explain these properties, including higher-order operators
\cite{Weinberg:1972ws}, new symmetries \cite{Froggatt:1978nt}, extra dimensions
\cite{ArkaniHamed:1999dc, Grossman:1999ra}, and the Renormalization Group (RG) flow of SM couplings \cite{Pendleton:1980as,Hill:1980sq}. Here, we explore a simple modification of the latter, as proposed in \cite{Shaposhnikov:2009pv, Harst:2011zx, Eichhorn:2017ylw, Eichhorn:2017lry, Eichhorn:2018whv}, motivated by the necessity of an ultraviolet (UV) completion.

Modern particle physics is dominated by the Effective Field Theory (EFT) paradigm: at each energy scale, one is content to deal with a model that describes all the observed degrees of freedom and reflects all the observed symmetries only with
a finite number of free parameters, which are relevant to describe experimental results. 
On the way towards higher energies, one has to update the model whenever a mass threshold is encountered.
There is not much need to discuss the ultraviolet limit, because new physics has regularly been found every time the energy frontier has been pushed further up. Conversely, any proposal for a truly microscopic (fundamental) theory is expected to be ``shielded" from a direct confrontation with experimental results by thresholds at which heavy (and as of yet unknown) degrees of freedom decouple. The predominant paradigm has been the existence of many such thresholds at a priori unknown energies, such that the link between the fundamental theory and LHC physics is tenuous at best.
The results of the LHC could prove to be paradigm-changing: not only is there no direct indication for physics beyond the SM at the TeV scale, there is even an intriguing indirect hint for the scale of new physics from the Higgs sector; as $m_H=125$ GeV \cite{Aad:2012tfa, Chatrchyan:2012xdj}, the Higgs particle's mass lies in the relatively narrow range of masses where a consistent extrapolation of the SM to much higher scales is theoretically viable, avoiding a low-scale Landau pole at higher 
Higgs masses \cite{Maiani:1977cg} and vacuum instability at low masses \cite{Krive:1976sg, Krasnikov:1978pu, Hung:1979dn}.
For the most recent indirect measurement of the top pole mass by ATLAS and CMS \cite{Sirunyan:2019zvx, Aad:2019ntk}, the scale of theoretical viability surpasses the Planck scale even if one demands absolute vacuum stability.
Nevertheless, the SM remains an EFT due to the presence of transplanckian Landau poles in the Abelian hypercharge and Higgs-Yukawa sector, indicating a nonperturbative triviality problem. New physics at high scales, therefore, has to exist.

In this paper, we parameterize the new physics in a very simple form, by adding a linear (dimension-like) term to the beta function of each coupling. We assume independence of internal symmetries, and therefore parameterize the contribution to all gauge couplings by a parameter $f_g$ and the contribution to all Yukawa couplings by a second parameter $f_y$. A key message of this paper is that such a simple addition could suffice to induce an ultraviolet completion of SM-like quark-gauge systems. Moreover, such a UV-completion could exhibit an enhanced predictive power compared to that of an effective field theory with the same couplings.

In particular, a correction of the beta function of the Abelian gauge coupling $g_Y$ of the form
\be
\beta_Y=\beta_Y^{SM}- f_g\, g_Y,
\label{eq:betaY}
\ee
would generate a nontrivial Fixed Point (FP),
that can be taken as a well-defined UV starting point for the flow of $g_Y$, thereby solving the problem of its Landau pole, or, more accurately, its triviality problem. Further, we consider a similar modification for the Yukawa couplings $y_i$, schematically
\be
\beta_{y_i} = \beta_{y_i}^{SM}- f_y\, g_i.\label{eq:betayiintro}
\ee
The two quantities $f_g$ and $f_y$ parameterize new physics in a minimal form. 
For the results in this paper, the specific new-physics interpretation of these terms is not critical. The most obvious new physics that could give rise to such terms is gravity\footnote{Gravity is among the ``oldest'' physics that one can think of, but here we stick to the particle-physics--centric view that everything that is not in the SM counts as ``new''.}.
Indeed, there are hints that gravity changes the RG of the SM by universal
terms as in Eq.~\eqref{eq:betaY} and Eq.~\eqref{eq:betayiintro}. In this context, universality refers to independence from any internal symmetries, i.e., these terms are the same, e.g., for Abelian and non-Abelian gauge couplings. Due to the dimensionful nature of the Newton coupling, it is well-known that universality in the sense of scheme-independence does not hold for these terms even at leading order (in contrast to the dimensionless SM couplings, where scheme-dependence sets in at three loops). The lack of universality also implies that the physical momentum dependence of the theory is not necessarily reflected in a corresponding dependence on the RG scale. This has been discussed, e.g., in \cite{Anber:2010uj,Donoghue:2019clr}, following a debate in the literature regarding the physical implications of gravity in one-loop perturbation theory, see also \cite{Ellis:2010rw,Gonzalez-Martin:2017bvw}. 
\\
For the gauge sector, the quantum-gravity contribution comes in the form in Eq.~\eqref{eq:betaY}, see, e.g., \cite{Robinson:2005fj,Daum:2009dn,Daum:2010bc,Harst:2011zx,Folkerts:2011jz,Eichhorn:2017lry}. A similar modification has been found to occur also for the Yukawa couplings  \cite{Zanusso:2009bs,Oda:2015sma,Eichhorn:2016esv,Hamada:2017rvn,Eichhorn:2017eht,Eichhorn:2017ylw,Rodigast:2009zj,Gonzalez-Martin:2017bvw,Salvio:2017qkx}. Within a Wilsonian approach to the RG, such contributions have important implications for the number of free parameters of the theory.
Considerable evidence has been gathered, indicating that Euclidean quantum gravity might be asymptotically safe.
Following the pioneering work of \cite{Reuter:1996cp}, the occurrence of an interacting, asymptotically safe, fixed point in the UV has been studied extensively in the last two decades, see, e.g., 
\cite{Reuter:2001ag,
Lauscher:2001ya,
Litim:2003vp,
Codello:2008vh,
Benedetti:2009rx,
Falls:2013bv,
Becker:2014qya,
Gies:2016con,
Christiansen:2017bsy,
Eichhorn:2018ydy} 
as well as
\cite{Percacci:2017fkn,
Eichhorn:2017egq,
Eichhorn:2018yfc,
Reuter:2019byg} for introductory  literature.
The corresponding results make asymptotically safe quantum gravity a candidate for a description of quantum gravity within a quantum-field theoretic framework and without the need to introduce additional fields except for the metric. 
The impact of quantum fluctuations of matter on gravity has been explored, e.g., in  
\cite{Dou:1997fg,
Dona:2013qba,
Dona:2015tnf,
Meibohm:2015twa,
Meibohm:2016mkp,
Christiansen:2017cxa,
Alkofer:2018fxj,
Alkofer:2018baq},
providing evidence that the matter content of the SM affects the gravitational FP without destroying it.
Conversely, there are indications that quantum gravity
fluctuations can affect matter couplings, see, e.g., 
\cite{Shaposhnikov:2009pv,Daum:2009dn,Daum:2010bc,Harst:2011zx,Folkerts:2011jz,Oda:2015sma,Eichhorn:2016esv,Eichhorn:2017eht,Eichhorn:2017ylw,Christiansen:2017cxa,Hamada:2017rvn,Eichhorn:2017lry,Eichhorn:2018whv,deBrito:2019umw} and generate a predictive UV completion, as indicated above.

In the context of asymptotically safe gravity, the key challenge is to determine fixed-point values for $f_g$ and $f_y$.
$f_g\geq 0$ holds in the entire gravitational parameter-space in the Einstein-Hilbert truncation, unless one chooses an unphysical, negative value of the Newton coupling \cite{Folkerts:2011jz,Christiansen:2017cxa}, see \cite{deBrito:2019umw} for the generalization beyond Einstein-Hilbert. $f_y>0$ is a significant restriction of the parameter space \cite{Eichhorn:2016esv,Eichhorn:2017eht,Eichhorn:2017ylw}. One can view this as an observational restriction of the microscopic gravitational parameter space that arises from observations of nonzero fermion masses at the electroweak scale. A quantitatively precise determination of the two parameters $f_g$ and $f_y$ from first principles has not been possible so far.

Although asymptotically safe gravity is a potential motivation for our ans\"atze for a UV completion of the SM, for the purpose of the present investigations we can remain agnostic about the origin of the new terms in the beta functions.
It is not necessary to specify the novel new-physics generating them.
The new physics can be parameterized in a very minimalistic form by $f_g$ and $f_y$.
Whatever their origin, these terms give rise to nontrivial FPs and the RG trajectories that originate from them are said to be ``renormalizable'' or ``asymptotically safe''. 
This is relevant to low-energy physics because the couplings that run along such trajectories are more constrained than those of a generic EFT or even a perturbatively renormalizable theory such as the SM.
As we shall discuss, this can give rise to predictions for some of 
the free parameters of the SM.

Apart from the postulated linear terms, our analysis of the RG flow is based on the standard perturbative one- and two-loop beta functions. 

It is important that we do not make any assumptions about the actual values of the couplings. The differences between the various low-energy values of Yukawa couplings arise as a consequence of the properties of the RG flow. To explain this point let us call ``theory space'' the space parameterized by all the SM couplings\footnote{In principle, as in EFT, one should consider also the perturbatively non-renormalizable couplings. For the present purposes, it will be enough to consider only the dimensionless ones.}.
The permutation group of three elements, $S_3$, acts on theory space by interchanging the Yukawa couplings of the three up-type quarks and those of the three down-type quarks among themselves, and the beta functions are invariant under this action, see Sec.~\ref{sec:3gen}. In particular, the pattern of FPs and the flows emanating from them are permutation-symmetric. It is the choice of an RG trajectory that necessarily breaks the symmetry described by this group.
There are actually two distinct ways in which this can happen. The Yukawa couplings could all be degenerate at the FP,
for example, they could all be asymptotically free. Then, the FP that describes the UV theory is permutation symmetric, and in order to be phenomenologically viable, the symmetry must be broken by the flow to the infrared. In practice, this amounts to choosing non-symmetric initial conditions for the flow in the vicinity of the fixed point at some very high scale. 
A more interesting possibility is that the FP which provides the UV completion of the theory is itself not permutation-symmetric, but is part of a multiplet of FPs that are interchanged under permutations.
In this case, it is the choice of UV completion that breaks
generation symmetry and this remains imprinted on the couplings all the way to the infrared.

A separate issue is the breaking of the degeneracy within each weak doublet. Since the up- and down-type quarks have different values of the hypercharge, this occurs automatically when the FP lies at nonvanishing hypercharge coupling $g_Y$.

Throughout the paper, we consider only the running of the Yukawa couplings and the mixing of quarks. We expect that analogous phenomena will happen also in the lepton sector, but we leave this for future investigation.
However, the contribution of leptons to the beta functions of the gauge couplings is taken into account, 
so that the beta functions that we consider in the gauge sector are the physically relevant ones, except for tiny higher-loop corrections due to lepton Yukawa couplings that we set to zero. 
We also neglect any self-interactions of the Higgs field. This is expected to be a robust approximation for the fixed-point structure in the gauge-Yukawa system, but it leaves open the question of whether the fixed-point structures observed here can be combined with a phenomenologically viable RG flow for the Higgs self-interactions.
Furthermore, we do not take into account any new physics between the electroweak and Planck scale, which may be required to explain other puzzles, such as dark matter or the matter-antimatter asymmetry of the universe. Our qualitative results are expected to extend to settings with minimal extensions of the SM, such as, e.g., put forward in \cite{Asaka:2005pn} to account for the aforementioned observations.\newline

This paper is structured as follows: In Sec.~\ref{sec:UVcomp}, we explain how the addition of linear terms to the beta functions for a single family solves the Landau-pole problems by inducing interacting fixed points. We further review how the difference between top and bottom Yukawa coupling becomes calculable at all scales, as first proposed in \cite{Eichhorn:2018whv}. 

In Sec.~\ref{sec:betas}, we provide the full beta functions with the new-physics contribution, and we eliminate the redundant
couplings to derive the
beta functions for the physically essential couplings only.

In Sec.~\ref{sec:2gen}, we discuss a toy model with two generations. This serves as a preparation for the phenomenologically relevant three-generation case. We highlight differences in fixed-point structures between the case with and without mixing, and discuss an example of phenomenologically viable flows, which reproduce the qualitative flavor-structure of the two heaviest generations of quarks in the SM. We also point out that unitarity of the CKM matrix is a key piece of nonperturbative information that imposes further restrictions on the flow of couplings, thereby generating flavor structure at low energies. Specifically, pole-structures in the flow of the CKM matrix elements result in upper bounds on otherwise unconstrained Yukawa couplings. In that section, we also discuss the theoretical viability of the fixed points in light of the fact that we work with one-loop beta functions throughout the rest of the paper.

Sec.~\ref{sec:3gen}, contains our results on three generations. We first discuss fixed points of the CKM matrix, and then use these as input for the gauge-Yukawa system. We find that the results from  \cite{Eichhorn:2018whv} can be extended to three generations and qualitatively reproduce the observed flavor and mixing structure. The low-energy value of the Abelian gauge coupling, the top Yukawa and the bottom Yukawa coupling are fixed uniquely at all scales in terms of 
the parameters $f_g$ and $f_y$. A precise quantitative agreement with the values of these couplings inferred from measurements cannot be achieved for an exact fixed-point trajectory emanating from a fixed point at non-vanishing Yukawa couplings. Additionally, the values of all the other quark Yukawa couplings are bounded from above. 

We draw conclusions and point out avenues for future work in Sec.~\ref{sec:conclusions}.
Several technical aspects have been deferred to appendices including a study of a special case of flavor-sensitive new physics in \ref{app:genbeta}.

\section{UV completion and predictive fixed points}\label{sec:UVcomp}
The SM is not UV complete due to Landau poles in the U(1) gauge coupling \cite{GellMann:1954fq} and the Higgs-Yukawa sector \cite{Maiani:1977cg}.
Beyond perturbation theory, there are no indications for a nonperturbative UV completion \cite{Gockeler:1997dn, Gockeler:1997kt, Gies:2004hy}, so that the SM suffers from the triviality problem. 
This is a consequence of the marginally irrelevant nature of the U(1) and Yukawa couplings. 

A minimal modification of the RG of the SM that renders it UV complete consists in adding to the beta function of each coupling a term linear in the coupling itself. 
These terms dominate over the one- (and higher-loop) term for small couplings. 
Since such terms are not present in the SM in four dimensions, we must assume that they are only present beyond some mass threshold $M_{\rm NP}$. 
They parameterize the effect of new physics in a general form, without specifying the nature and number of the additional fields. 

In the rest of this section, we discuss the beta functions for 
a drastically simplified version of the SM,
consisting only of the top-bottom-$U(1)$ system, including the new terms.
More precisely, we assume the full particle content of the SM when determining the running of the $U(1)$,
but we set to zero the $SU(2)$ and $SU(3)$ gauge couplings as well as
$y_u$, $y_d$, $y_c$ and $y_s$ and the lepton Yukawa couplings.
This is sufficient to highlight the salient features of the new-physics setting we explore here, extending the study in \cite{Eichhorn:2018whv}.
The beta functions at one loop are
\bea
\beta_{y_t}&=&\frac{y_t}{16\pi^2}\left( \frac{9}{2}y_t^2 + \frac{3}{2}y_b^2 -\frac{17}{12} g_Y^2\right)-f_y\, y_t,\label{eq:betayt}\\
\beta_{y_b}&=&\frac{y_b}{16\pi^2}\left( \frac{9}{2}y_b^2 + \frac{3}{2}y_t^2 -\frac{5}{12} g_Y^2\right)-f_y\, y_b,\label{eq:betayb}\\
\beta_{g_Y}&=&\frac{g_{Y}^3}{16\pi^2}\frac{41}{6}-f_{g}\, g_Y \label{eq:betagy},
\eea
Herein, $f_y$ and $f_g$ parameterize the new physics, and we assume that
\bea
f_y=
\begin{cases}
0,\quad\quad\,  k<M_{\rm NP}\\
{\rm const}, \, \, k \geq M_{\rm NP},
\end{cases}
\eea
\bea
f_g=
\begin{cases}
0,\quad\quad \, k<M_{\rm NP}\\
 {\rm const}, \, \, k \geq M_{\rm NP}.
\end{cases}
\eea
$M_{\rm NP}$ is the mass-threshold at which the new physics becomes important and $k$ is the RG scale. In line with the view advocated in the introduction, $M_{\rm NP}\gg 246\, \rm GeV$ is of particular interest, and we will choose $M_{\rm NP} = M_{\rm Planck}$ throughout this paper. The requirement that $f_g= \rm const$, $f_y=\rm const$ beyond $M_{\rm NP}$ can be realized if the new physics is scale-invariant. The idea that scale invariance could be crucial to understand the SM goes back to work in the 1980's 
\cite{Pendleton:1980as,Hill:1980sq,Wetterich:1981ir,Paschos:1984xa,Babu:1987im}, but has been gaining new traction since the discovery of the Higgs at 125 GeV, which can be interpreted as a hint for a scale-invariant dynamics at high scales \cite{Meissner:2006zh,Shaposhnikov:2008xi,tHooft:2016uxd,Helmboldt:2016mpi,Lewandowski:2017wov,Ferreira:2018itt,Shaposhnikov:2018xkv,Shaposhnikov:2018nnm,Wetterich:2019qzx}, see \cite{Alekhin:2017kpj,Moch:2018las} for an extrapolation of the current experimental status to the Planck-scale Higgs-potential. Scale symmetry, dynamically achieved at an interacting fixed point of the RG, may lead to an enhanced predictive power: 
the breaking of scale-symmetry by the RG flow is encoded in just a few free parameters, namely the relevant couplings, while all other couplings are (in principle) calculable.

If the new physics antiscreens the gauge and Yukawa couplings, i.e., $f_g>0$ and $f_y>0$, it renders the model UV complete. The competition between the antiscreening new-physics terms and the screening one-loop terms leads to additional structure in the beta functions, namely interacting fixed points. This provides an option for a UV completion beyond asymptotic freedom. It is achieved at a non-vanishing value of some of the SM couplings, where the new contribution and the one-loop  (or resummed higher-loop) contribution can balance out. Specifically, the above system possesses eight
fixed-point solutions admitting non-negative values for suitable choices of $f_g$ and $f_y$. 
The four with an asymptotically free hypercharge coupling are \footnote{Here and in the following fixed-point values for couplings are denoted by a $\ast$ in the subscript.}
\begin{alignat}{3}
y_{t\,\ast}&= 0\,, \quad &&y_{b\, \ast}=0\,, \quad &&g_{Y\, \ast}=0\,,\label{eq:GFP}\\
y_{t\,\ast}&=  \frac{4\pi}{3}\sqrt{ 2f_y}\,, \quad &&y_{b\, \ast}=0\,, \quad &&g_{Y\, \ast}=0\,,\label{eq:fpyt}\\
y_{t\,\ast}&= 0\,, \quad &&y_{b\, \ast}=\frac{4\pi}{3}\sqrt{ 2f_y}\,, \quad &&g_{Y\, \ast}=0\,,\label{eq:fpyb}\\
y_{t\,\ast}&=  \frac{4\pi}{3} \sqrt{\frac{3 f_y}{2}}\,, \quad &&y_{b\, \ast}=\frac{4\pi}{3} \sqrt{\frac{3 f_y}{2}}\,, \quad &&g_{Y\, \ast}=0\, .\label{eq:fpysym}
\end{alignat}
The first fixed point is the free fixed point. For the case $f_g=0=f_y$, it is the only fixed point and is IR attractive in all three directions, rendering the system trivial.
The second fixed point has been explored in \cite{Eichhorn:2017ylw}.
The four fixed points with an asymptotically safe hypercharge coupling are given by
\begin{alignat}{3}
y_{t\,\ast}&= 0\,, \quad &&y_{b\, \ast}=0\,, \quad &&g_{Y\, \ast}=4\pi\sqrt{\frac{6\,f_g}{41}}\,,\\
y_{t\,\ast}&= \frac{4\pi}{3}\sqrt{2f_y+\frac{17\,f_g}{41}}\,, \quad &&y_{b\, \ast}=0\,, \quad &&g_{Y\, \ast}= 4\pi\sqrt{\frac{6\,f_g}{41}}\,,\\
y_{t\,\ast}&=0\,, \quad &&y_{b\, \ast}= \frac{4\pi}{3}\sqrt{2f_y+\frac{5\,f_g}{41}}\,, \quad &&g_{Y\, \ast}=4\pi\sqrt{\frac{6\,f_g}{41}}\,,\\
y_{t\,\ast}&=\frac{4\pi}{3} \sqrt{\frac{3 f_y}{2} + \frac{69\, f_g}{164}}\,, \quad &&
y_{b\, \ast}= \frac{4\pi}{3}\sqrt{\frac{3 f_y}{2}-\frac{3f_g}{164}}\,, \quad &&g_{Y\, \ast}=4\pi\sqrt{\frac{6\,f_g}{41}}\,.\label{eq:fptby}
\end{alignat}
The first of these four fixed points has been discussed in \cite{Eichhorn:2017lry} (and its counterpart in QED in \cite{Harst:2011zx}) and the last fixed point in \cite{Eichhorn:2018whv}.\\
We aim at further exploring whether a nontrivial fixed-point structure can generate the differences between the quark masses in the SM.  Therefore, we draw attention to two sources of symmetry-breaking in the $y_t,\, y_b$ subsystem:\\
i) At vanishing $g_Y$, $y_t$ and $y_b$ can be different only if one of them vanishes while the other is non-zero,
but the \emph{combined} fixed-point structure in Eq.~\eqref{eq:GFP}-\eqref{eq:fpysym} is symmetric under the mapping $y_t \leftrightarrow y_b$. In particular, at the fixed point in  Eq.~\eqref{eq:fpysym}, where $y_t$ and $y_b$ are both non-vanishing, their values are equal. Nevertheless, the choice of fixed point can break the $y_t \leftrightarrow y_b$ symmetry.\\
ii) As the top and bottom quarks have different Abelian charges, any fixed point with non-vanishing value for the hypercharge coupling, $g_{Y\, \ast}$, necessarily leads to different values of $y_{t\,*}$ and $y_{b\,*}$. In particular, this holds for the fixed point in Eq.~\eqref{eq:fptby}, where the three fixed-point values satisfy the following relation \cite{Eichhorn:2018whv}
\be
y_{t\,\ast}^2- y_{b\, \ast}^2 = \frac{1}{3}g_{Y\, \ast}^2\,.
\ee
This relation shows that a finite fixed-point value for the Abelian gauge coupling automatically results in a difference between the two Yukawa couplings, with the top Yukawa coupling being larger (unless it is set to zero at the fixed point) due to its larger Abelian charge.
This ordering of the two Yukawa couplings is realized phenomenologically. 

Realizing such a difference between Yukawa couplings in the fixed-point regime does not automatically entail that it is necessarily imprinted on IR physics. To decide whether this is the case, one needs to explore the flow away from the fixed point.

Any of the above fixed points can be chosen as the UV starting point of an RG flow, i.e., they all define theoretically viable microscopic theories. Depending on the number of IR attractive directions, the fixed points can provide a first-principles derivation of three, two, one or none of the IR values, either in agreement or disagreement with observation. Specifically, the linearized flow in the vicinity of a fixed point takes the form
\be
z_I= c_I\, \left(\frac{k}{k_0}\right)^{-\theta_I},\label{eq:linflow}
\ee
where $z_I$ are suitable linear combinations of the distance of couplings to their fixed-point values,
$k_0$ is a reference scale, and the $\theta_I$ scaling exponents.  For a weak-coupling fixed point, the matrix relating the $z_I$ to the couplings is to a good 
approximation diagonal and the $z_I$ essentially correspond to $g_{Y}- g_{Y\, \ast}$, $y_{t}- y_{t\, \ast}$ and $y_b-y_{b\, \ast}$.
The scaling exponents are (minus) the eigenvalues of the stability matrix, which consists of the first derivatives of the beta functions at the fixed point, i.e.,
\be
\theta_I = -{\rm eig} \left( \frac{\partial \beta_{g_i}}{\partial g_j}\right)\Big|_{\vec{g}= \vec{g}_{\ast}},
\ee
where we have summarized all couplings of the system in the vector $\vec{g}$. 
The linear combinations  $z_I$ of the deviations $g_i-g_{i*}$ are correspondingly related
to the eigenvectors of the 
stability matrix.

If an RG trajectory approaches a FP
in the IR limit, then every IR attractive (``irrelevant'')
direction provides one prediction of an IR value of a coupling. 
This follows as quantum fluctuations drive the coupling towards
its fixed-point value.
This is the mechanism that was invoked in
\cite{Pendleton:1980as, Hill:1980sq}.
However, irrelevant directions give predictions also
when a trajectory emanates from a FP in the UV.
This is because a trajectory can only depart from a FP if its tangent vector has overlap with a relevant 
(IR repulsive) directions. More intuitively, a small deviation from the fixed-point value along a relevant direction grows under the RG flow to the IR, as one sees from Eq.~\eqref{eq:linflow} with $\theta_I>0$. Conversely, a small deviation from the fixed-point value along an irrelevant (IR attractive) direction is driven back to zero by the RG flow, as an inspection of Eq.~\eqref{eq:linflow} with $\theta_I<0$ reveals.
Thus, the deviation from the fixed-point value for a relevant coupling can be chosen freely, introducing a free parameter into the low-energy dynamics. Conversely, the deviation from the fixed-point value for an irrelevant coupling cannot be chosen freely. In some cases, it has to stay at its fixed-point value for all scales. Typically, irrelevant couplings deviate from their fixed-point values, but only as they are driven away from the fixed point due to the flow in the relevant directions\footnote{Note that these considerations apply separately to
the UV and IR behavior and even if an RG-trajectory passes close to an intermediate fixed point. Thus, the predictive power of irrelevant directions of several fixed points can accumulate.}.

For the system at hand, the signs of the scaling exponents follow a simple pattern: If $f_g$ and $f_y$ are both positive, i.e., the new-physics effect is anti\-screening, then the free fixed point is IR repulsive in all three directions. Accordingly, for a trajectory emanating from this FP, no predictions can be made about the IR values of the couplings. 
For the FPs considered here, 
every nonzero coupling gives rise to
an IR-attractive direction. This follows from the mechanism that generates these fixed points: An antiscreening contribution dominates for $g_i<g_{i\,\ast}$, whereas a screening one dominates for $g_{i}>g_{i\,\ast}$. This implies a growth of the coupling during the RG flow to the IR for $g_{i}<g_{i \, \ast}$ and a decrease of the coupling for $g_{i}>g_{i \, \ast}$, rendering $g_{i\,\ast}$ an IR attractive point. Therefore, a fixed point with one (two) [three] non-zero couplings entails one (two) [three] predictions.

In Eqs.~\eqref{eq:betayt}-\eqref{eq:betagy}, the predictions for the IR values are exactly the fixed-point values. Once the strong and $SU(2)$ gauge coupling are added to the system, there is a slow flow already beyond $M_{\rm NP}$, driven by the non-Abelian gauge couplings which emanate from their free fixed point and increase towards the IR. The value of a coupling with nonzero fixed-point value is fixed at $M_{\rm NP}$, providing a unique initial condition for the RG flow in the SM below $M_{\rm NP}$. This translates into a calculation of the corresponding value of the coupling at the electroweak scale from first principles.

\begin{figure}[!t]
\centering
\includegraphics[width=0.6\linewidth]{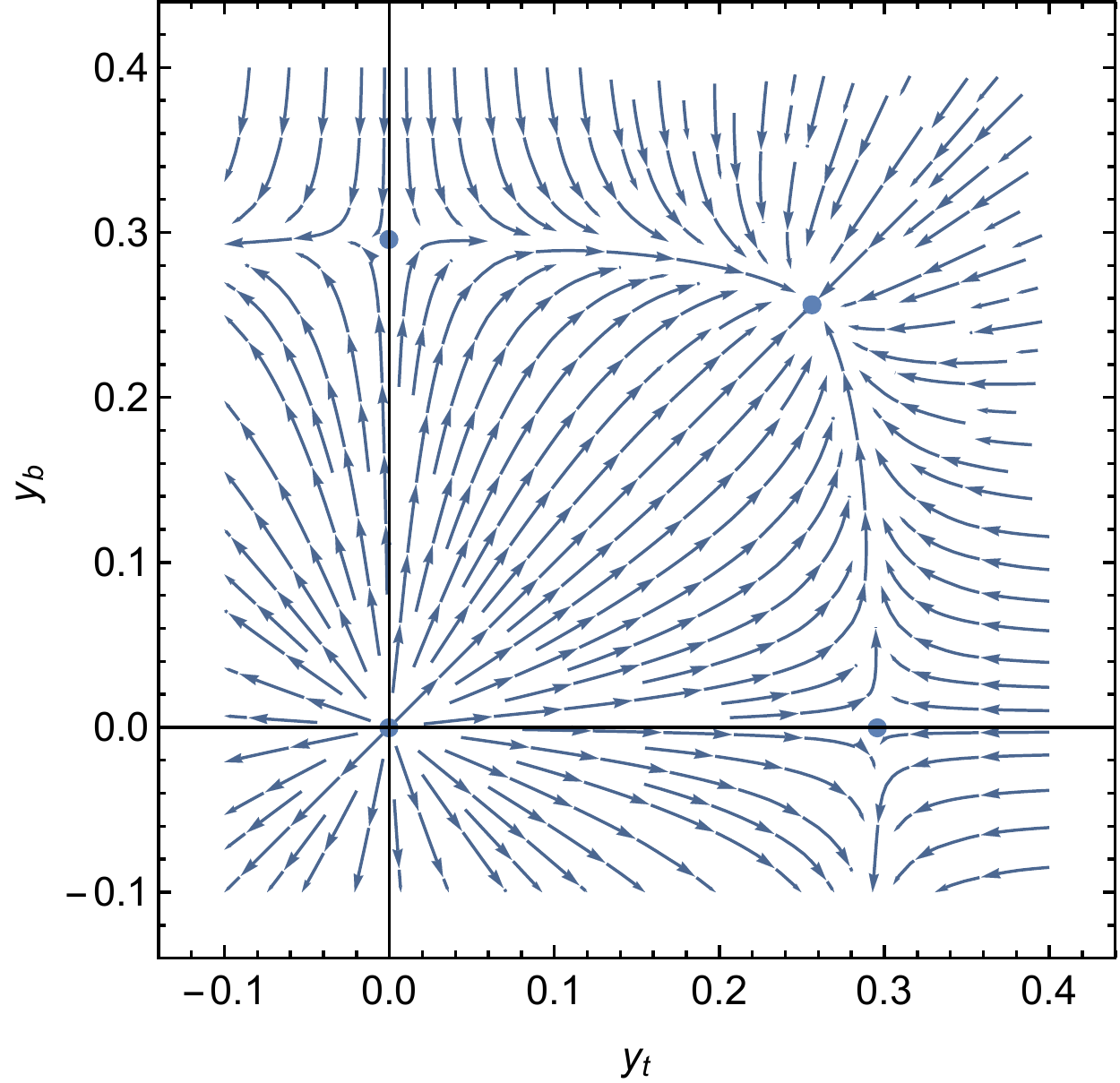}
\caption{\label{fig:Yukawa_schematic_plot} We choose $f_y=1/400$ and set $g_Y=0$ to show the flow in the $(y_t,\,y_b)$ plane. The separatrices connecting the partially interacting to the fully interacting fixed point result in upper bounds on the low-energy values of the Yukawa couplings even for trajectories starting from the free fixed point, where the IR values of both Yukawa couplings correspond to free parameters.}
\end{figure}

Beyond the predictions of IR values, the fixed-point structure in Eq.~\eqref{eq:GFP}-\eqref{eq:fptby} has additional consequences: The individual fixed points cannot be considered in isolation with regard to their consequences in the IR. For instance, by exploring only the Gaussian fixed point Eq.~\eqref{eq:GFP} while neglecting the existence of further fixed points, one would draw the erroneous conclusion that no statements can be made about achievable IR values. Yet, the (partially) interacting fixed points Eqs.~\eqref{eq:fpyt}-\eqref{eq:fptby} actually shield all IR values $y_{t}>y_{t\, \rm crit}$, $y_b>y_{b\, \rm crit}$ and $g_Y>g_{Y\, \rm crit}$ from the free fixed point, as the RG flow cannot cross the separatrices connecting the various fixed points, cf.~Fig.~\ref{fig:Yukawa_schematic_plot}.  Here, the (interconnected) critical IR values $g_{Y\, \rm crit}$ etc are those IR values reached on a trajectory emanating from the interacting fixed point.
Accordingly, the fixed-point structure imposes globally valid upper bounds, as has been discussed in \cite{Eichhorn:2017ylw, Eichhorn:2017lry}.

It has already been observed in \cite{Pendleton:1980as, Hill:1980sq, Wetterich:1981ir, Paschos:1984xa}  that it could be possible to explain structures in the SM from fixed points of the RG. For instance, a minimally mixing form of the CKM matrix actually corresponds to an IR attractive fixed point, at least for the phenomenologically relevant ordering of Yukawa couplings \cite{Pendleton:1980as}. The predictive power of partial IR attractive fixed points in the SM has also been noted for the ratio of the top-Yukawa coupling to the strong gauge coupling in \cite{Pendleton:1980as, Hill:1980sq, Wetterich:1981ir}.

More recently, the idea of interacting RG fixed points with and without quantum gravity has received more attention. In particular, \cite{Litim:2014uca} has triggered studies of gauge-Yukawa systems in four dimensions which exhibit asymptotic safety due to a cancellation of the one-loop versus the higher-loop terms \cite{Litim:2015iea,Esbensen:2015cjw,Bond:2016dvk,Bond:2017lnq,Bond:2018oco}. Our study is complementary, in the sense that we explore a different mechanism for asymptotic safety, see also \cite{Eichhorn:2018yfc} for an overview of fixed-point mechanisms including examples.

\section{Beta functions for Yukawa couplings and CKM matrix elements}\label{sec:betas}

The beta functions for the Yukawa couplings are calculated 
in the flavor basis,  in which the interactions of the fermions
with the Higgs field have the form 
$$
-q_L\, H\, Y_D\, d_R-q_L\, \tilde H\, Y_U\, u_R,
$$
where $q_L$ are the left-handed quark doublets,  $d_R$ and $u_R$ are the right-handed down-type and up-type quarks,
$H$ is the Higgs doublet and $\tilde H=i\sigma_2 H^*$ with the second Pauli matrix $\sigma_2$. 
In this basis, the Yukawa couplings are represented by the
$N_g\times N_g$ matrices $Y_U$ and $Y_D$,
where $N_g$ is the number of generations, and we are suppressing generation indices.

The one-loop beta functions for these Yukawa matrices are given by
(the two-loop beta functions are given in \ref{app:twoloop}):
\bea
\beta_{Y_U}&=&\frac{1}{16\pi^2}\Biggl[\frac{3}{2}Y_U Y_U^{\dagger}Y_U - \frac{3}{2}Y_D Y_D^{\dagger}Y_U + 3 \,{\rm Tr}\left(Y_U Y_U^{\dagger} + Y_D Y_D^{\dagger} \right)Y_U \nonumber \\
&& \qquad \qquad - \left(\frac{17}{12}g_Y^2 + \frac{9}{4}g_2^2 +8g_3^2\right)Y_U \Biggr] - f_y\, Y_U,
 \label{eq:YukawamatrixflowU} \\
\beta_{Y_D}&=&\frac{1}{16\pi^2}\Biggl[\frac{3}{2}Y_D Y_D^{\dagger}Y_D - \frac{3}{2}Y_U Y_U^{\dagger}Y_D + 3 \,{\rm Tr}\left(Y_U Y_U^{\dagger} + Y_D Y_D^{\dagger} \right)Y_D \nonumber \\
&& \qquad \qquad - \left(\frac{5}{12}g_Y^2 + \frac{9}{4}g_2^2 +8g_3^2\right)Y_D \Biggr]  - f_y\, Y_D\ .
\label{eq:YukawamatrixflowD}
\eea
Here, $g_2$ and $g_3$ are the SU(2) and SU(3) gauge couplings, respectively. We also included the linear terms coming from new physics, which we assume to be completely independent of
the generation indices. This is natural in the case of gravity,
but it is obviously not the most general case\footnote{See \ref{app:genbeta} for a discussion of more general ans\"atze for the new-physics parameterizing terms.}.

Some of these couplings are
redundant, in the sense that they can be eliminated from the Lagrangian by field redefinitions.
Being inessential, they are not required to reach a FP in order to achieve UV completeness.
It is, therefore, necessary, in the discussion of the RG, to separate the essential from the redundant couplings.
This is achieved by going to the mass basis.

To diagonalize the matrices $M_U=Y_U Y_U^{\dagger}$ and $M_D=Y_D Y_D^{\dagger}$, we introduce two unitary matrices $U$ and $D$, such that
\bea
{\rm diag}(y_u^2, y_c^2, y_t^2)&=& U\, M_U\, U^{\dagger} \, , \label{eq:defnyu}\\
{\rm diag} (y_d^2, y_s^2, y_b^2)&=&D\, M_D\, D^{\dagger}\, .\label{eq:defnyd}
\eea
This provides us with the CKM-matrix elements which enter in the electroweak currents, i.e., the couplings of $W^{\pm}$ to the quarks, defined by
\be
V=U\, D^{\dagger} =   
\begin{bmatrix}
V_{ud} & V_{us}& V_{ub}\\
V_{cd}& V_{cs} & V_{cb}\\
V_{td}& V_{ts}& V_{tb}
  \end{bmatrix}.
\ee
From Eqs.~\eqref{eq:YukawamatrixflowU} and \eqref{eq:YukawamatrixflowD}, 
we can derive the scale-dependence of the CKM-matrix elements as well as that of the Yukawa couplings $y_i$. 

The key point to use in the derivation of $\beta_{y_i}$ is the unitarity of $U$ and $D$, which implies that 
\be
U\partial_t U^{\dagger} + (\partial_t U)U^{\dagger} =0, \quad D\partial_t D^{\dagger} + (\partial_t D)D^{\dagger} =0.\label{eq:unitarityUD}
\ee

From Eq.~\eqref{eq:defnyu}, we obtain that
\be
[{\rm diag}(y_u^2, y_c^2, y_t^2),(\partial_t U)U^{\dagger}]+ \partial_t \, {\rm diag}(y_u^2, y_c^2, y_t^2) = U ( \partial_t M_U ) U^{\dagger},\label{eq:YUflow}
\ee
where we used Eq.~\eqref{eq:unitarityUD} to obtain the commutator. As the matrix ${\rm diag}(y_u^2, y_c^2, y_t^2)$ is 
diagonal, 
the commutator term is purely off-diagonal and therefore does not contribute to the beta functions of the Yukawa couplings $y_i$.

Next, we need to rewrite $U ( \partial_t M_U) U^{\dagger}$ 
for which we make use of the following identities
\bea
U{\rm Tr} (M_U)U^{\dagger}&=& \mathbb{I}\,{\rm Tr} M_U
=\mathbb{I}\,{\rm Tr} (U^{\dagger}U M_U)=\mathbb{I}\,\left(y_u^2+y_c^2+y_t^2\right),
\\[2mm]
U(M_U)^2U^{\dagger}&=&U M_U U^{\dagger}U M_UU^{\dagger} 
=\left( {\rm diag}(y_u^2, y_c^2, y_t^2)\right)^2,
\\[2mm]
U M_D M_U U^{\dagger} &=& U D^{\dagger} {\rm diag}(y_d^2, y_s^2, y_b^2)D U^{\dagger} {\rm diag}(y_u^2, y_c^2, y_t^2) 
\nonumber\\
&=& V{\rm diag}(y_d^2, y_s^2, y_b^2) V^{\dagger} {\rm diag}(y_u^2, y_c^2, y_t^2).
\eea

From the diagonal entries in Eq.~\eqref{eq:YUflow}, we obtain the beta functions for the Yukawa couplings as
\bea
\beta_{y_i}&=& \frac{y_i}{16\pi^2}\Bigl(3 \sum_j y_j^2+ 3 \sum_{\rho}y_{\rho}^2- \left(\frac{17}{12}g_Y^2+\frac{9}{4}g_2^2+8g_3^2 \right)+ \frac{3}{2}y_i^2 \nonumber \\
&& \qquad \qquad - \frac{3}{2} \sum_{\rho}y_{\rho}^2 |V_{i\rho}|^2\Bigr) - f_y\, y_i,
\label{buptype} \\
\beta_{y_{\rho}}&=& \frac{y_{\rho}}{16\pi^2}\Bigl(3 \sum_j y_j^2+ 3 \sum_{\alpha}y_{\alpha}^2- \left(\frac{5}{12}g_Y^2+\frac{9}{4}g_2^2+8g_3^2 \right) + \frac{3}{2}y_{\rho}^2 \nonumber \\
&& \qquad \qquad - \frac{3}{2}\sum_i y_i^2 |V_{i\rho}|^2 \Bigr) - f_y\, y_{\rho},
\label{bdowntype}
\eea
where a Latin index on a Yukawa coupling denotes an up-type Yukawa coupling, i.e., $u,c,t$, and a Greek index signals a down-type Yukawa coupling, i.e., $d,s,b$. The fact that the beta function for each Yukawa coupling is proportional to the respective Yukawa coupling is a consequence of the chiral 
symmetry in the quark sector. 
It signals that Yukawa couplings are not generated by the RG flow if they are set exactly to zero at some scale.

Next, we derive the flow of the CKM matrix elements, as in 
\cite{Sasaki:1986jv,Babu:1987im,Barger:1992pk,Kielanowski:2008wm}. To that end, we consider the off-diagonal elements in Eq.~\eqref{eq:YUflow}, which can be written as
\bea
&{}&\left((\partial_t U)U^{\dagger}\right)_{ij} \label{eq:ptUUdagger}\\
& =& -\frac{1}{16\pi^2}\frac{3}{2} \left(V^{\dagger}\cdot{\rm diag}(y_u^2, y_c^2, y_t^2)\cdot V \right)_{ij} \frac{ {\rm diag}(y_d^2, y_s^2, y_b^2)_j+ {\rm diag}(y_d^2, y_s^2, y_b^2)_i}{ {\rm diag}(y_d^2, y_s^2, y_b^2)_i- {\rm diag}(y_d^2, y_s^2, y_b^2)_j},\nonumber
\eea
where no summation over the repeated indices $i,j$ is performed. The nontrivial structure in the denominator arises from the commutator term in Eq.~\eqref{eq:YUflow}, which follows directly from the unitarity of $U$. We observe that the diagonal entries of the right-hand side of Eq.~\eqref{eq:ptUUdagger} are purely imaginary, due to Eq.~\eqref{eq:unitarityUD}. Therefore, the corresponding terms drop out when $\partial_t |V|^2$ is evaluated. Using  Eq.~\eqref{eq:ptUUdagger}, its hermitian conjugate, and the corresponding equation for $(\partial_t D)D^{\dagger}$ as well as its hermitian conjugate, we arrive at
\bea
\beta_{|V_{i\rho}|^2}&=& - \frac{3}{2}\Biggl(\sum_{\sigma, j\neq i} \frac{y_i^2+y_j^2}{y_i^2-y_j^2} y_{\sigma}^2 \left(V_{i\sigma}V_{j\sigma}^{\ast}V_{j\rho}V_{i\rho}^{\ast}+V_{i\sigma}^{\ast}V_{j\sigma} V_{j\rho}^{\ast}V_{i\rho}\right) \nonumber \\
&& \qquad + \sum_{j, \sigma\neq \rho}\frac{y_{\rho}^2+y_{\sigma}^2}{y_{\rho}^2 - y_{\sigma}^2}y_j^2 \left(V_{j\sigma}^{\ast}V_{j\rho}V_{i\sigma}V_{i\rho}^{\ast}+V_{j\sigma}V_{j\rho}^{\ast}V_{i\sigma}^{\ast}V_{i\rho}  \right) \Biggr).
\label{eq:betaV} 
\eea
We obtain the result that the new-physics contribution vanishes from the flow of the CKM-matrix elements. This is a consequence of the flavor-universality that we assume in Eqs.~\eqref{eq:YukawamatrixflowU} and \eqref{eq:YukawamatrixflowD}.

The parameters $y_i$, $y_\rho$ and $|V_{i\rho}|^2$
form a coordinate system for a subset of the original 
theory space, defined by $y_i\not= y_j$ and $y_\rho\not= y_\sigma$.
When one approaches a point with $y_i=y_j$ or $y_\rho=y_\sigma$
for some $i$, $j$, $\rho$, $\sigma$, the beta function \eqref{eq:betaV}
diverges.
The only exceptions are the points where the degenerate Yukawas
are zero. Such points can be continuously approached from
almost any direction. For example, the hypersurface defined by $y_u=y_c=0$ (with all other
Yukawas nonzero and nondegenerate) can be approached along any
line of the form $y_u(t)=a\, t$, $y_c(t)=b\,t$, as long as $a\not=b$.
This underlies, e.g., the analysis in \cite{Pendleton:1980as}, where this point is approached from a phenomenologically relevant direction.

If some up- or down-type Yukawa couplings are degenerate, 
there is no way to distinguish the corresponding quarks.
Then, there are additional unitary field redefinitions that can be used to remove some elements of the CKM matrix. Thus, the dimensionality of the space of essential couplings is actually \emph{smaller} on the hypersurfaces defined by these degeneracies.
In particular when all the up- or down-type Yukawas are equal,
the whole CKM matrix is redundant and can be set to one.
From Eqs.~\eqref{eq:YukawamatrixflowU} and \eqref{eq:YukawamatrixflowD}
we see that this condition is preserved by the flow.

The one-loop flow features two invariant combinations 
\cite{Harrison:2010mt,Feldmann:2015nia,W.:2017ciz}
\begin{equation}
I_{(1)}=\frac{{\rm Tr}(M_{U}M_{D})}{({\rm det}(M_{U}M_{D}))^{1/N_g}}, 
\quad  
I_{(2)}= {\rm Tr}((M_{U}M_{D})^{-1})({\rm det}(M_{U}M_{D}))^{1/N_g}.
\label{eq:rginv}
\end{equation}
As discussed in \ref{app:invariants}, the existence of these invariants 
implies linear relations between the beta functions for Yukawa couplings.
As we shall see in Sec.~\ref{sec:2gen}, these relations lead to the appearance of lines and planes of fixed points.

\section{ Two-generation model}\label{sec:2gen}

The two-generation system, which in our case consists of the second and the third generation, 
features only one independent mixing angle. 
It can be obtained by restricting
to the appropriate ``lower'' submatrix of the three-generation CKM matrix and implementing unitarity.
The matrix of the squared moduli of the CKM matrix elements is written,  leaving only one free parameter
$W$, as
\begin{equation}
 \Bigl[  \{ |V_{ij}|^{2} \} \Bigr] =
    \begin{bmatrix}
W & 1-W \\
1-W & W 
    \end{bmatrix} \, .
\label{CabibboWithW}
\end{equation}
This makes evident that $W$ is related to the mixing angle via $W=\cos^2 \theta_{23}$.

The beta function for the mixing parameter $W$ is given by
\bea
\beta_{W}&=&\frac{3}{16\pi^2} W \left (W-1\right)
\left[
(y_t^2+y_c^2)\frac{y_b^2-y_s^2}{y_t^2-y_c^2}
+(y_b^2+y_s^2)\frac{y_t^2-y_c^2}{y_b^2-y_s^2}\right] \, .
\nonumber \\ \label{eq:CKMrunning2gen}
\eea

\subsection{Fixed-point structure}
\label{sec:2GenFPstructure}

Assuming that the gauge couplings take fixed-point values at
$g_{Y \ast} = 4\pi \sqrt{4\, f_g \, / \, 41}$ and $g_{2 \ast} = 0 = g_{3 \ast}$,
we discuss the fixed-point structure of the Yukawa couplings and mixing parameter $W$ (details can be found in \ref{app:2gen}).
As was already observed in \cite{Pendleton:1980as},  
Eq.~\eqref{eq:CKMrunning2gen} implies an IR
attractive fixed point at $W=1$ as long as $y_t^2>y_c^2$ and $y_b^2>y_s^2$ (or $y_t^2<y_c^2$ and $y_b^2<y_s^2$). 
On the other hand, the fixed point at $W=0$ is IR repulsive.
This conclusion persists in our setting since the beta function for the CKM-matrix element does not contain the BSM contributions.

Due to the permutation symmetries $t \leftrightarrow c$ 
and $b\leftrightarrow s$, the fixed points always appear in quartets, the physically most interesting one being shown in
Table~\ref{tab:twogen}.

\begin{table}[ht]
\begin{center}
\begin{tabular}{@{}|c|cccc|c|@{}} \hline
\# & 
$y_{t \ast}^2/\left(\frac {15}{615} \pi^2\right) $  & 
$y_{c \ast}^2/\left(\frac {15}{615} \pi^2\right)$ & 
$y_{b \ast}^2/\left(\frac {15}{615} \pi^2\right)$ & 
$y_{s \ast}^2/\left(\frac {15}{615} \pi^2\right)$ & $W_{\ast} $  \\ \hline
 1a & $ 41 \left( f_g + 2 f_y \right) $ & $0$ & 
$  - 19 f_g + 82 f_y  $ & $0$ &  $0$  \\
 1b & $ 41 \left( f_g + 2 f_y \right) $ & $0$ &  
$0$ & $ - 19 f_g + 82 f_y  $ & $1$  \\
 1c & $0$ & $41 \left( f_g + 2 f_y \right) $ & 
$0$ & $  - 19 f_g + 82 f_y  $ &  $0$  \\
 1d & $0$ & $ 41 \left( f_g + 2 f_y \right) $ & 
$  - 19 f_g + 82 f_y  $ & $0$ &  $1$  \\
\hline
\end{tabular}
\end{center}
\caption{\label{tab:twogen} Displayed is an example for a quartet of fixed-point solutions, for details see
text. (Note that a common factor in the $y_{i \ast}^2$ has been divided out.)}
\end{table}

The solutions 1a, 1b, and the solutions 1c, 1d 
are related among themselves by the intergenerational permutation $b\leftrightarrow s$, which changes $W_{\ast} =0$ to $W_{\ast} =1$.
\\
Case 1a is the only phenomenologically viable one within this quartet, by the following argument: We find that 
$g_{Y \ast}^2 = \frac{96}{41} f_g \pi^2$ requires $f_g > 0$.
In order to avoid negative values for the $y_{\mathrm{i} \ast}^2$, it must hold that $f_y \geq \frac{19}{82} f_g \geq 0$. Due to the poles in the beta function for $W$, the ordering of Yukawas cannot change under the flow.
Thus, a phenomenologically viable fixed point must realize
$y_{t \ast}^2 > y_{c \ast}^2$ and 
$y_{b \ast}^2 > y_{s \ast}^2$.
The first condition excludes the solutions 1c and 1d, the second 1b and 1c.
As further discussed in \ref{app:2gen}, all other FPs can also
be discarded as phenomenologically uninteresting.
In the following subsection, we will discuss 
the phenomenology of the FP 1a in some more detail.

\subsubsection{Without mixing}

It is instructive to first analyze  the beta functions for the Yukawa couplings without mixing. For the present 
discussion we choose the case without permutation, i.e., $W=1$. The beta functions for the top and the
bottom Yukawa then read
\bea
\beta_{y_t}&\!\!\!=\!\!\!&\frac{y_t}{16\pi^2}\left( \frac{9}{2}y_t^2 + \frac{3}{2}y_b^2+ 3y_s^2 + 3y_c^2 - \frac{9}{4}g_2^2-8 g_3^2-\frac{17}{12} g_Y^2\right)-f_y\, y_t,\label{eq:betayt2nomix}\\
\beta_{y_b}&\!\!\!=\!\!\!&\frac{y_b}{16\pi^2}\left( \frac{9}{2}y_b^2 + \frac{3}{2}y_t^2+ 3y_s^2 + 3y_c^2 - \frac{9}{4}g_2^2-8 g_3^2-\frac{5}{12} g_Y^2\right)-f_y\, y_b.
\label{eq:betayb2nomix}
\eea
Correspondingly, those for the strange, $y_s$, and the charm, $y_c$, are obtained by the pairwise interchange $(t,b)\leftrightarrow (c,s)$.

Choosing $y_{s\, \ast}=0=y_{c\, \ast}$ admits the fixed points for the top and bottom Yukawa
couplings which have been discussed in Sec.~\ref{sec:UVcomp} for the one-generation case.
Concentrating on the non-trivial fixed point given in the last row of Eqs.~\eqref{eq:fptby}, $y_c$ corresponds to an IR repulsive direction, and $y_s$ 
to an IR attractive one. Combining this with the fact that vanishing mixing and no permutation already constitutes a partial IR attractive fixed point, this fixed point produces a roughly viable phenomenology: 
Top and bottom masses and the Abelian gauge coupling are predicted in the vicinity of their measured 
values as in Ref.~\cite{Eichhorn:2018whv} because the corresponding flows do not deviate significantly from 
those in the system with only the third generation.
The measured mass of the charm can be accommodated in such a setting but is not predicted. 
The above assumption of no mixing is self-consistent because the mixing matrix remains
exactly diagonal at all scales. Due to the irrelevance of the strange Yukawa coupling, this fixed point would imply a massless strange quark in contradiction to the observed small ratio 
$m_s/m_t\approx6\cdot 10^{-4}$ valid for scales well above one GeV\footnote{The present analysis is only a leading-order one in perturbation theory. Therefore, the dynamically generated quark masses present in the strongly interacting domain of QCD, {\it i.e.}, at sub-GeV scales, cannot be described at all. Thus, in the
present context, infrared values of scales still mean multi-GeV scales. 
For most calculations, we terminated the flow towards the infrared at a scale $k=173$ GeV.}.
As a leading-order result, the prediction $m_s/m_t =0$ might appear to be a reasonable first approximation of the tiny ratio $m_s/m_t\approx6\cdot 10^{-4}$. However, the additional global symmetry rotating the strange-quark at $y_s=0$ implies that $y_s=0$ is a symmetry-protected point. Thus, the values $m_s/m_t\approx6\cdot 10^{-4}$ and $m_s/m_t=0$ are \emph{qualitatively} distinct. To drive $y_s$ away from zero, additional new physics 
as, e.g., mixing, is necessary. Whether the present setting can be extended by a suitable new-physics correction that corrects the prediction $m_s/m_t=0$ to $m_s/m_t\approx6\cdot 10^{-4}$ is an intriguing open question.

\subsubsection{With mixing}
\label{sec:2genWithMixing}
Taking into account the running of the mixing parameter, the system of beta functions for the Yukawa
couplings is modified to
\bea
\beta_{y_t}&=&\frac{y_t}{16\pi^2}\Biggl( \frac{9}{2}y_t^2 + \frac{3}{2}y_b^2\left( 2- W \right)+ 
\frac{3}{2}y_s^2 \left( 1+ W \right)+ 3y_c^2 - \frac{9}{4}g_2^2-8 g_3^2
\nonumber \\ && \qquad \quad 
-\frac{17}{12} g_Y^2\Biggr)
 -f_y\, y_t,\label{eq:betayt2mix}
 \eea
 \bea
\beta_{y_b}&=&\frac{y_b}{16\pi^2}\Biggl( \frac{9}{2}y_b^2 + \frac{3}{2}y_t^2\left( 2- W \right)+ 3y_s^2 + \frac{3}{2}y_c^2\left( 1+ W \right) - \frac{9}{4}g_2^2-8 g_3^2
\nonumber \\  && \qquad \quad -\frac{5}{12} g_Y^2\Biggr)
 -f_y\, y_b.\label{eq:betayb2mix}
\eea
Again, the beta functions for the quarks of the second generation are obtained by the interchange $(t,b)\leftrightarrow (c,s)$.

Even at $y_{s\, \ast}=0=y_{c\, \ast}$ the inclusion of the mixing terms has a crucial consequence 
for the scaling exponents: At the $W=0$ fixed point, the size of the screening contribution of the top Yukawa to the 
beta function for the strange Yukawa is halved in comparison to the case with $W=1$.
This change is sufficient to make the fixed point at $y_s=0$, $W=0$ IR repulsive in $y_s$.
Accordingly, finite IR values of the strange Yukawa can be accommodated. 
Due to the mixing, the fixed-point values of the top and bottom Yukawa change to (in comparison with Eq.~\eqref{eq:fptby})
\bea
y_{t\, \ast}=\frac{4\pi}{\sqrt{15}}\sqrt{2f_y+f_g} , \quad 
y_{b\, \ast}= \frac{4\pi}{\sqrt{615}}\sqrt{82 f_y-19 f_g} \, ,
\label{eq:nonzerotb}
\eea
where $f_g$ is fixed in terms of the IR-value of the gauge coupling. 
Requiring $y_b$ to be relatively small, implies that $f_y$ is only marginally larger
than $(19/82)  f_g$.
This results in a fixed-point value 
$y_{t\, \ast}$ in the vicinity of
$\frac{4\pi \sqrt{f_g(1+19/41)}}{\sqrt{15}}\approx 0.39$. This should be compared to the fixed-point value 
$y_{t\, \ast}\approx 0.27$ in \cite{Eichhorn:2018whv}. 
While the flow is slightly different in the present case, all changes to the beta function for the top 
Yukawa coupling are sub-leading. Therefore, the larger fixed-point value $y_{t\, \ast}$
translates into a somewhat larger top mass.
\\

\begin{figure}
\centering
\includegraphics[width=0.7\linewidth]{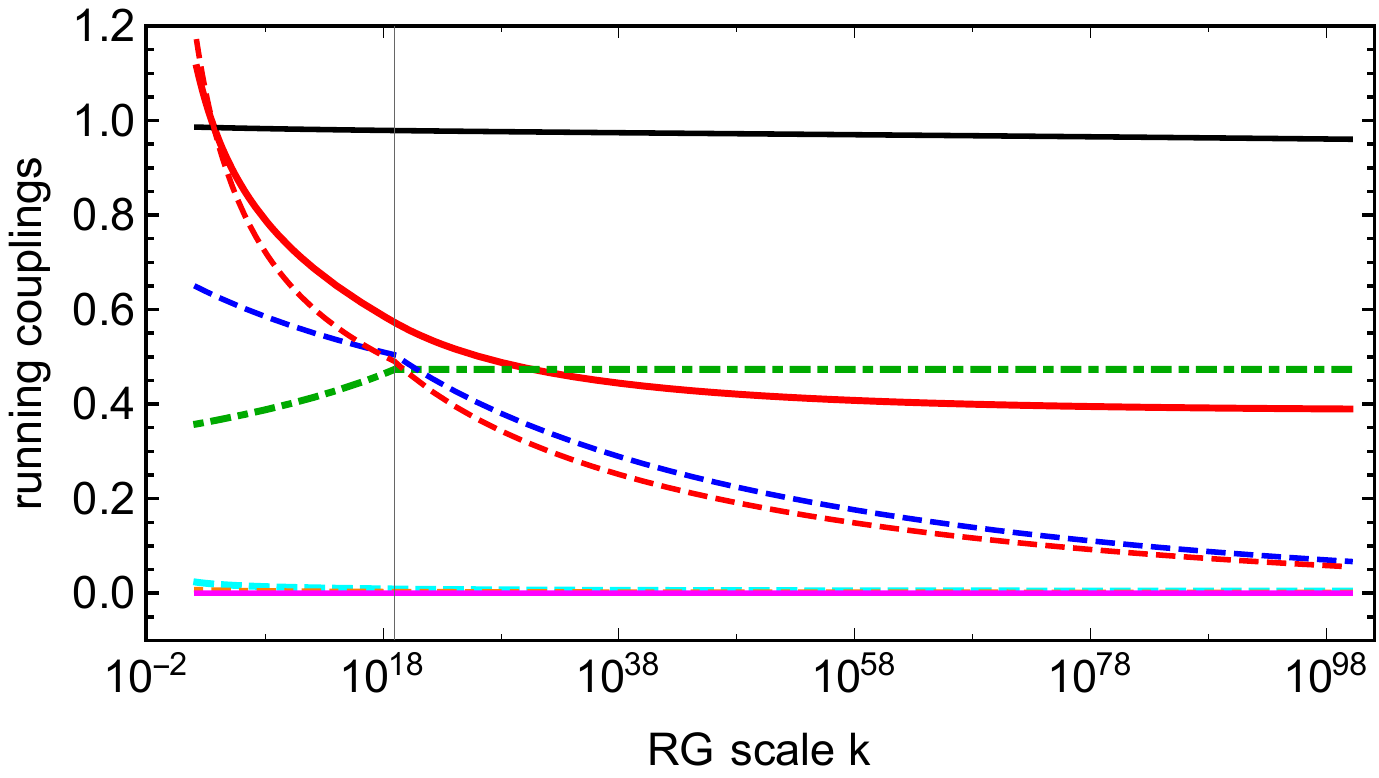}\
\includegraphics[width=0.7\linewidth]{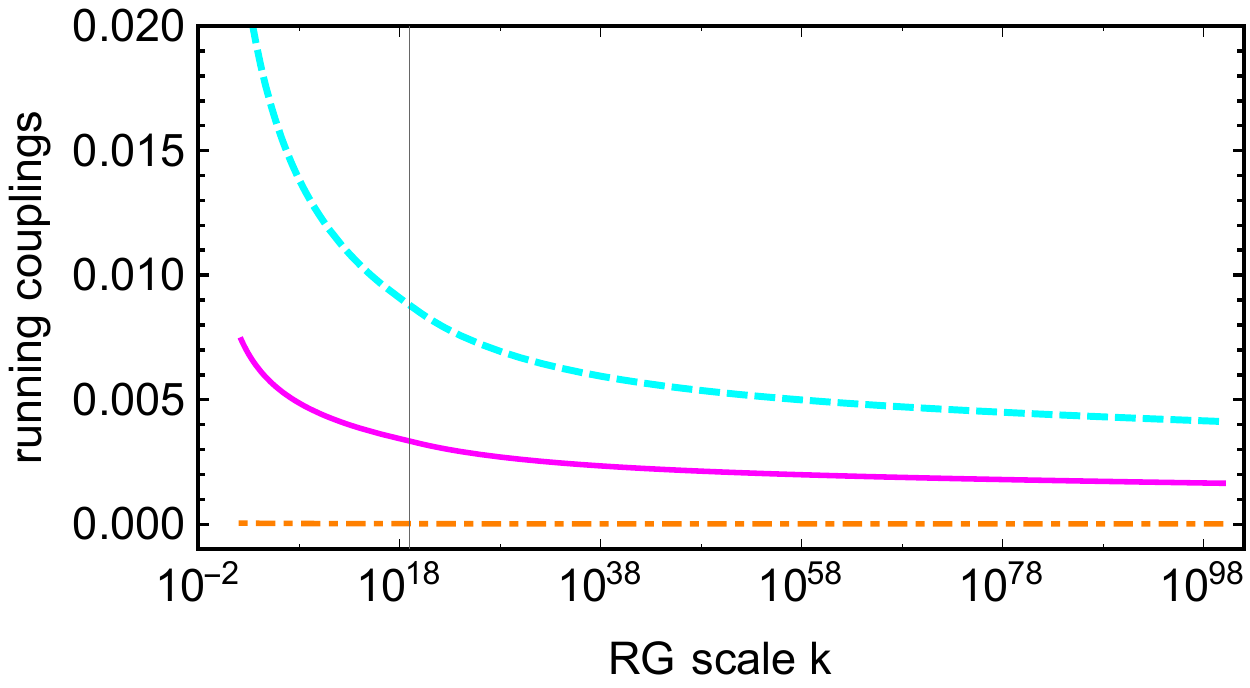}
\caption{\label{fig:running_two_gens}We show the running on a trajectory emanating from the 
asymptotically safe fixed point 1a of Table \ref{tab:twogen}, with top (red, continuous), bottom (cyan, dashed), charm (magenta), 
strange (orange, dot-dashed), Abelian gauge (green, dot-dashed), strong gauge (red, dashed) and 
weak gauge (blue, dashed) couplings. The mixing parameter $W$ (black, continuous) starts running from 
zero at very high scales to reach values close to $W=1$ at the electroweak scale. 
The top mass is overestimated significantly on this trajectory. }
\end{figure}

In Fig.~\ref{fig:running_two_gens} we show the  RG flow of the Yukawa couplings starting from
fixed point 1a of Table \ref{tab:twogen}, i.e., the fixed point with mixing discussed above.
The IR values of the couplings are extracted at $k_{\rm IR} = 173\, \rm GeV$. The SU(2) and SU(3) gauge coupling, which are aligned with relevant directions of the fixed point, take the values $g_2(k_{\rm IR})\approx 0.648$ and $g_{3}(k_{\rm IR})\approx 1.17$. We choose $f_g=9.7 \cdot 10^{-3}$, resulting in $g_Y(k_{\rm IR})\approx 0.358$. The IR value of the CKM matrix element is $W(k_{\rm IR})\approx 0.999$. Further, we choose $f_y =2.248 \cdot10^{-3}$. The IR values of the Yukawa couplings are $y_t(k_{\rm IR})\approx1.1,\, y_b(k_{\rm IR})\approx2.4\cdot 10^{-2},\,y_c(k_{\rm IR})\approx7.4\cdot 10^{-3},\,y_s(k_{\rm IR})\approx5.6 \cdot 10^{-4}$. A tree-level conversion to current quark masses suffices for our purposes, with the exception of the top mass, for which we do a one-loop matching  \cite{Tarrach:1980up,Bohm:1986rj,Hempfling:1994ar}. 
It  gives $M_t \approx 193\, \rm GeV$, $M_b \approx 4.2\, \rm GeV$, $M_c \approx 1.3\, \rm GeV$ and $M_s \approx 97\, \rm MeV$. The quark masses agree with their measured SM-values, except for the top quark. In summary, allowing mixing in the quark sector provides the possibility to accommodate the measured strange-mass, while changing the fixed-point value for the top so that the prediction of its mass increases.

\subsection{Upper bound on the strange mass}\label{subsec:strangeupperbound}
\begin{figure}[!t]
\centering
\includegraphics[width=0.4\linewidth]{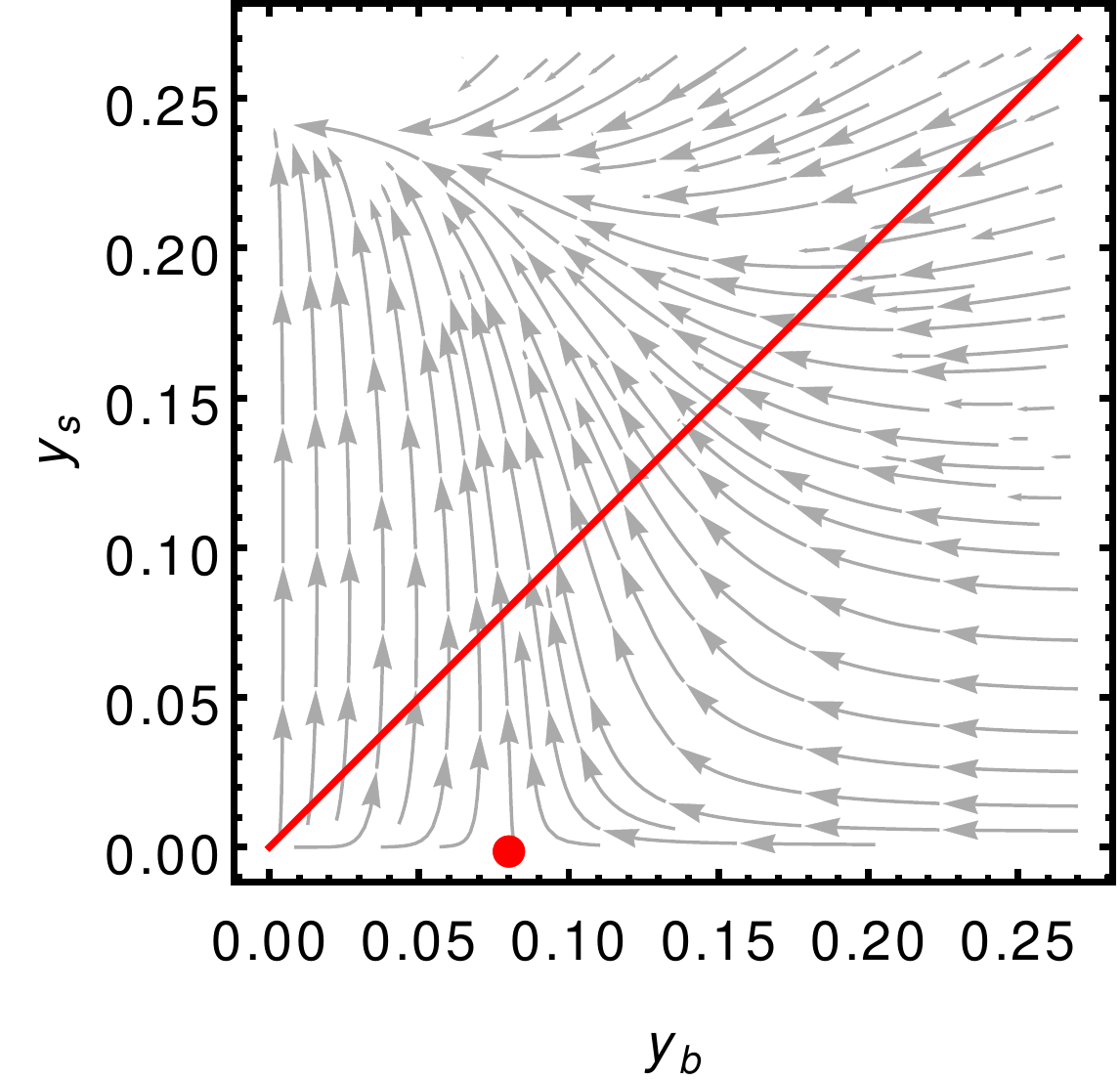}
\quad\quad\quad
\includegraphics[width=0.4\linewidth]{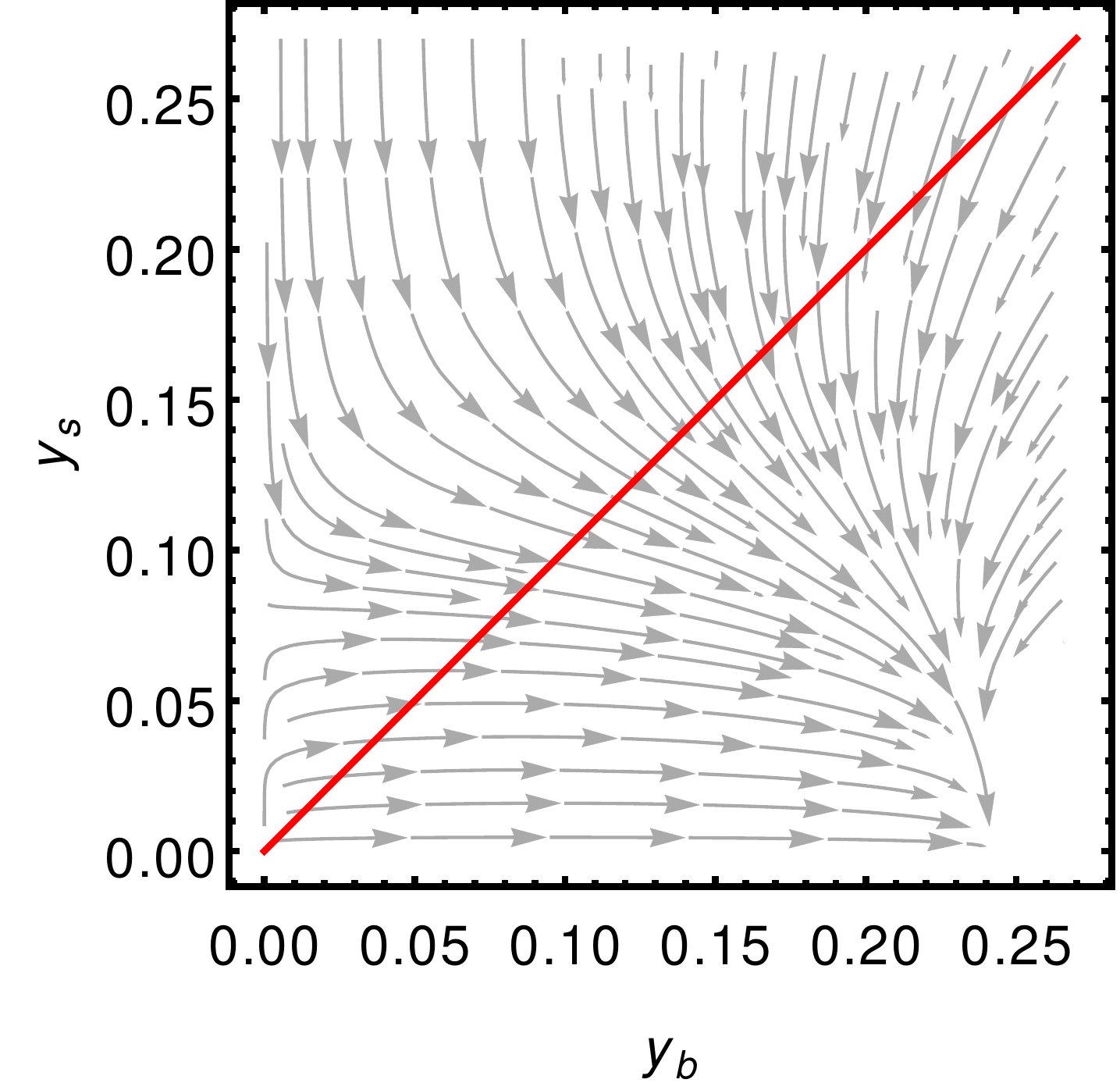}
\caption{\label{fig:streamplot} 
RG flow (towards the IR) in the $y_b, y_s$-plane for $W=0$ (left panel) and for $W=1$ (right panel). All other couplings are set to their values at 
fixed point 1a. We choose
$f_y = 2.55\times10^{-3}$
such that the fixed point 1a (red dot) in $y_{b/s}$ is clearly visible. For values close to the 
red line (pole in $\beta_W$), the flow transitions from the left to the right panel which gives an upper bound for $y_s$, even if the flow emanates from fixed point 1a.
}
\end{figure}
Since the flow of the CKM matrix element  has a pole at $y_s=y_b$, the strange-Yukawa coupling is bounded by the bottom Yukawa coupling at all scales, $y_s(k)<y_b(k)$. The equivalent  statement holds for the charm and the top Yukawa coupling. Interestingly, at fixed $f_y$, an even stronger bound on $y_s$ (and $y_c$) arises once the other relevant parameters of the system are fixed.

This is a consequence of an interplay between two fixed points. In fact, the flow exhibits a crossover behavior from an ultraviolet fixed point to an infrared fixed point, with the latter limiting the IR value of $y_s$.
For the purpose of demonstrating the underlying mechanism, we consider a flow
starting from fixed point 1a with 
$y_{b\, \ast} \neq 0, y_{s\, \ast}  =0, W_\ast=0$ in the UV, and reaching the fixed point 1b, with 
$y_{b\, \ast}  = 0, y_{s\, \ast} \neq 0, W_\ast=1$ in the IR.
For the mechanism, it is a key feature that the former fixed point is IR attractive in $y_b$ and the latter in $y_s$ and $W$. 

At the above two fixed points 1a and 1b, $g_{Y\,\ast1a}\equiv g_{Y\,\ast1b}$ and $y_{t\,\ast1a}\equiv y_{t\,\ast1b}$ correspond to irrelevant directions. To simplify the discussion, we can therefore assume that $g_Y$ and $y_t$ remain constant for RG flows transitioning between the two fixed points.

As fixed point 1a is IR repulsive in $y_s$, the flow towards the IR results in a growth of $y_s$.
At an intermediate scale, $y_s$ becomes comparable to $y_b$ (close to the diagonal in Fig.~\ref{fig:streamplot}). 
At this point, the denominator in the beta function of $W$ becomes small and $W$ flows towards 1. 
One is now close to the diagonal in the right panel of
Fig.~\ref{fig:streamplot}, and $y_s$ can only decrease.
Due to the transition of $W$ from 0 to 1, the flow of $y_s$ changes its character and becomes dominated by an \emph{IR attractive} fixed point. In other words, one approaches the domain of attraction of fixed point 1b which is IR 
attractive for $y_s$. Hence, the value for $y_s$ `freezes'. 
As a first conclusion, this results in an upper bound on $y_s (k_{\rm IR})$.

Next, we consider the value of this bound: Neglecting contributions from the non-Abelian gauge couplings, $\left. y_{s\, \ast} \right|_{1b} = \left. y_{b\, \ast} \right|_{1a}$, the low-energy value of the strange Yukawa coupling is bounded by the high-energy value of the bottom Yukawa coupling. Taking into account the non-Abelian gauge couplings results in sub-leading corrections to this equality.

\begin{figure}[!t]
\centering
\includegraphics[width=0.85\linewidth]{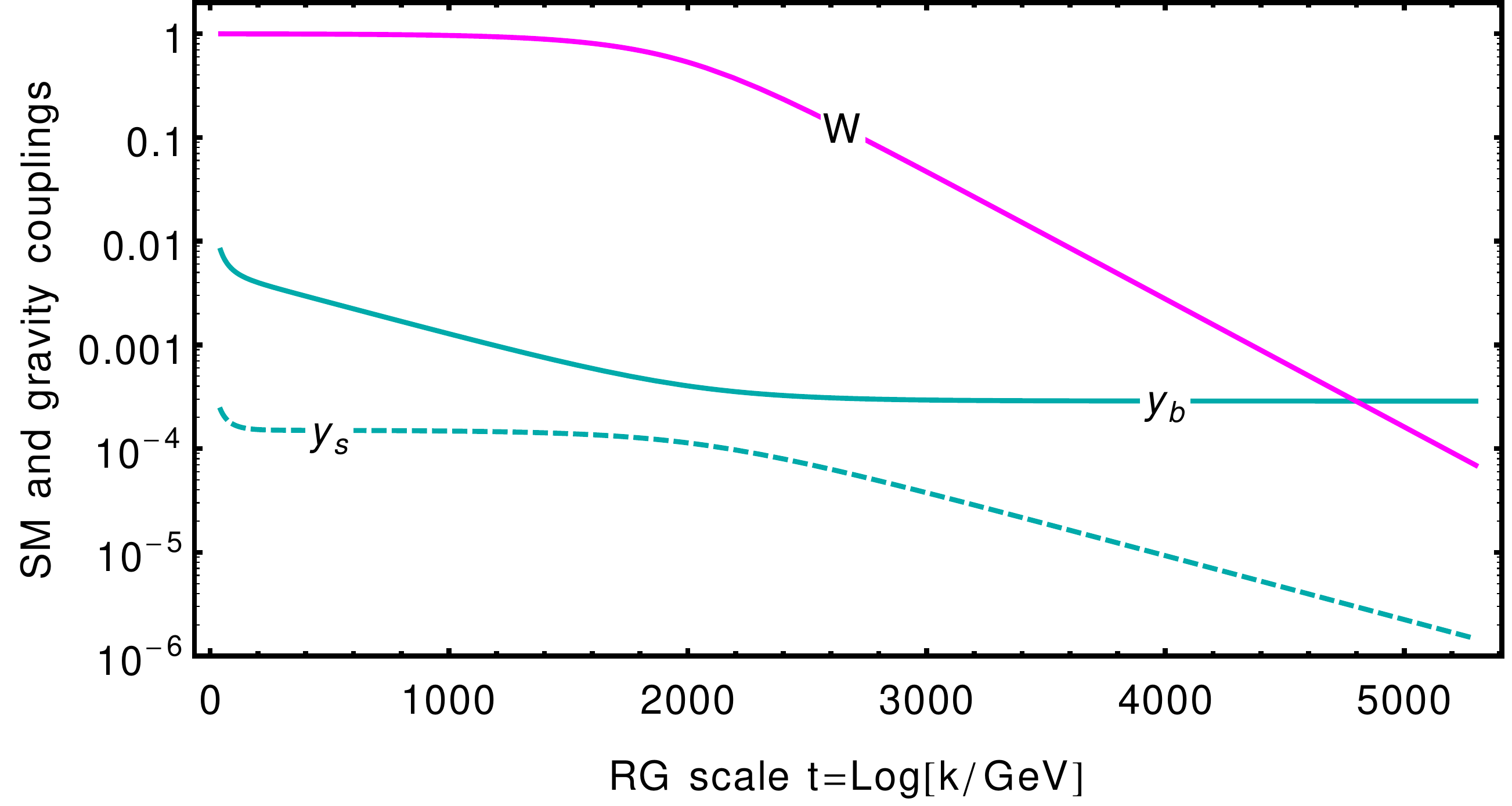}
\caption{\label{fig:upperBoundOnYs} We fix $f_y = 2.247566\times10^{-3}
$, $f_g=9.7\times 10^{-3}$ and choose the non-Abelian gauge couplings such that their IR values match observations. Among the three couplings shown, $y_s$ and $W$ are relevant at the UV fixed point. As $y_s$ increased towards the IR, it triggers a transition of $W$ to the neighborhood of the fixed point at $W_{\ast}$. In turn, this results in a change of the flow of $y_s$, which is drawn towards an IR fixed point. The same fixed point is IR repulsive in $y_b$. 
As a consequence,  below the transition at roughly $k = e^{2000}\, \rm GeV$, $y_s$ freezes out and $y_b$ starts to increase, until both reach phenomenologically viable values in the IR. 
This leads to an upper limit for $y_s$ which is close to the UV fixed point value of $y_b$.
}
\end{figure}

To exemplify
this behavior,
we plot the flow of $y_b$, $y_s$, and $W$ emanating from the fixed point (1a) in Fig.~\ref{fig:upperBoundOnYs}. $g_Y$, $y_b$, and $y_t$ are (closely aligned with) 
IR-attractive directions of fixed point (1a) and are thus fixed in terms of $f_g$ and $f_y$. We choose $f_g=9.7\times 10^{-3}$ and 
$f_y = 2.247566\times10^{-3}
$ 
 such that $g_Y$ and $y_b$ deviate less than $1\%$ from the infrared tree-level values \cite{Tanabashi:2018oca}. As a 
consequence, the one-loop matched top quark is about 10\% too heavy, cf.~Sec.~\ref{sec:2genWithMixing}. We also fix the IR-repulsive directions of fixed point 1a, i.e., $y_c$ and $W$ (as well as $g_2$ 
and $g_3$), such that they deviate less than $1\%$ from the IR tree-level values. These values entail a running of $y_s$ close to the upper bound and a transition between the two fixed points, $W_{\ast}=0$ and $W_{\ast}=1$, at around $k = 10^{1000}\, \rm GeV$.
\\

\subsection{Theoretical viability of the fixed-point solutions}
\label{sec:theoreticalViability}
\begin{figure}[!]
\includegraphics[width=0.45\linewidth]{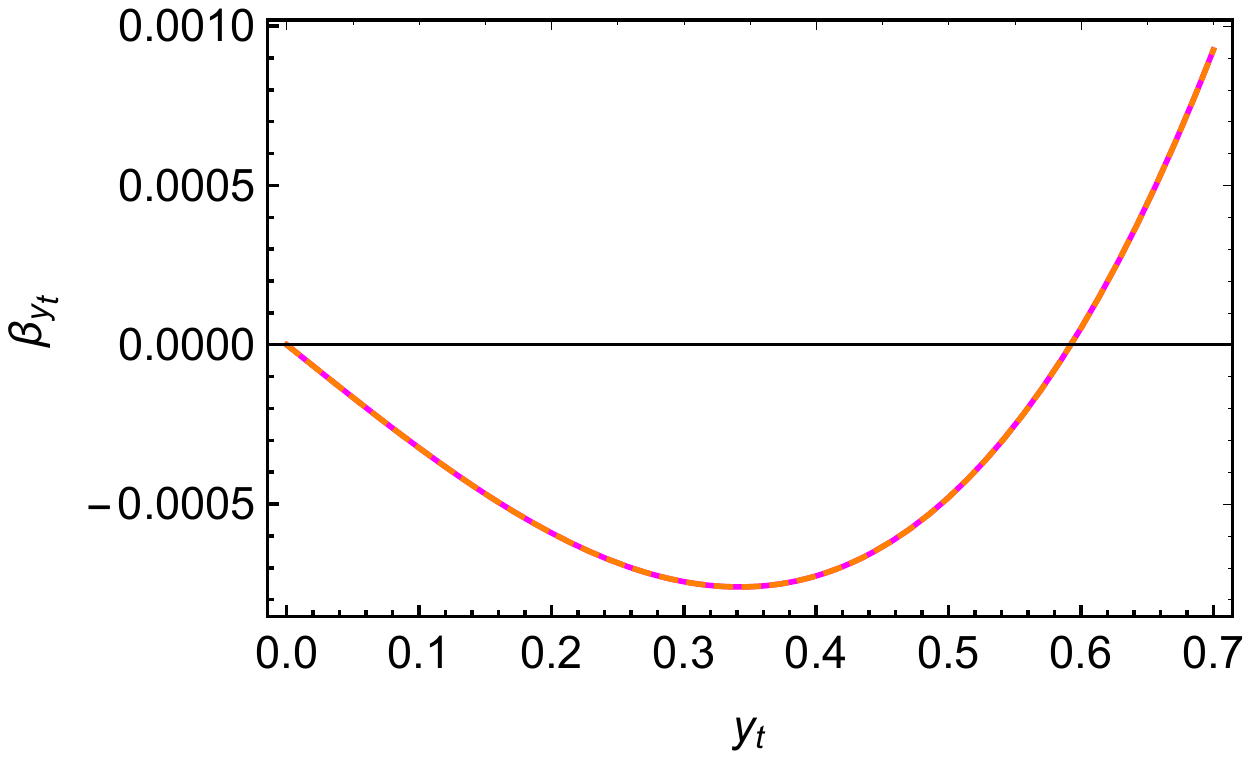}\quad \includegraphics[width=0.45\linewidth]{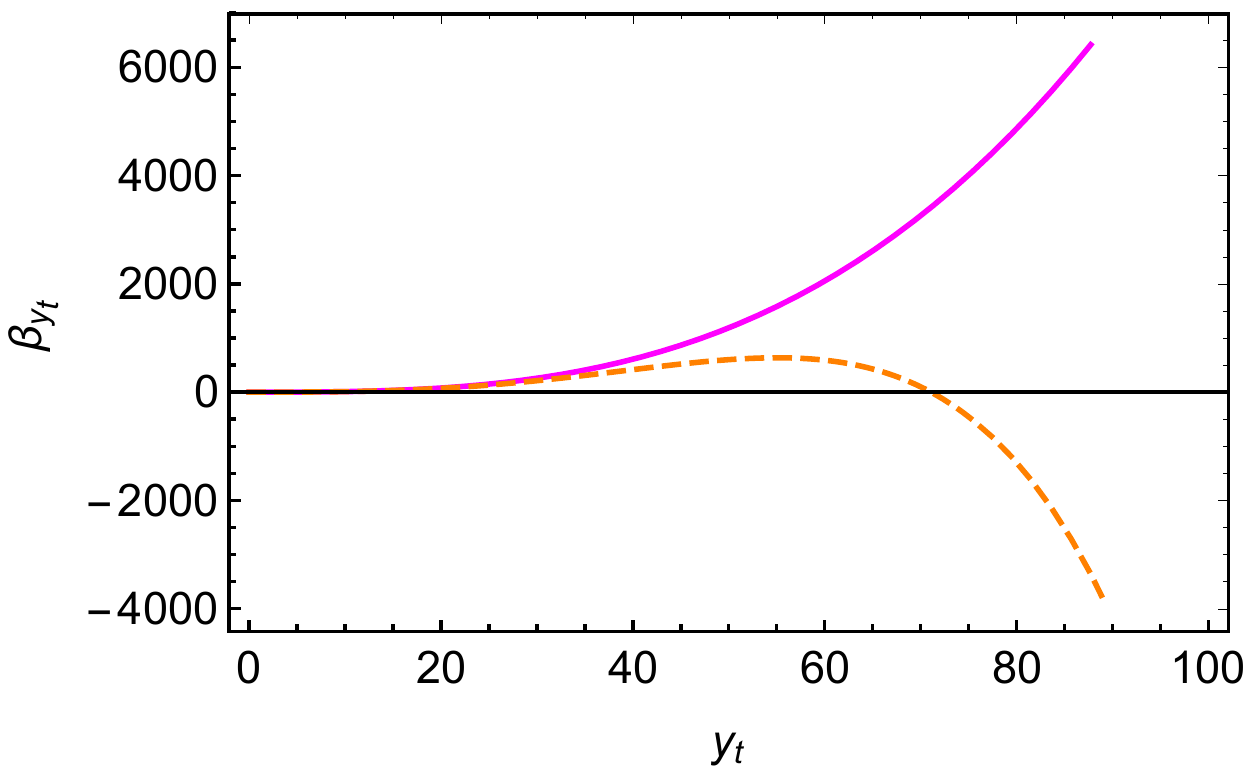}
\caption{\label{fig:betatopillu}
We show the one- and two-loop top beta function
$\beta_{y_t}$ (magenta and orange-dashed, respectively), with all other couplings vanishing, and $f_y=1/300$. Near the one-loop FP, the one- and two-loop results are actually indistinguishable, and the difference only arises at much larger values of the coupling. The additional zero that arises from the two-loop beta function is far beyond the validity of the perturbative approximation.
}
\end{figure}

Polynomial beta functions generically admit many real solutions, some of which may be artifacts of the approximation, whereas others are genuine features of the theory. Understanding the mechanism that generates the fixed points may help to distinguish the latter from the former.

For example, in the case of the Yukawa couplings,
we see from Eqs. (\ref{eq:betayt2nomix},\ref{eq:betayb2nomix})
that the FPs arise from a balance of
an antiscreening contribution, which is a combination of the term $f_y$ and the contribution of the Abelian gauge coupling, with a screening contribution, arising from quantum fluctuations of the fermions and the Higgs. 
If we take $f_y\to 0$, then for fixed gauge couplings there could
still be a cancellation. However, when we also let $f_g\to 0$,
the gauge couplings themselves go to zero.
Thus also the fixed point for the Yukawas must
go to zero in this limit.
Indeed this is the case for all fixed points studied here. As we take $f_y\rightarrow 0$, $f_g\rightarrow 0$, all fixed points merge with the free fixed point, as can be seen, e.g., from the expressions provided in \ref{app:2gen}.

As long as $f_g\ll1$ and $f_y \ll1$, these fixed points lie close enough to the free fixed point, so that higher-order loop corrections play a subleading role. In particular, additional zeroes of the beta functions might arise but are likely to be artifacts of the approximation, as they must lie at larger values of the couplings and are therefore more likely to be unstable under the extension to higher loop orders. 
For the top Yukawa, the situation is illustrated in Fig.~\ref{fig:betatopillu}.
\\
The situation is different in the case of lines of fixed points. These occur at one loop as a consequence of the two quantities conserved under the RG flow, cf.~Eq.~\eqref{eq:rginv}. Beyond one loop, the absence of conserved quantities breaks the corresponding degeneracy of couplings and the lines will reduce to isolated fixed points or vanish completely. 
We will, therefore, exclude lines of fixed points from our analysis.
In summary, throughout this paper, we will report only the one-loop results. For the phenomenologically interesting fixed points and all plotted RG-flows towards the IR, we have checked that two-loop corrections are indeed subleading, see \ref{app:twoloop} for the full set of two-loop beta functions.

\section{Three families}\label{sec:3gen}
We now turn to the phenomenologically interesting case of three generations.
In Sec.~\ref{sec:hierarchiesFromRGflow}, we discuss to what extent RG flows can generate hierarchy patterns of Yukawa couplings. In Sec.~\ref{sec:mathematicalFPstructure}, we analyze the overall mathematical fixed-point structure, while Sec.~\ref{sec:topDrive} is devoted to RG flows in the heavy-top limit. The latter is approximately realized by the actual RG trajectory describing the Standard-Model and is sufficient to understand most aspects of scale invariance for the CKM sector of the SM. We discuss the resulting phenomenological implications in Sec.~\ref{sec:threeGenPheno}.

\subsection{Hierarchies from the Renormalization Group flow}
\label{sec:hierarchiesFromRGflow}
Without any mixing or distinct charges, the fixed-point structure of an $N_g$-generation system of Yukawa couplings (i.e., $2N_g$ fermions) straightforwardly extends the one-generation case, cf.~Fig.~\ref{fig:Yukawa_schematic_plot}: Each Yukawa coupling can either be vanishing or non-vanishing at the fixed point. This results in $2^{2N_g}$ possible fixed points. While a single fixed point can exhibit distinct values of the different Yukawa couplings, the overall fixed-point structure is symmetric under the exchange of any two fermions and their respective Yukawa couplings, at least at $g_{Y\, \ast}=0$. At  $g_{Y\, \ast}\neq0$, the overall fixed-point structure is symmetric under exchanges of up-type and of down-type quarks, only.
\\
The possible RG-flows (cf.~Fig.~\ref{fig:Yukawa_schematic_plot} for the example of $N_g=1$) connecting the various fixed points at which at least one Yukawa coupling is non-vanishing, constitute a $(2N_g - 1)$-dimensional hypersurface in coupling space. This hypersurface is IR-attractive and cannot be crossed by the RG flow. 
It, therefore, separates the part of theory space in which UV-complete theories are possible (i.e., the interior of the hypersurface) from non-fundamental theories (outside the boundary). Perturbative initial conditions are generically attracted to the boundary surface towards the IR. 
\\

\begin{figure}[!t]
\centering
\includegraphics[width=0.49\linewidth]{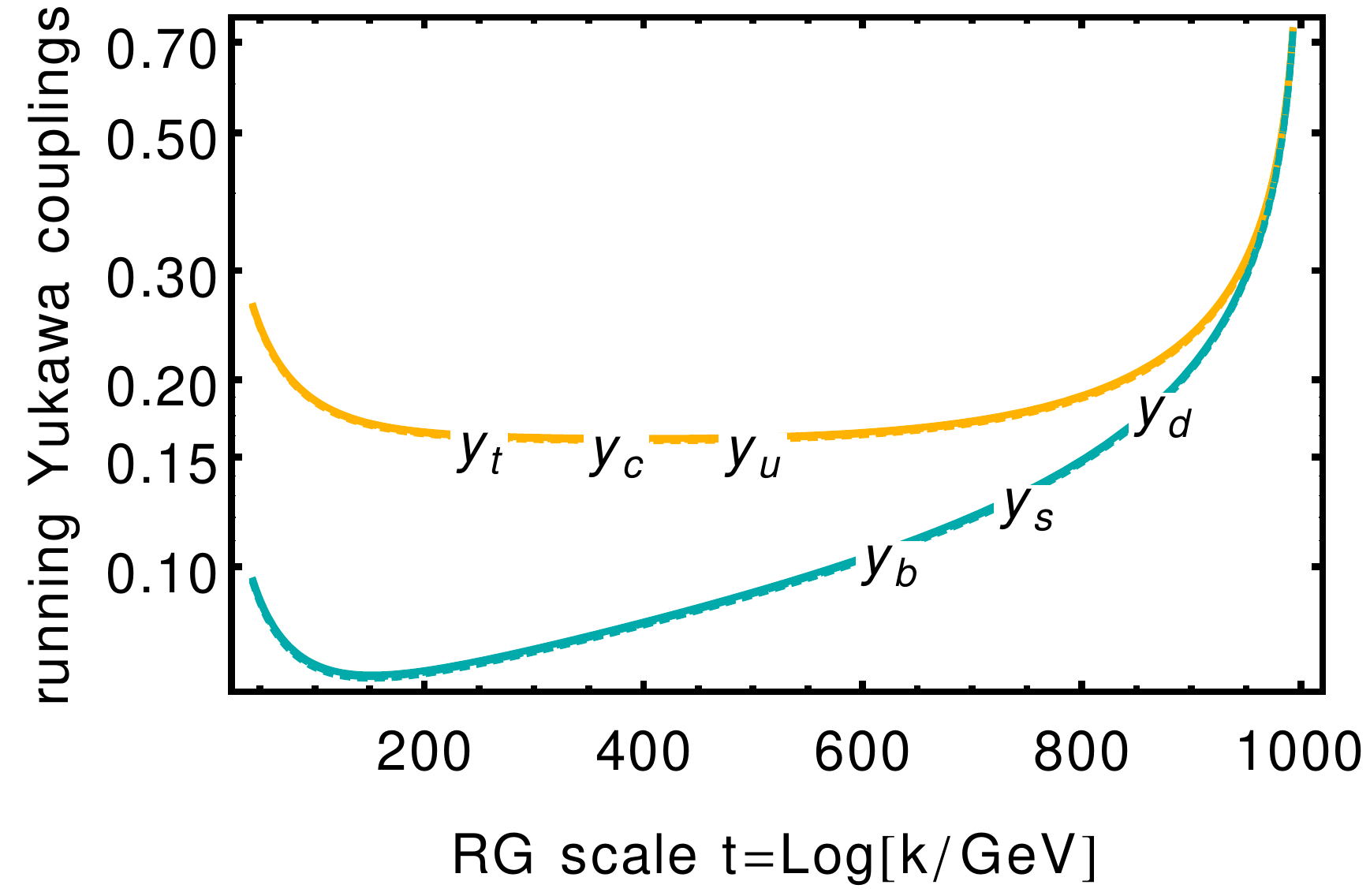}
\hfill
\includegraphics[width=0.49\linewidth]{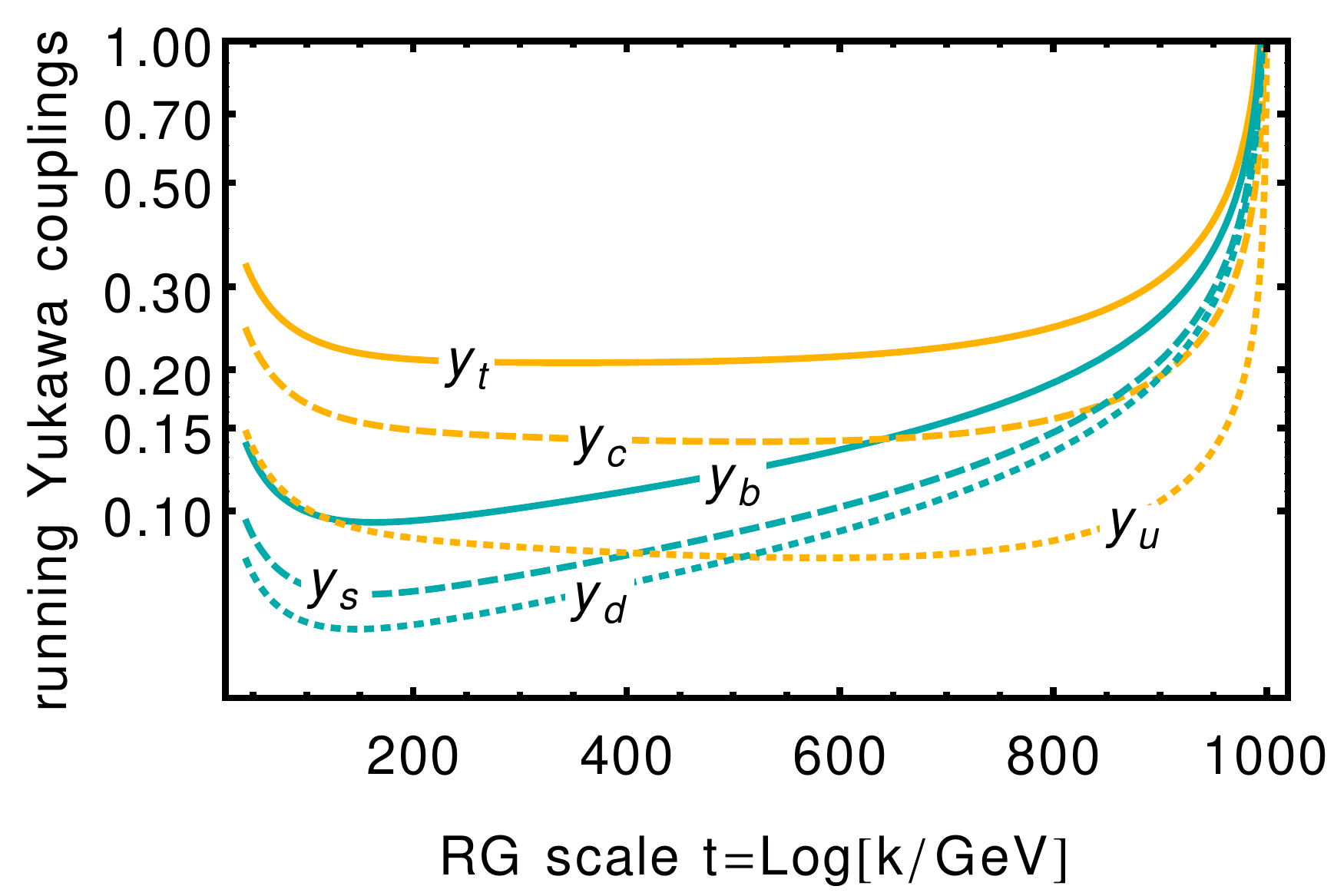}
\caption{\label{fig:3genRunning_developOrder} RG flow of Yukawa couplings for $f_y=0$ and a diagonal CKM matrix. Left-hand panel: no initial hierarchy, i.e., $y_i=2$ for all quarks. Right-hand panel: initial hierarchy in the up-type couplings only, i.e., $y_t=4>y_c=2>y_u=1$ and $y_b=y_s=y_d=3$.}
\end{figure}

The boundary hypersurface is rendered non-symmetric under the exchange of up-type (hypercharge $+2/3$) and down-type (hypercharge $-1/3$) quarks by their distinct hypercharges. For non-vanishing U(1) gauge coupling, this results in a dynamically generated hierarchy between the two differently charged types of Yukawa couplings \cite{Eichhorn:2018whv}.  Even for initial conditions outside the boundary hypersurface with (approximately) equal Yukawa couplings, the RG-flow develops a hierarchy between the up-type and the down-type Yukawas, cf.~left-hand panel in~Fig.~\ref{fig:3genRunning_developOrder}.

In general, the RG flow drives Yukawa couplings of fermions with a larger charge (in magnitude) to larger values in comparison to those of fermions with a smaller charge, cf.~gauge contributions in Eqs.~\eqref{buptype}-\eqref{bdowntype} and, in particular, the different hypercharge contributions\footnote{
The dynamical generation of hierarchies generalizes to non-Abelian gauge groups. Their 1-loop contributions to the beta functions of Yukawa couplings are proportional to the second Casimir of the fermionic representation. Hence, non-Abelian gauge groups can imprint hierarchies amongst Yukawa couplings whenever the second Casimirs of the respective fermionic representations differ. In the SM this can only occur between lepton and quark Yukawas and will be discussed elsewhere.
}.

The transmission of hierarchies between quark-Yukawa couplings of different generations depends on the CKM elements.
Without any initial hierarchy, i.e., for near-degenerate Yukawa couplings\footnote{The exactly degenerate case features inessential CKM elements and will not be discussed here.}, no hierarchy between Yukawa couplings of equally-charged quarks is generated by the RG flow. This holds independently of the values of the CKM elements. It is a consequence of unitarity, cf.~Eq.~\eqref{buptype}-\eqref{bdowntype}, where the CKM-dependent terms sum to similar values for different down-type Yukawas if the up-type Yukawas are all approximately equal.
\\
If this is not the case, partial hierarchies can be transmitted by the RG flow. For instance, if only the up-type Yukawas exhibit an initial hierarchy while the down-type Yukawas are (approximately) equal, the hierarchy is transmitted to the down-type sector by the RG flow. The efficiency of such a transmission depends on the values of the CKM-matrix elements. For instance, the special CKM configuration for which all $|V_{t\rho}|^2$, all $|V_{c\rho}|^2$, and all $|V_{u\rho}|^2$ are equal for $\rho=(d,s,b)$, respectively, does not mediate any hierarchies.
However, such special CKM configurations are generically not fixed points of the RG flow themselves, cf.~Secs.~\ref{sec:mathematicalFPstructure} - \ref{sec:threeGenPheno}. %
\\

In conclusion, the RG flows of the Yukawa sector can result in the emergence of nontrivial structures in the IR.
The following mini\-mal initial structure is required so that the RG flow can generate a full phenomenologically interesting hierarchy pattern in the quark sector: At least one initial hierarchy between arbitrary pairs of Yukawa couplings for each pair of generations is necessary. The right-hand panel in Fig.~\ref{fig:3genRunning_developOrder} depicts initial conditions for which the up-type Yukawas already exhibit a hierarchy while the down-type Yukawas do not. The up-type hierarchy is transmitted to the down-type sector by the RG flow. We note that we do not observe a mechanism to invert the hierarchy within the lightest generation.

\subsection{Fixed-point structure}
\label{sec:mathematicalFPstructure}

We do not attempt a full solution of the fixed-point equations for Yukawa couplings and CKM elements simultaneously, as the corresponding system of equations is too large to be easily tractable by standard computer algebra systems. Instead, our strategy will be to factorize the system. We first derive six special fixed-point solutions for the CKM-matrix elements which are \emph{independent} of the values of the Yukawa couplings. These six fixed-point solutions each give rise to a family of fixed-point solutions for the Yukawa couplings.  We will discuss only the phenomenologically relevant nontrivial solution.
Using a numerical grid search, we could not find any further non-trivial fixed points at which the factorization condition does not hold. It goes without saying that this does not exclude their existence.
\\

\subsubsection{Mixing parameters}

For three generations of quarks, the CKM matrix contains $4$ physical parameters, which we parameterize by
$X=|V_{ud}|^{2}$, $Y=|V_{us}|^{2}$, $Z=|V_{cd}|^{2}$ and $W=|V_{cs}|^{2}$. Then, the matrix of the squared CKM elements takes the form
\begin{equation}
 V_{2}=\Bigl[  \{ |V_{ij}|^{2} \} \Bigr] =
    \begin{bmatrix}
        X & Y & 1-X-Y \\
Z & W & 1-Z-W \\
1-X-Z & 1-Y-W & X+Y+Z+W-1 
    \end{bmatrix} \, .
\label{CKMwithWXYZ}
\end{equation}
From Eq.~\eqref{eq:betaV}, we find the beta functions for these parameters. As previously stressed, the new-physics contribution cancels in the running of CKM parameters and we recover the standard beta functions, which we write below for the sake of completeness.
{\scriptsize
\begin{align}
\frac{dX}{dt}&=-\frac{3}{(4\pi)^{2}}\left[\frac{y^{2}_{u}+y^{2}_{c}}{y^{2}_{u}-y^{2}_{c}}\left\{(y^{2}_{d}-y^{2}_{b})XZ+\frac{(y^{2}_{b}-y^{2}_{s})}{2}(W(1-X)+X-(1-Y)(1-Z))\right\}\right.
\label{betaX} 
\\
&\left.\phantom{\frac{1}{2}} +\frac{y^{2}_{u}+y^{2}_{t}}{y^{2}_{u}-y^{2}_{t}}\left\{(y^{2}_{d}-y^{2}_{b})X(1-X-Z)+\frac{(y^{2}_{b}-y^{2}_{s})}{2}((1-Y)(1-Z)-X(1-2Y)-W(1-X)) \right\} \right. 
\nonumber \\
&\left.\phantom{\frac{1}{2}} +\frac{y^{2}_{d}+y^{2}_{s}}{y^{2}_{d}-y^{2}_{s}}\left\{ (y^{2}_{u}-y^{2}_{t})XY+\frac{y^{2}_{t}-y^{2}_{c}}{2}(W(1-X)+X-(1-Y)(1-Z)) \right\} \right. 
\nonumber\\
 &\left.\phantom{\frac{1}{2}} +\frac{y^{2}_{d}+y^{2}_{b}}{y^{2}_{d}-y^{2}_{b}}\left\{(y^{2}_{u}-y^{2}_{t})X(1-X-Y)+\frac{y^{2}_{t}-y^{2}_{c}}{2}((1-Y)(1-Z)-X(1-2Z)-W(1-X))\right\} \right],
\nonumber
\end{align}

 \begin{align}
     \frac{dY}{dt}&=-\frac{3}{(4\pi)^{2}}\left[\frac{y^{2}_{u}+y^{2}_{c}}{y^{2}_{u}-y^{2}_{c}}\left\{\frac{(y^{2}_{b}-y^{2}_{d})}{2}(W(1-X)+X-(1-Y)(1-Z))+(y^{2}_{s}-y^{2}_{b})YW\right\}\right. \nonumber \\
               &\left.\phantom{\frac{1}{2}} +\frac{y^{2}_{u}+y^{2}_{t}}{y^{2}_{u}-y^{2}_{t}}\left\{ \frac{(y^{2}_{b}-y^{2}_{d})}{2}((1-Y)(1-Z)-W(1-X)-X(1-2Y))+(y^{2}_{s}-y^{2}_{b})Y(1-Y-W) \right\} \right. \nonumber\\
        &\left.\phantom{\frac{1}{2}} +\frac{y^{2}_{s}+y^{2}_{d}}{y^{2}_{s}-y^{2}_{d}}\left\{(y^{2}_{u}-y^{2}_{t})XY+\frac{y^{2}_{t}-y^{2}_{c}}{2}(W(1-X)+X-(1-Y)(1-Z))\right\} \right. \nonumber \\
       & \left.\phantom{\frac{1}{2}} +\frac{y^{2}_{s}+y^{2}_{b}}{y^{2}_{s}-y^{2}_{b}}\left\{(y^{2}_{u}-y^{2}_{t})Y(1-X-Y)
    +\frac{(y^{2}_{c}-y^{2}_{t})}{2}(W(1-X-2Y)+X-(1-Z)(1-Y)) \right\}\right],
     \label{betaY}
\end{align}
 \begin{align}
     \frac{dZ}{dt}&=-\frac{3}{(4\pi)^{2}}\left[\frac{y^{2}_{c}+y^{2}_{u}}{y^{2}_{c}-y^{2}_{u}}\left\{(y^{2}_{d}-y^{2}_{b})XZ+\frac{(y^{2}_{b}-y^{2}_{s})}{2}(W(1-X)+X-(1-Z)(1-Y))\right\}  \right. \nonumber\\
      & \left.\phantom{\frac{1}{2}}+\frac{y^{2}_{c}+y^{2}_{t}}{y^{2}_{c}-y^{2}_{t}}\left\{(y^{2}_{d}-y^{2}_{b})Z(1-X-Z)
         +\frac{(y^{2}_{s}-y^{2}_{b})}{2}(W(1-X-2Z)+X-(1-Y)(1-Z))\right\} \right. \nonumber\\
         &\left.\phantom{\frac{1}{2}} +\frac{y^{2}_{d}+y^{2}_{s}}{y^{2}_{d}-y^{2}_{s}}\left\{\frac{(y^{2}_{u}-y^{2}_{t})}{2}((1-Y)(1-Z)-X-W(1-X))+(y^{2}_{c}-y^{2}_{t})ZW\right\} \right. \nonumber \\
     & \left.\phantom{\frac{1}{2}}+\frac{y^{2}_{d}+y^{2}_{b}}{y^{2}_{d}-y^{2}_{b}}\left\{\frac{(y^{2}_{t}-y^{2}_{u})}{2}((1-Z)(1-Y)-W(1-X)-X(1-2Z))+(y^{2}_{c}-y^{2}_{t})Z(1-Z-W) \right\}\right],
      \label{betaZ}
\end{align}
 \begin{align}
     \frac{dW}{dt}&=-\frac{3}{(4\pi)^{2}}\left[\frac{y^{2}_{c}+y^{2}_{u}}{y^{2}_{c}-y^{2}_{u}}\left\{(y^{2}_{s}-y^{2}_{b})WY+\frac{(y^{2}_{b}-y^{2}_{d})}{2}((1-X)W+X-(1-Y)(1-Z))\right\}  \right. \nonumber \\
     &  \left.\phantom{\frac{1}{2}} +\frac{y^{2}_{c}+y^{2}_{t}}{y^{2}_{c}-y^{2}_{t}}\left\{(y^{2}_{s}-y^{2}_{b})W(1-Y-W)
         +\frac{(y^{2}_{b}-y^{2}_{d})}{2}((1-Y)(1-Z)-X-W(1-X-2Z))\right\} \right. \nonumber\\
         &\left.\phantom{\frac{1}{2}} +\frac{y^{2}_{s}+y^{2}_{d}}{y^{2}_{s}-y^{2}_{d}}\left\{(y^{2}_{c}-y^{2}_{t})WZ+\frac{(y^{2}_{t}-y^{2}_{u})}{2}Z((1-X)W+X-(1-Y)(1-Z))\right\} \right. \nonumber \\
     & \left.\phantom{\frac{1}{2}}+\frac{y^{2}_{s}+y^{2}_{b}}{y^{2}_{s}-y^{2}_{b}}\left\{(y^{2}_{c}-y^{2}_{t})W(1-Z-W)+\frac{(y^{2}_{t}-y^{2}_{u})}{2}((1-Y)(1-Z)-X-W(1-X-2Y)) \right\}\right].
      \label{betaW}
\end{align}
}
The standard parametrization of the quark mixing is generally given in terms of the mixing angles $\theta_{12}$, $\theta_{13}$, $\theta_{23}$ and $\delta$. The variables $X, Y, Z$ and $W$ are related to the mixing angles, 
\bea
\theta_{12}&=&\arctan\sqrt{\frac{Y}{X}},
\\
\theta_{13}&=&\arccos\sqrt{X+Y},
\\
\theta_{23}&=&\arcsin\sqrt{\frac{1-W-Z}{X+Y}},
\eea
and to the CP-violating phase,
{\small
\be
\delta=\arccos
\frac{(X+Y)^2Z-Y(X+Y+Z+W-1)-X(1-W-Z)(1-X-Y)}
{2\sqrt{XY(1-X-Y)(1-Z-W)(X+Y+Z+W-1)}}.
\ee
}

\subsubsection{Fixed points of the CKM-matrix}

To factorize the full set of fixed-point equations, we consider the case
in which each factor inside the curly brackets in Eqs.~\eqref{betaX}-\eqref{betaW}
vanishes. The resulting fixed-points of the matrix  $V_{2}$, cf.~Eq.~\eqref{CKMwithWXYZ}, correspond to the matrices
\bea
&&
M_{123}=\begin{bmatrix}
1 & 0 & 0\\
0 & 1 & 0\\
0 & 0 & 1
\end{bmatrix}\ ,\quad
M_{132}=\begin{bmatrix}
1 & 0 & 0\\
0 & 0 & 1\\
0 & 1 & 0
\end{bmatrix}\ ,\quad
M_{321}=\begin{bmatrix}
0 & 0 & 1\\
0 & 1 & 0\\
1 & 0 & 0
\end{bmatrix}\ ,
\nonumber\\
&&
M_{213}=\begin{bmatrix}
0 & 1 & 0\\
1 & 0 & 0\\
0 & 0 & 1
\end{bmatrix}\ ,\quad
M_{312}=\begin{bmatrix}
0 & 0 & 1\\
1 & 0 & 0\\
0 & 1 & 0
\end{bmatrix}\ ,\quad
M_{231}=\begin{bmatrix}
0 & 1 & 0\\
0 & 0 & 1\\
1 & 0 & 0
\end{bmatrix}\ .\quad
\label{eq:ckmconf}
\eea
These matrices provide a faithful representation
of the permutation group of three objects.
The second, third and fourth are odd permutations
corresponding to interchanging two families, whereas
the other three correspond to cyclic permutations. The matrix $M_{123}$ represents the case of no mixing,
where each up-type quark interacts only with the corresponding
down-type quark. 
In the standard terminology, the other cases are also
referred to as ``no mixing'', since each up-type quark
interacts only with one down-type quark,
although possibly belonging to a different family.

The critical exponents in the CKM sector feature a dependence on the fixed-point values of the Yukawa couplings. We limit ourselves to the phenomenologically most interesting case where $y_{t\, \ast}\gg y_{i\, \ast}$, with $i=b,c,s,u,d$. As this will later turn out to be the two interesting cases, we provide the critical exponents for $M_{123}$, which are 
\be
\theta_I\vert_{M_{123}} =\left\{-\frac{3}{16\pi^2}y_t^2, -\frac{3}{16\pi^2}y_t^2,-\frac{3}{16\pi^2}y_t^2,0 \right\},
\ee
 and for $M_{321}$ as well as $M_{231}$, which are 
 \be
\theta_I\vert_{M_{321}} = \theta_I\vert_{M_{231}} =\left\{\frac{3}{16\pi^2}y_t^2, \frac{3}{16\pi^2}y_t^2,\frac{3}{16\pi^2}y_t^2,0 \right\}. 
 \ee
 Including a finite, but small value for the bottom-Yukawa coupling renders all four critical exponents of $M_{321}$ relevant while $M_{231}$ obtains a single irrelevant direction. 
\\
Hence, an RG-trajectory connects the fixed-point configuration $M_{321}$ in the UV with $M_{123}$ in the IR. The latter resembles the measured structure of the CKM matrix, where off-diagonal elements are small compared to the diagonal ones. The IR attractive properties of the fixed-point configuration $M_{123}$ provide a mechanism to explain why the SM is driven towards a near-diagonal mixing matrix. The measured value of the CP-violating phase remains unexplained in this context - although its measured value can be realized due to it being a relevant parameter at the UV fixed point.
\\

\subsubsection{Fixed-point structure of the gauge and Yukawa couplings}

We can now insert the CKM fixed-point matrices of Eq.~\eqref{eq:ckmconf} into the beta functions for the Yukawa couplings and solve them, avoiding the degenerate points at finite values of the Yukawa couplings.
The resulting fixed-point equations can be solved analytically. Altogether we obtain 17 fixed-point solutions
for each choice of CKM matrix.
Of these, six are isolated fixed points,
nine are lines of fixed points and
one is a plane of fixed points. We include the fixed-point solution where four Yukawa couplings vanish, as it can be approached from any direction with unequal up- and down-type Yukawa couplings. The lines and planes are discarded. We conjecture them to either reduce to fixed points or vanish entirely beyond the one-loop order\footnote{As the corresponding analysis of fixed-point solutions at two-loop order is technically challenging due to the large number of solutions (already the one-loop system features 392 solutions of the Yukawa-system before degeneracies of the Yukawa couplings are excluded) we do not attempt it here.}.
Regarding the isolated solutions, we discard those at which more than two Yukawa couplings are finite. At $f_g>0$ (which must be required to avoid the U(1) Landau pole), these fixed points feature at least one negative $y_i^2$ and are thus unphysical.
Further, we only keep fixed points for which the CKM sector contains four relevant directions. It appears that an irrelevant direction in the CKM sector always results in incompatibility with the measured values.

Applying the above criteria results in only a single potentially interesting fixed-point. It corresponds to the CKM configuration of $M_{321}$, with non-vanishing top and bottom Yukawa couplings given in Eq.~\eqref{eq:nonzerotb}. All the other Yukawa couplings vanish. 
Before discussing the phenomenology of this fixed-point solution in Sec.~\ref{sec:fundamentalPredictiveButNotViable}, we detail the limit of a heavy top quark in which the phenomenologically relevant RG flows can be understood most easily.

\subsection{The top quark driving the RG flow}
\label{sec:topDrive}

The key features of the realistic RG flow of the three-generation system can be understood in the heavy-top limit, i.e., for $y_t\gg y_{i\neq t}$. In the far UV, the non-Abelian gauge couplings are still close to their asymptotically free fixed point and can, therefore, be neglected. While the following discussion employs these approximations, the respective plots are obtained by evolving the full 1-loop running (without any of the above approximations).
\\

Taking the above limits, the running of the top-Yukawa coupling $y_t$ itself is independent of all other Yukawas and the CKM parameters, i.e.,
\begin{align}
    16\pi^2\beta_{y_t}=
    y_t\left[\frac{9}{2}\,y_t^2 - \frac{17}{12}g_Y^2 - 16\,\pi^2\,f_y\right]\;.
\end{align}
Given a fixed point for the Abelian gauge coupling, i.e., $g_{Y\,\ast}$, the top Yukawa exhibits an IR-attractive fixed point at
\begin{align}
\label{eq:topFP}
    y_{t\,\ast}^2 = \frac{2}{9}\left[\frac{17}{12}g_{Y\,\ast}^2 + 16\,\pi^2\,f_y\right]\;.
\end{align}
In the following, we eliminate $f_y$ in favor of $y_{t\,\ast}^2$.
\\

The running of all down-type Yukawas $y_\rho$ (with $\rho = d,\,s,\,b$) is driven by the top Yukawa (and the gauge as well as the new-physics) contribution, i.e., in the above heavy-top limit,
\begin{align}
\label{eq:betaDownType}
    16\pi^2\beta_{y_\rho} = y_\rho\left[
        \frac{9}{2}y_\rho^2     +3y_t^2 - \frac{5}{12}g_Y^2 - 16\,\pi^2\,f_y
        -\frac{3}{2}|V_{t\rho}|^2\,y_t^2
    \right]\;,
\end{align}
with (partial) fixed points $y_{\rho\,\ast}^2$ and associated critical exponents $\theta_\rho$ at
\begin{align}
\label{eq:dtypeThetaForHeavyTopLimit}
    y_{\rho\,\ast\,\text{(int)}}^2 &= \frac{1}{3}\left[y_{t\,\ast}^2 - \frac{2}{3}g_{Y\,\ast}^2 + |V_{t\rho}|^2\,y_{t\,\ast}^2\right]\;,
    \quad\quad
    \theta_{\rho\,\text{(int)}} = -\frac{9}{16\pi^2}\;y_{\rho\,\ast\,\text{(int)}}^2\;,
    \\
    y_{\rho\,\ast\,\text{(free)}}^2 &= 0\;,
    \quad\quad\quad\quad\quad\quad\quad\quad\quad\quad\quad\quad\quad\,
    \theta_{\rho\,\text{(free)}} = \frac{9}{32\pi^2}\;y_{\rho\,\ast\,\text{(int)}}^2\;.
\end{align}
While the top Yukawa and the Abelian gauge coupling remain at their interacting fixed-point values, the respective down-type Yukawas scale away from their free fixed point with $k^{-\theta_{\rho\,\text{(free)}}}$ as long as $y_\rho\ll y_{\rho\,\ast\,\text{(int)}}$. If the corresponding CKM element $|V_{t\rho}|^2$ transitions from (close to) $0$ to (close to) $1$, or vice versa, the (partial) fixed-point scaling of the respective $\rho$-type quark changes. \\

In contrast to the running of the down-type Yukawas, the running of the up-type Yukawas does not depend on the CKM elements $V_{ti}$.
Moreover, since the up-type Yukawas share the same hypercharge with the top Yukawa, their (partial) non-vanishing fixed points in the heavy-top limit lie at fixed $y_{j\neq t\,\ast\,\text{(int)}}=y_{t\,\ast}/3$.
Whenever $y_{j\neq t}\ll y_{t\,\ast}/3$, the other up-type Yukawas are asymptotically free and simply scale away from zero under the RG flow (with $k^{-\theta_{j\neq t\,\text{(free)}}}$ where $\theta_{j\neq t\,\text{(free)}}=3y_{t\,\ast}^2 / (32\pi^2)$).
\\

Regarding the RG evolution of the CKM-matrix elements, it is important to take the limits $y_{j\neq t}\rightarrow 0$ and $y_{\rho}\rightarrow 0$, in such an order as to preserve the phenomenological quark-mass ordering for the quarks of equal hypercharge. As observed in \cite{Pendleton:1980as}, this results in an IR-attractive fixed point at diagonal CKM matrix. We note that, within the heavy-top limit, this fixed point is part of a fixed line. In the given parameterization and the heavy-top limit, the running of CKM-matrix parameters is given by
\begin{align}
    \beta_X &= \frac{3\,y_t^2}{16\pi^2}\left[X (X + Z-1)\right]\;,
    \\
    \beta_Y &= \frac{3\,y_t^2}{16\pi^2}\left[W (X + Y-1) + X (2 Y-1) + (Y-1) (Y + Z-1)\right]\;,
    \\
    \beta_Z &= \frac{3\,y_t^2}{16\pi^2}\left[Z (X + Z-1)\right]\;,
    \\
    \beta_W &= \frac{3\,y_t^2}{16\pi^2}\left[(W-1) (W + X + Y-1) + (2 W + Y-1) Z\right]\;.
\end{align}
This set of $\beta$-functions exhibits three fixed lines and one isolated fixed point with corresponding critical exponents, i.e.,
\begin{align}
    \text{fixed line:}\quad
    & Z_\ast = 1-X_\ast\;,\quad W_\ast=Y_\ast=0\;,
    \notag\\  
    &\frac{3\,y_t^2}{16\pi^2}\theta_i = (-1,\;0,\;0,\;1)\;,\label{eq:CKMline1}
    \\
    \text{UV-attractive fixed line:}\quad
    & X_\ast = 0\;,\quad Z_\ast=0\;,\quad W_\ast= 1-Y_\ast\;,
    \notag\\
    \label{eq:IRattractivefixedLine}
    &\frac{3\,y_t^2}{16\pi^2}\theta_i = (0,\;1,\;1,\;1)\;,
    \\
    \text{IR-attractive fixed line:}\quad
    & X_\ast = 1-Y_\ast\;,\quad Z_\ast=Y_\ast\;,\quad W_\ast= 1-Y_\ast\;,
    \notag\\
    \label{eq:UVattractivefixedLine}
    &\frac{3\,y_t^2}{16\pi^2}\theta_i = (-1,\;-1,\;-1,\;0)\;,
    \\
    \text{fixed point:}\quad
    & X_\ast = 0\;,\quad Z_\ast=0\;,\quad Y_\ast= 1\;,\quad W_\ast=1\;,
    \notag\\
    &\frac{3\,y_t^2}{16\pi^2}\theta_i = (-1,\;-1,\;1,\;1)\;.\label{eq:CKMFP}
\end{align}
To link up with the general results in Eq.~\eqref{eq:ckmconf}, we note that the fixed line Eq.~\eqref{eq:CKMline1} contains the fixed points $M_{312}$ and $M_{132}$, the fixed line Eq.~\eqref{eq:IRattractivefixedLine} contains the fixed points $M_{231}$ and $M_{321}$ and the fixed line Eq.~\eqref{eq:UVattractivefixedLine} contains the fixed points $M_{213}$ and $M_{123}$. 
The fixed point Eq.~\eqref{eq:CKMFP} only exists in the heavy-top limit, as it is destroyed by contributions proportional to the other Yukawas.

The two-loop running does not modify the CKM fixed-point structure in the heavy-top limit. It merely adds a subleading correction to the overall size of the critical exponents.
The IR- and UV-attractive fixed lines are of particular interest.
Their persistence at two-loop order suggests that these lines of fixed points are not an artifact of the one-loop approximation.
Rather, they seem to arise due to the degeneracy of the lower Yukawa couplings in the heavy-top limit. This is in contrast to the lines of fixed points which were discarded in Sec.~\ref{sec:mathematicalFPstructure} and renders them of phenomenological interest.
\\
Both of the lines are characterized by a common relation $W_\ast= 1-Y_\ast$. RG-flows can transition from the IR-repulsive to the IR-attractive fixed line, cf.~Fig.~\ref{fig:3genRunning_toTBYextension}. Points on the IR-repulsive fixed line can be connected to different points on the IR-attractive fixed line depending on the dynamics of the full flow. In particular, this depends on the choice of initial conditions for the relevant CKM parameters near the IR-repulsive fixed line and the resulting full dynamics.
\\

In addition to the above non-trivial fixed points, the CKM running is absent whenever all Yukawas (including the top) are in the asymptotically free regime. This asymptotically free form of scale invariance occurs at values of the CKM parameters that are determined by the ratios of the Yukawas -- we remind the reader that the CKM flow only makes sense if the Yukawas are held at finite ratios when the limit of vanishing Yukawas is taken.
\\

A special situation arises if the top and Abelian-gauge contribution (nearly) cancel out, i.e., if $3\,y_{t\,\ast}^2 \approx 2\,g_{Y\,\ast}^2$. In this case, the $\rho$-type Yukawas, for which $|V_{t\rho}|^2\approx 0$, can nearly be held constant. Similarly, if $3\,y_{t\,\ast}^2 \approx g_{Y\,\ast}^2$, the $\rho$-type Yukawas for which $|V_{t\rho}|^2\approx 1$ are (nearly) frozen. If one of the above near-cancellations is realized, the respective down-type Yukawas are (nearly) frozen -- irrespective of their particular value $0<y_\rho<y_{\rho\,\ast}$.

\subsection{Predictivity and UV completions of the Standard Model}
\label{sec:threeGenPheno}

Returning to the main motivation for postulating scale-invariant new-physics contributions at some high-energy scale, i.e., removing Landau poles and UV-completing the SM gauge-Yukawa sector, we will now detail different fixed points and discuss whether they allow to UV-complete or UV-extend the SM. To decide whether a given set of IR values can or cannot be reached from a starting point in the UV requires to select a very high value for the UV scale. This is a consequence of the unusually slow (compared to the typical speed of the RG flow in the SM) running of the CKM matrix elements. Our somewhat unusual discussion of RG flows over hundreds of orders of magnitude is possible since the new-physics contribution removes the Landau pole in the U(1) coupling. 

In the following, we will discuss three different scenarios which are distinguished by the following criteria:
\begin{enumerate}
\item[i)] UV completion vs. UV extension: The presence of a fixed point can allow a UV completion for a given model. This occurs when the IR values of the model's couplings can be reached on a ``true" fixed-point trajectory. The presence of a fixed point can also allow a UV extension for a given model when the IR values of the model's couplings can be reached on a trajectory that does \emph{not} emanate from the fixed point but instead passes very close to it. In that case, the flow exhibits near-scale invariance over a large range of scales. The ``true" UV completion is in this case given by a different theory, and the fixed point merely extends the regime of validity of the EFT by (potentially very many) orders of magnitude.
\item[ii)] Predictive power: For both cases (UV complete and UV extended), the critical hypersurface of the fixed point determines the degree of predictivity of the model. For a UV complete trajectory, the predictions for the IR values of the irrelevant couplings are ``sharp". For a UV extended trajectory, the predictions for the IR values of the irrelevant couplings are not ``sharp", but are instead given by mapping UV windows to IR windows of possible values. The longer the trajectory spends in the vicinity of the fixed point, the more the IR window closes around the prediction from a UV complete trajectory.
\end{enumerate}

\subsubsection{UV completion 1): Free fixed point for Yukawa couplings}
\label{sec:viableFundamentalButNotPredictive}
Setting $f_g=0.097$ and $f_y>-2.1886\times10^{-4}$, there is a trajectory emanating from the fixed point at $g_{Y\, \ast}^2= 96\pi^2 f_g/41$ and $y_{i\, \, \ast}=0$, that reaches IR values in agreement with those inferred from SM measurements. We fix the RG trajectory by matching all Yukawa couplings and CKM matrix elements to agree with the tree-level matched PDG values \citep{Tanabashi:2018oca}. For the top-Yukawa coupling, there is a significant difference between running mass and pole mass. To perform a correct matching, we include sizable loop effects equivalent to tree-level matching to a running mass of $163$ GeV \cite{Tarrach:1980up,Bohm:1986rj,Hempfling:1994ar,Chetyrkin:1999ys,Chetyrkin:1999qi,Melnikov:2000qh,Buttazzo:2013uya}.

\begin{figure}[!t]
\centering
\includegraphics[width=0.58\linewidth]{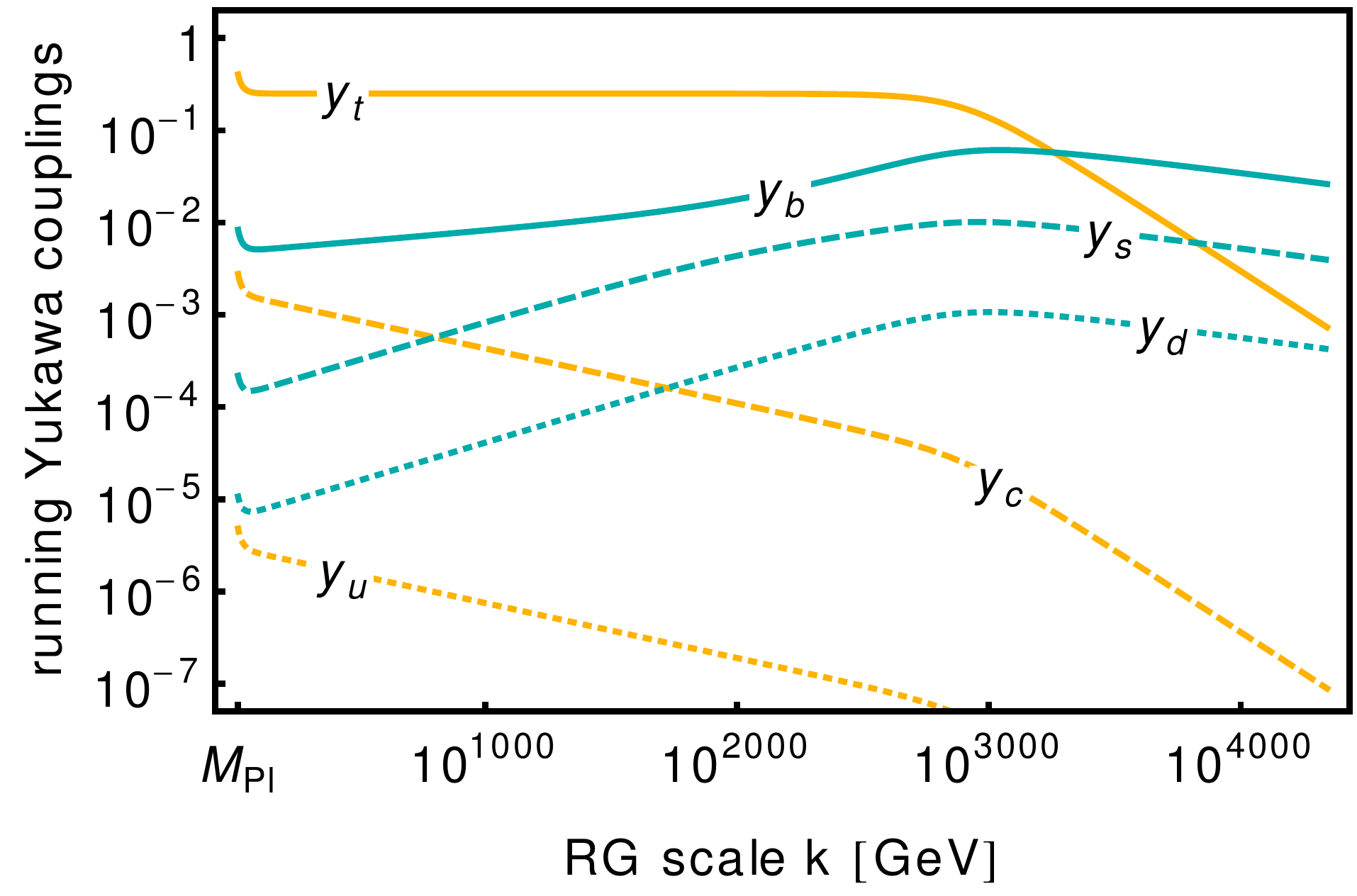}
\hfill
\includegraphics[width=0.395\linewidth]{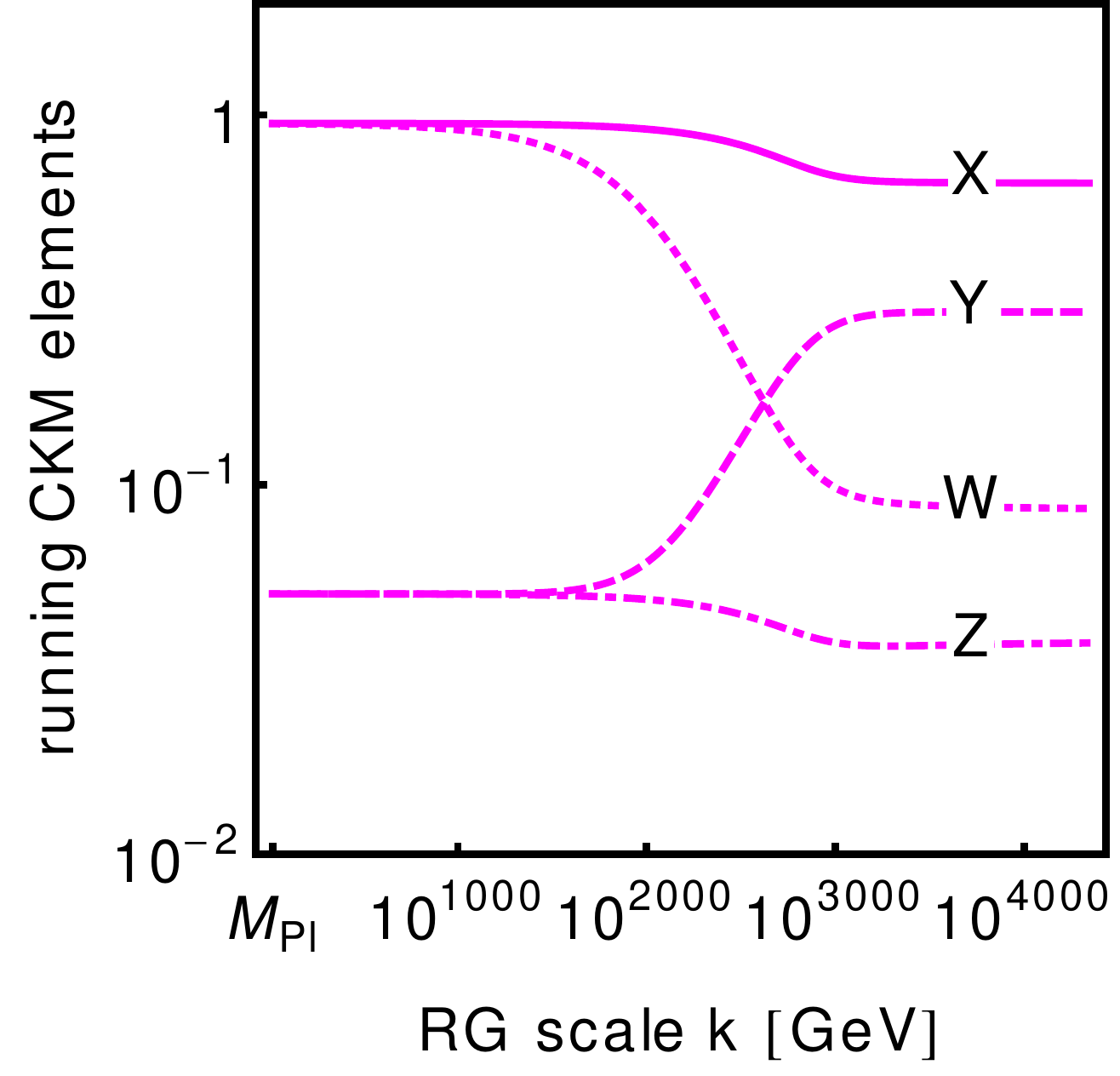}
\caption{\label{fig:3genRunning_toAF} RG flow of Yukawa couplings (left-hand panel) and CKM parameters (right-hand panel). All Yukawas are tree-level matched to the respective PDG values \citep{Tanabashi:2018oca} at $k=173\,\text{GeV}$ (including sizable loop effects for the top-Yukawa, cf.~main text). For $f_y=-2.1886\times10^{-4}$, as well as for larger values of $f_y$, the corresponding trajectory realizes asympototic freedom for all Yukawa couplings.}
\end{figure}

At the FP all Yukawa couplings are relevant and the U(1) gauge coupling is irrelevant so that the IR values of the latter can be calculated, cf.~left panel of Fig.~\ref{fig:3genRunning_toAF}.
For vanishing Yukawa couplings, the CKM matrix elements are automatically do not run. Their values are tied to the ratio with which Yukawa couplings emanate from the free fixed point.
Thus the CKM elements can start running from a point on this hyperplane in the UV and reach values in agreement with SM measurements in the IR, cf.~right panel of Fig.~\ref{fig:3genRunning_toAF}

Within our parameterization for NP, we have two free parameters, $f_g$ and $f_y$ and the fixed point results in one prediction. Thus the predictive power of this general setting is insufficient to rule it out since $f_g$ can always be adjusted to produce the correct IR value of $g_Y$. For any particular NP model in this general setting, $f_g$ is expected to become a calculable quantity -- gravity is an example -- allowing to meaningfully test the particular model, defined by the fixed point, by comparing the resulting prediction for $g_Y$ to experiment.

\subsubsection{UV completion 2): Interacting fixed point for Yukawa couplings}
\label{sec:fundamentalPredictiveButNotViable}
Having seen that a UV completion can be achieved, we next ask how much the predictive power can be increased. A number of highly predictive fixed points can be excluded based on phenomenological considerations, as they result in IR values for the Yukawa couplings of the first and/or second generation which are much too large.

\begin{figure}[!t]
\centering
\includegraphics[width=0.58\linewidth]{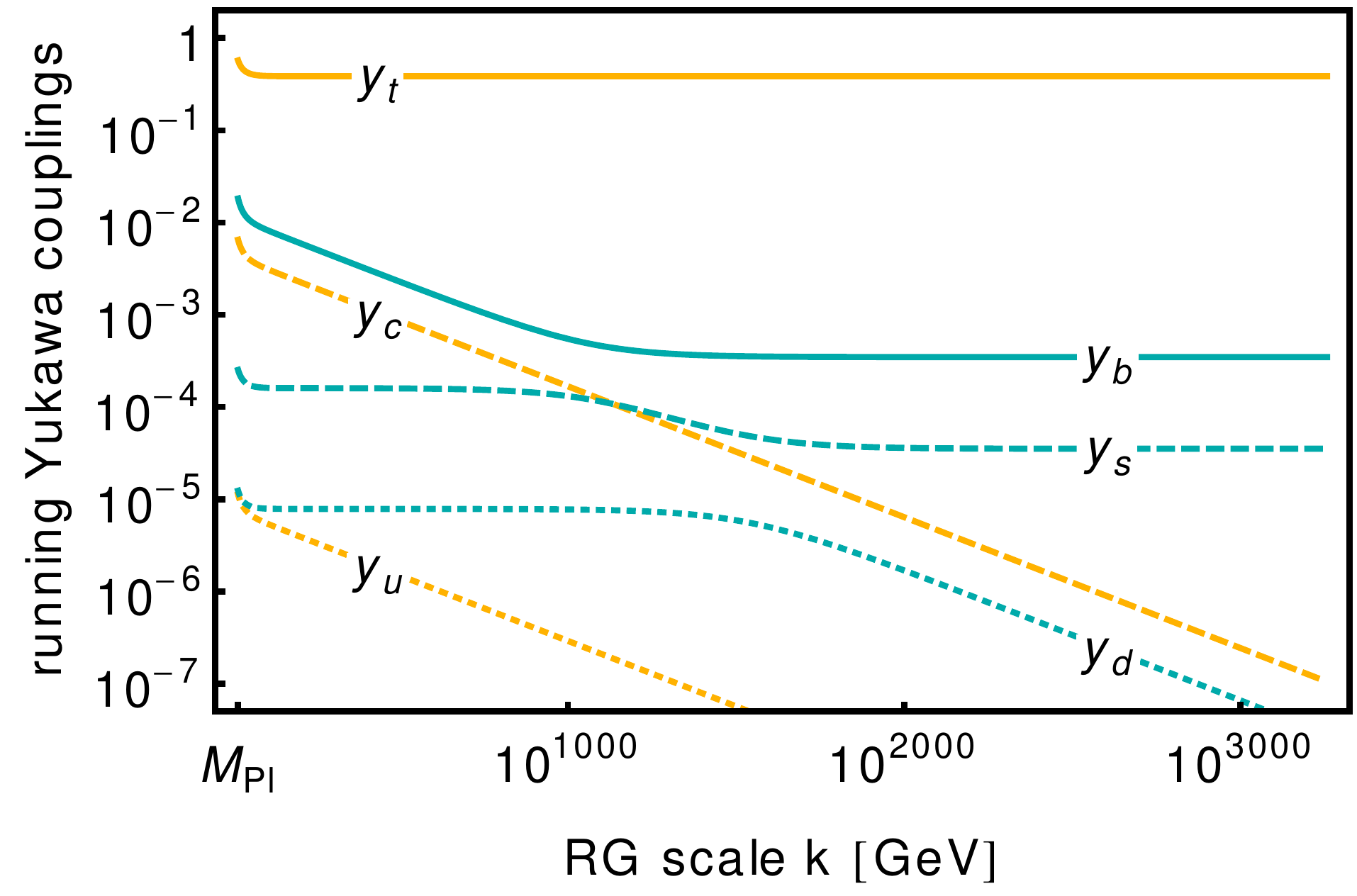}
\hfill
\includegraphics[width=0.395\linewidth]{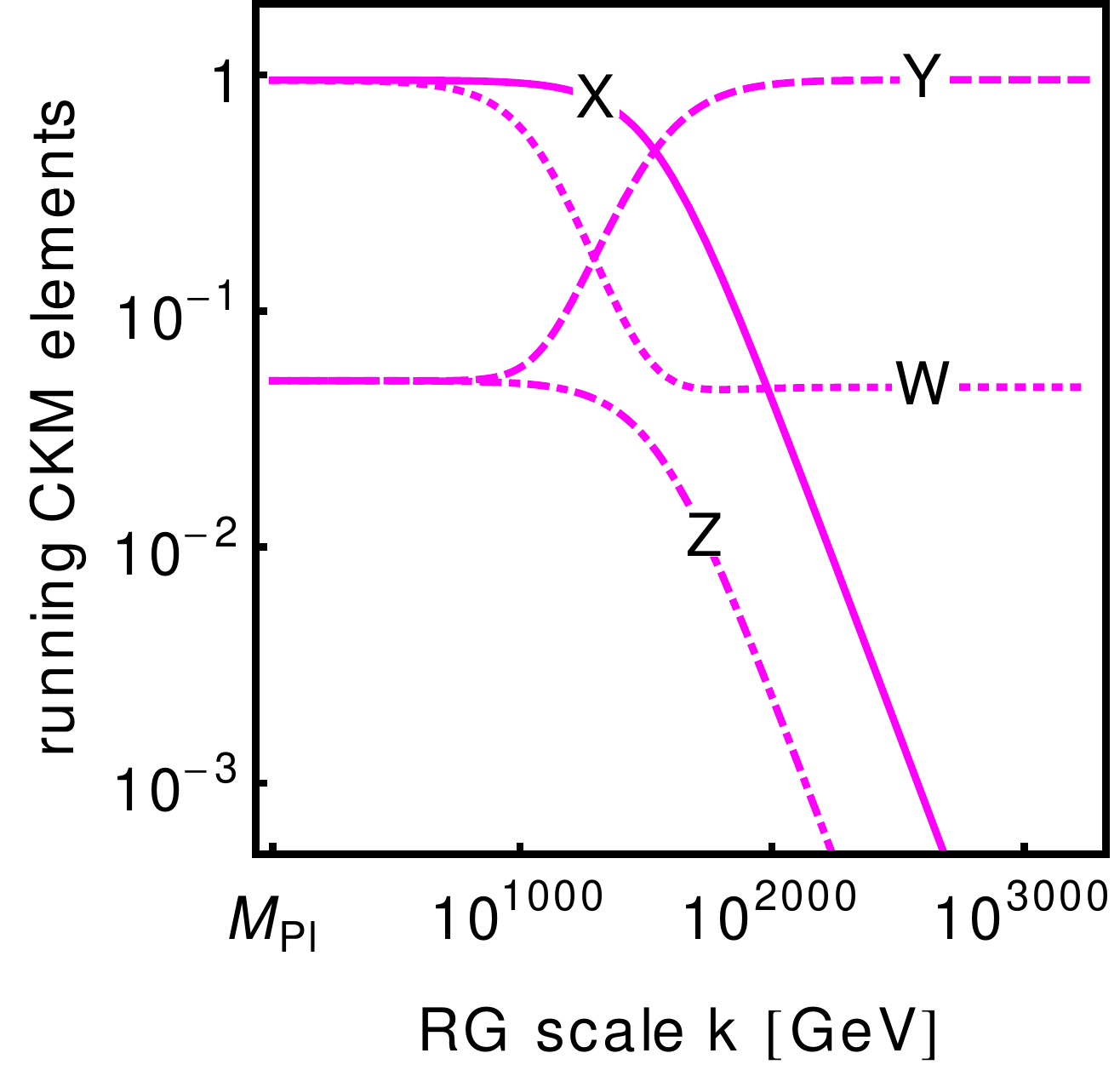}
\caption{\label{fig:3genRunning_toTBYextension} We choose $f_y=2.2476\times10^{-3}$ and the corresponding IR value for the top-Yukawa coupling to realize asympototic safety. All other coupling values are chosen to agree with tree-level matched PDG values \citep{Tanabashi:2018oca} at $k=173\,\text{GeV}$.}
\end{figure}

The most promising candidate appears to be the fixed point at which $g_{Y\, \ast}^2= 96\pi^2 f_g/41$, $y^2_{t\, \ast}= 16\pi^2/15(f_g+2f_y)$ and $y_{b\, \ast}^2 = 16\pi^2/615(-19 f_g+82 f_y)$, all other Yukawa couplings vanish, and the CKM matrix elements $Y_{\ast}=1$, $X_{\ast}=Z_{\ast}=W_{\ast}=0$, cf.~Sec.~\ref{sec:mathematicalFPstructure}. We choose $f_y=2.2476\times10^{-3}$ for which we are able to match the flow with the SM values of all couplings but the top, cf.~Fig.~\ref{fig:3genRunning_toTBYextension}. The flow is characterized by a transition from the approximate UV fixed line in Eq.~\eqref{eq:UVattractivefixedLine} to the approximate IR fixed line in Eq.~\eqref{eq:IRattractivefixedLine} with a transition scale of $\sim 10^{1500}$ GeV.
Above this transition scale $V_{tb}\approx 0$ and $V_{ts}\approx 0$ and consequently $y_b$ and $y_s$ are (nearly) scale invariant. Below the transition scale, $V_{ts}\approx 0$ and $V_{tu}\approx 0$ and therefore not just $y_d$ but also $y_s$ appear as (nearly) scale invariant, cf.~Eq.~\eqref{eq:dtypeThetaForHeavyTopLimit}. We have verified that such RG flows can emanate from the fixed point at which all Yukawa couplings are asymptotically free. Determining whether a similar flow can also emanate directly from the interacting fixed point found in Sec.~\ref{sec:mathematicalFPstructure} remains to be explored in the future.

In any case, this fixed point is IR attractive in $g_Y$, $y_t$ and $y_b$, resulting in three predictions. Therefore, this general class of NP models, parameterized by $f_g$ and $f_y$, can actually be ruled out. Let us stress that here we compare tree-level matched IR values calculated in a toy model (neglecting the lepton sector of the SM) to values extracted from measurements at first loop order. This comparison results in the conclusion that the top quark comes out about 10\% too heavy, compared to experiment. 

Despite the above disagreement with experiment, the fixed point results in IR structure in the Yukawa sector which is qualitatively similar to the SM. In particular, the fixed point imprints a significant difference between the top mass and the masses of the other quarks.

\subsubsection{UV extension: Near scale invariance above $M_{\rm NP}$}
\label{sec:viablePredictiveButNotFundamental}
The fixed point in the previous subsection \ref{sec:fundamentalPredictiveButNotViable} results in IR values which are not in precise agreement with measured values of SM couplings. This motivates us to explore trajectories that are not UV complete but nevertheless exist over a very large range of scales. This is enabled by them passing close to the fixed point such that the model becomes nearly scale invariant, i.e., shows very little RG evolution over a large range of scales. In this regime, all irrelevant parameters (in particular the top and bottom Yukawa) are attracted to their FP values, and this also greatly restricts their values in the IR limit.

\begin{figure}[!t]
\centering
\includegraphics[width=0.7\linewidth]{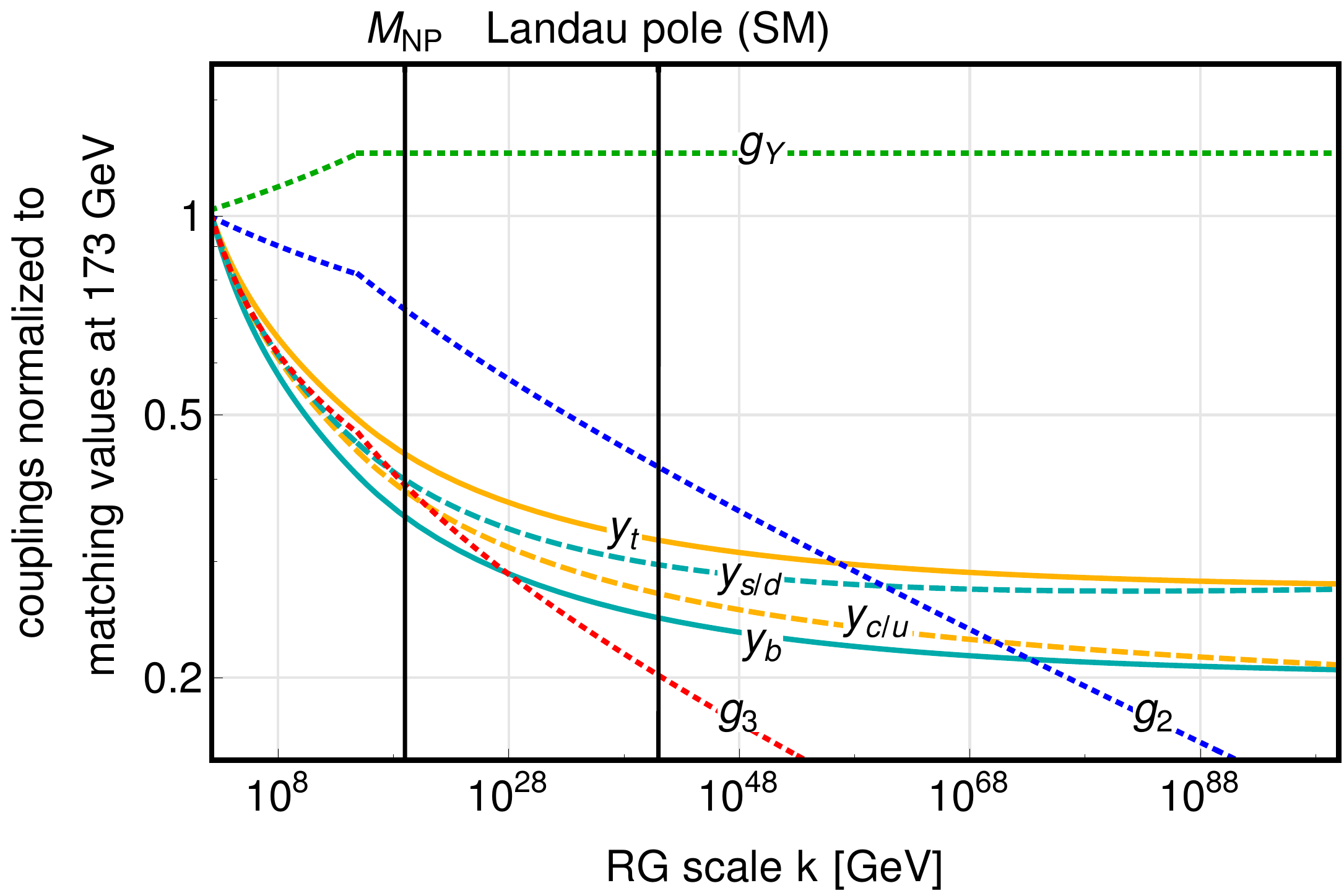}
\caption{\label{fig:3genRunning_correct1LoopIRValues} RG-flow for the SM gauge and Yukawa couplings, normalized to their IR values, including parameterized new physics with $f_g=9.7\times10^{-3}$ and $f_y=-2.1886\times10^{-4}$ above $M_\text{NP}=10^{19}$ GeV. The CKM-matrix elements are omitted because they exhibit no significant running over the given range of scales. All Yukawa couplings and CKM-matrix elements agree with tree-level matched PDG values \citep{Tanabashi:2018oca} at $k=173\,\text{GeV}$.
}
\end{figure}
To present how the new-physics contributions can realize nearly scale-invariant behavior, we choose $f_g=9.7\times10^{-3}$ and $f_y=-2.1886\times10^{-4}$ to realize non-vanishing fixed points for $g_Y$ and $y_t$, respectively. Due to the extremely slow running of the CKM-matrix elements and because of the relation $3\,y_{t\,\ast}^2 \approx g_{Y\,\ast}^2$ which the Standard-Model coupling values approximately obey \cite{Eichhorn:2018whv}, this realizes a nearly scale-invariant but interacting regime above the Planck scale, cf.~Fig.~\ref{fig:3genRunning_correct1LoopIRValues}.
This allows us to extend the SM to scales significantly above the Landau pole, cf.~Fig.~\ref{fig:3genRunning_correct1LoopIRValues}.

The trajectory in Fig.~\ref{fig:3genRunning_correct1LoopIRValues} again realizes the tree-level matched PDG values of the SM couplings. This is nontrivial, since the fixed point that the trajectory passes by very closely features three IR attractive directions, and we only have two free parameters in our general parameterization of new physics.

This near-scale invariant regime can actually be realized on a trajectory that emanates from the free-Yukawa UV fixed point in Sec.~\ref{sec:viableFundamentalButNotPredictive}. Indeed, Fig.~\ref{fig:3genRunning_correct1LoopIRValues} and Fig.~\ref{fig:3genRunning_toAF} show different regimes of the same RG trajectory. The latter emanates from the free-Yukawa fixed point and passes very close to the more predictive, interacting fixed point.

Let us emphasize that determining the degree of predictivity of such a UV-extended model is more challenging than for UV complete models. For the latter, the degree of predictivity can be quantified as the number of IR attractive directions minus the number of free parameters. For a UV extended model, a range of IR values, i.e., an IR window, is achievable even along IR attractive directions, instead of a single, ``sharp" prediction. The smaller the IR window, the more predictive the model. The IR window accessible for the coupling shrinks, as the UV cutoff scale of the model is pushed to higher values. 

\subsection{Discussion}\label{sec:discussionsFPs}

From the examples given in this subsection, we infer that a generalization of the results in \cite{Eichhorn:2018whv} to the rather more involved case of three generations with nontrivial mixing angles is possible, and the observed flavor structure and mixing angles of the SM can be accommodated in a UV complete setting. The $U(1)$ gauge coupling, suffering from triviality in the absence of the new-physics contribution, instead becomes a calculable quantity.
Concerning the Yukawa sector, the removal of the Landau pole is sufficient to allow for RG trajectories connecting a fixed point that is asymptotically free in all Yukawa couplings to the exact SM values of couplings in the IR. As a consequence of the extremely slow running of CKM matrix elements, such trajectories stretch over a rather astonishing range of scales.

\begin{figure}
\centering
\includegraphics[width=0.9\linewidth]{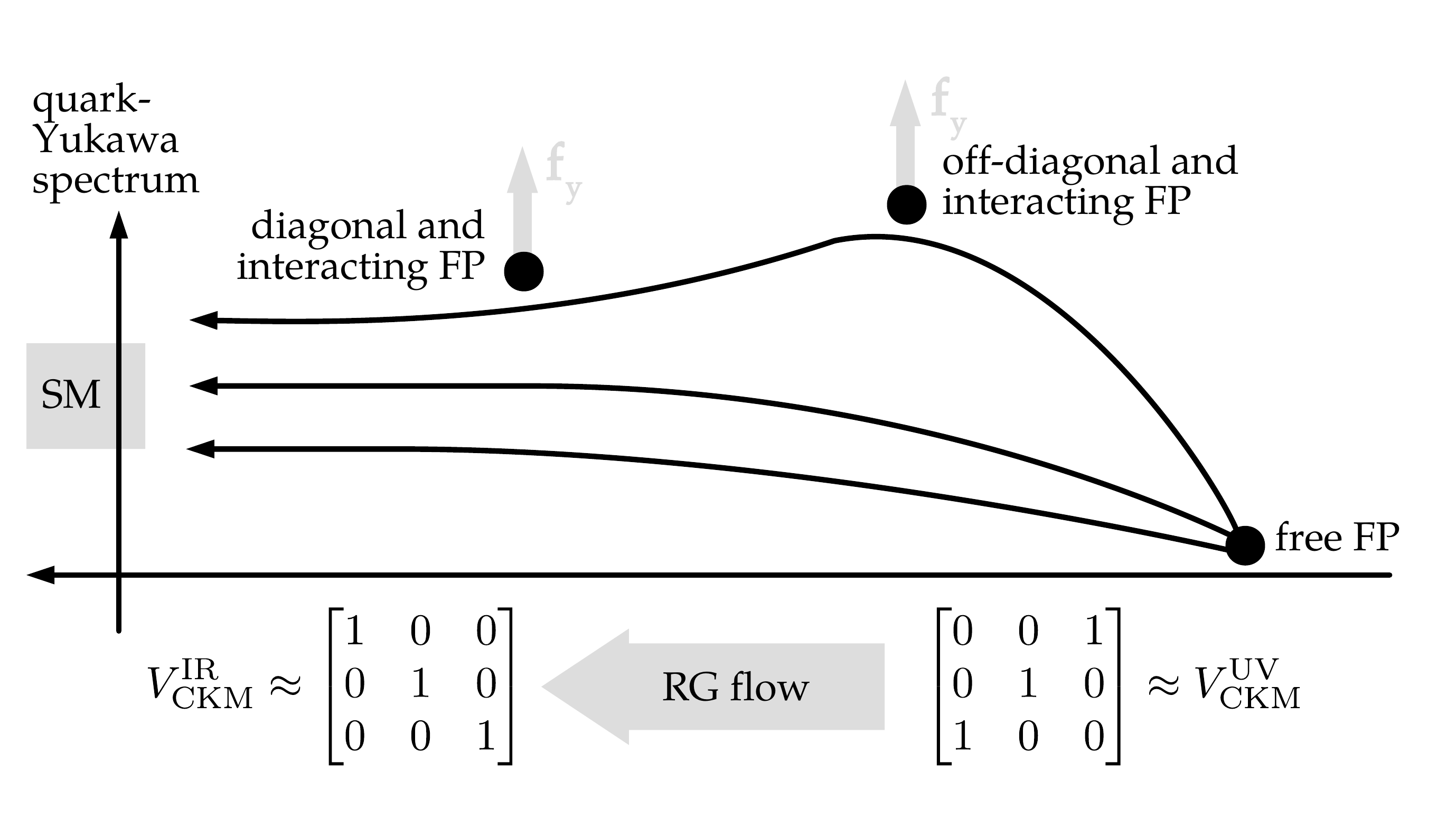}
\caption{
Schematic RG flow of Yukawa couplings and mixing parameters. Depending on the specific value of $f_y$, trajectories resulting in an IR spectrum of Yukawa couplings in the vicinity of the SM can be dominated by intermediate scaling regimes of one or several interacting fixed points. For the diagonal and interacting FP see Eq.~\eqref{eq:fptby}. For the off-diagonal and interacting FP see Eq.~\eqref{eq:nonzerotb}.}
\label{fig:conclusions}
\end{figure}

Most intriguingly and depending on the specific value of $f_y$ such trajectories can pass quite close to two interacting fixed points, each with two additional infrared attractive directions.
These fixed points are distinguished by a transition in the CKM matrix elements, cf.~Fig.~\ref{fig:conclusions} for a schematic overview. 
\begin{itemize}
\item
The interacting and  anti-diagonal fixed point is given by Eq.~\eqref{eq:nonzerotb}  and dominates the flow at approximately  anti-diagonal CKM matrix, cf.~$M_{321}$ in Eq.~\eqref{eq:ckmconf}.
\item
The interacting and diagonal fixed point is given by Eq.~\eqref{eq:fptby} and dominates the flow at approximately diagonal CKM matrix, cf.~$M_{123}$ in Eq.~\eqref{eq:ckmconf}.
\end{itemize}

For $f_y\approx -2.1886\times 10^{-4}$, a crossover trajectory emanates from the fixed point at vanishing Yukawa couplings towards the interacting and anti-diagonal fixed point and is dominated by the diagonal and interacting fixed point \cite{Eichhorn:2018whv} over a large range of scales, cf.~middle trajectory in the schematic Fig.~\ref{fig:conclusions} and Fig.~\ref{fig:3genRunning_toAF} for the explicit RG flow. Due to its two IR attractive directions in the Yukawa sector, this fixed point imprints a significant mass difference between top quark mass and the down-type quark masses on the system. The IR endpoint of this trajectory lies at values for the Yukawa couplings and CKM which are in agreement with measurements.
\\

Alternatively, for $f_y\approx 2.2476\times 10^{-3}$, a trajectory emanates from the fixed point at vanishing Yukawa couplings and first passes close to the off-diagonal and interacting fixed point. It then crosses over from this off-diagonal and interacting fixed point towards the diagonal and interacting fixed point, cf.~upper trajectory in the schematic Fig.~\ref{fig:conclusions} as well as Fig.~\ref{fig:3genRunning_toTBYextension} for the explicit RG flow. The current approximation, however, suggests that the intermediate scaling relations of the anti-diagonal and interacting fixed point result in a 10\% too large top mass in comparison with experiment. Whether similar trajectories can also emanate directly from the anti-diagonal and interacting fixed point remains as an intriguing open question.
\\

For any $f_y$ that is larger than $f_y\approx -2.1886\times 10^{-4}$, one can always accommodate the measured IR values of all quark masses. This is a consequence of the relevance of all Yukawa couplings at the free fixed point. For rather large $f_y$, the corresponding trajectories no longer pass in the vicinity of any of the interacting fixed points, as these move towards larger values of the Yukawa couplings, as $f_y$ is increased.

\section{Conclusions and outlook}\label{sec:conclusions}

In this paper, we have explored possible patterns of the quark masses and mixings in a new physics setting such that the quark-gauge sector of the SM is rendered UV complete. Adding only two parameters to the beta functions of the SM, we have found a complex pattern of Renormalization Group fixed points. We have explored the idea that
interacting fixed points of the Renormalization Group can lead to a
quantum field theory with a higher 
predictive power than one would infer from canonical power 
counting. At interacting fixed points, couplings of the renormalizable SM, corresponding to free parameters according to canonical power-counting, can in principle turn into calculable quantities, reducing the number of free parameters of the theory. 

We have explored this general set of ideas for the quark-gauge sector of the SM. We have met the minimal requirement of solving the Landau-pole problem
of the Abelian gauge coupling by the addition of an antiscreening contribution. The new term in the beta functions is linear such that the new physics is parameterized by one parameter, $f_g$. 
In a given new-physics setting, $f_g$ should be calculable from first principles, turning the Abelian gauge coupling into a predictable quantity.
Here, we have fixed the parameter $f_g$ by matching the calculated value of $g_Y$ to the observed one.

Similarly, we have added a flavor-independent new-physics contribution $f_y$ to the flow of the general matrices of Yukawa couplings. This 
is logically independent of the effect giving rise to
asymptotic safety of the $U(1)$ gauge coupling. As a matter of fact, triviality in the Yukawa sector could potentially also be avoided by a completely asymptotically free fixed point, as explored, e.g., in 
\cite{Cheng:1973nv,Giudice:2014tma,Holdom:2014hla,Gies:2016kkk,Gies:2018vwk,Gies:2019nij},
or by an asymptotically safe fixed point induced by higher-dimensional operators 
\cite{Gies:2009hq,Gies:2014xha}, see, however, \cite{Gies:2017zwf}.
On the other hand, since these contributions could have a common source, we included both. In particular, the parameter $f_y$ can give rise to additional calculable structures in the quark sector.
For the parameter $f_y$, we have explored different values, since it influences which UV fixed points can be connected to (approximately) viable IR physics. In particular, there are fixed points at which fixing $f_y$ by matching the calculated value of a quark mass, e.g., $m_b$, to the observed one leads to a genuine prediction, e.g., for the mass of the top.

We come to three important conclusions
\begin{itemize}
\item The new-physics contribution drops out of the running of the CKM-matrix elements. The IR attractive fixed point, discovered in \cite{Pendleton:1980as} therefore persists. Interestingly, the measured values are rather close to this fixed point. Given that the running of the CKM matrix elements is in general very slow (such that hundreds of orders of magnitude need to be crossed for a significant change to occur in the CKM elements), this is compatible with the idea that it is this fixed point that dominates the IR physics of the CKM matrix.
\item By matching to the exact SM values in the IR, any $f_y$ above a critical value allows to construct UV-complete trajectory emanating from an asymptotically free fixed point for all Yukawa couplings. For specific values of $f_y$, trajectories with IR endpoints in close vicinity to the SM-values can be dominated by one or several intermediate scaling regimes of interacting fixed points over hundreds of orders of magnitude, cf.~Fig.~\ref{fig:conclusions} for a schematic overview.
\item Extending the results of \cite{Eichhorn:2018whv}, we find that the slow running of CKM parameters modifies the scaling relation, switching between the anti-diagonal and the diagonal interacting fixed point. 
On the one hand, this transition results in a top mass that is about 10 \%  too large. Even at the current level of approximation which is only a one-loop study, this is in strong contradiction with experimental values. On the other hand, the CKM transition is crucial in accommodating non-vanishing masses for the strange and the down quark, which only become relevant once the CKM matrix assumes an anti-diagonal configuration. Additionally, the interplay of CKM-matrix elements and Yukawa couplings leads to upper bounds on those Yukawa couplings that remain free parameters. 
\end{itemize}

A peculiar feature of the flows that we have
studied is the extremely slow evolution of the CKM parameters.
In some cases, in order to come close to a fixed point in the UV,
one has to go to energies of order 10$^{10000}$GeV,
that even dwarf the scale of the $U(1)$ Landau pole.
This behavior can be traced to the fact that the scaling exponents
of some fixed points are tiny. 
Additionally, the flows that we study actually pass at least one intermediate scaling regime in the vicinity of an interacting fixed point, which slows down the flows further.
If one insists on asymptotic safety in the strict sense, namely trajectories that precisely start from a fixed point in the UV, then this seems unavoidable.
The physical meaning of these ultra-high scales is
questionable, so for the time being we merely record
them as mathematical curiosities. 
Since the meaning of a true UV completion is to enable extensions to \emph{arbitrarily high} scales, one could
take these RG trajectories at face value at scales even much higher than that. On the other hand, one may note that for all practical purposes,
above the Planck scale one is very near a fixed point.
We can, therefore, identify trajectories that have the correct
IR limit and can be greatly (though not infinitely) extended in the UV.
The predictive power of these nearly-asymptotically safe trajectories
cannot be quantified in an integer number of predictions. Instead, it consists of the map of large UV intervals to small IR intervals.
Less predictive trajectories can be truly UV-complete.

In the main text, we have restricted our attention to
flavor-blind new physics. A study of a special case of generation-breaking new physics, see \ref{app:genbeta}, where a distinct new-physics parameter is introduced for each generation of quarks reveals,  that contrary to what one might expect, the larger number of free parameters does not allow us to improve the agreement with experiments. 

The best agreement with phenomenology is already achieved for the case of ``minimal" new physics which does not distinguish the generations and only comes with one free parameter in the quark sector.

As a specific example of the new physics that can generate the contributions $f_g$ and $f_y$, we reviewed the related results within asymptotically safe gravity, where in fact the fixed-point structure of such gauge-Yukawa systems and the corresponding predictions was first discussed in \cite{Eichhorn:2018whv}. 
Nevertheless, let us emphasize that any kind of new physics 
which would provide such 
contributions to the beta functions of the SM couplings will enable the predictive scenario we discussed here. In the future, an extension to the lepton sector of the SM is of obvious interest, as is a broadening of our work to (minimal) extensions of the SM. Lastly, the exploration of possible BSM candidates that could generate terms of the form we include here is of strong interest.

\section*{Acknowledgements}
We thank  Axel Maas for helpful discussions.

 A.~E.~and A.~H.~are supported by the DFG under grant no.~Ei-1037/1. A.~E.~is also supported by a research grant (29405) from VILLUM FONDEN and partially supported by a visiting fellowship at the Perimeter Institute for Theoretical Physics. A.~H.~is supported by a Royal Society International Newton fellowship and also acknowledges support by the German Academic Scholarship Foundation. A.~H.~and C.~N.~thank CP3-Origins for hospitality during a part of this project. M.~S.~is grateful to the Paul-Urban foundation for financial support.

\newpage

\appendix

\section{Lines and planes of fixed points from one-loop RG invariants}\label{app:invariants}

Given the structure of the Yukawa beta functions at 1-loop, we can construct two quantities that are invariant along the RG flow. For $Y_{U}$ and $Y_{D}$ the up and down Yukawa matrices, we have the following two invariants in the case of three generations of quarks
\begin{equation}
    I_{(1)}=\frac{{\rm Tr}(M_{U}M_{D})}{({\rm det}(M_{U}M_{D}))^{1/3}}, \quad  I_{(2)}={\rm Tr}((M_{U}M_{D})^{-1})({\rm det}(M_{U}M_{D}))^{1/3},
    \label{eq:rginv}
\end{equation}
where $M_{U}=Y_{U}Y_{U}^{\dagger}$ and $M_{D}=Y_{D}Y_{D}^{\dagger}$. In the diagonalized basis, we have
\begin{equation}
    I_{(1)}=\sum_{i\rho}\frac{y_{i}^{2}y_{\rho}^{2}|V_{i\rho}|^{2}}{(y_{t}y_{c}y_{u}y_{b}y_{s}y_{d})^{2/3}}, \quad  I_{(2)}={(y_{t}y_{c}y_{u}y_{b}y_{s}y_{d})^{2/3}}\sum_{i\rho}y_{i}^{-2}y_{\rho}^{-2}|V_{i\rho}|^{2}.
    \label{eq:rginv2}
\end{equation}

These are invariants for the flow in the 10-dimensional
space of the Yukawas and $X$, $Y$, $Z$, $W$.
If we put $X$, $Y$, $Z$, $W$ on a fixed point, then these
are invariant for the flow in the 6-dimensional space
of the Yukawas only.

For example when the mixing matrix is the identity 
($X=1$, $Y=0$, $Z=0$, $W=1$), the invariants in (\ref{eq:rginv2}) become
\begin{equation}
    I_{(1)}=\frac{(y_{t}^{2}y_{b}^{2}+y_{c}^{2}y_{s}^{2}+y_{u}^{2}y_{d}^{2})}{(y_{t}y_{c}y_{u}y_{b}y_{s}y_{d})^{2/3}}, \quad I_{(2)}=(y_{t}y_{c}y_{u}y_{b}y_{s}y_{d})^{2/3}\left(\frac{1}{y_{t}^{2}y_{b}^{2}}+\frac{1}{y_{c}^{2}y_{s}^{2}}+\frac{1}{y_{u}^{2}y_{d}^{2}}\right),
\end{equation}
where, for simplicity, we have omitted the fixed-point symbol $*$ in each Yukawa coupling.
However one can show that each term in these sums is an invariant
by itself:
\begin{equation}
    V_{1}=\frac{y_{u}^{2}y_{d}^{2}}{y_{t}y_{c}y_{b}y_{s}}, \quad V_{2}=\frac{y_{c}^{2}y_{s}^{2}}{y_{t}y_{u}y_{b}y_{d}}, \quad V_{3}=\frac{y_{t}^{2}y_{b}^{2}}{y_{c}y_{u}y_{s}y_{d}}.
\end{equation}
Let us see how these invariants arise and how they are related to
the lines and planes of fixed points.

For all of the $6$ particular CKM configurations we have considered, and the gauge fixed point ($g_{1}=\sqrt{96f_{g}\pi^{2}/41}$, $g_{2}=0$, $g_{3}=0$), the
Yukawa beta functions take the form
\begin{equation}
    \beta_{y_{j}^{2}}=y_{j}^{2}h_{j}(y_{k}^2),
\end{equation}
where $h_{j}$ are linear functions of the couplings $y_{k}^2$. When looking for non-trivial fixed points, we have to solve the system of equations $h_{j}=0$. Thus, when 
some of the $h_i$'s
are linearly dependent we have infinitely many solutions. Hence, surfaces of fixed points appear.

In general, an RG invariant is a quantity $I$ that satisfies $\frac{d}{dk}I=0$. In terms of the beta functions of the couplings
\begin{equation}
    0=\beta_{y_{j}^{2}}\partial_{j}I=y_{j}^{2}h_{j}\partial_{j}I.
    \label{eq:inv}
\end{equation}
We take for the moment the case of $n$ couplings $y_{j}$. Then, if there are some dependent function $h_{j}$, e.g., $h_{n-1}$ and $h_{n}$, we have that
\begin{equation}
    h_{n-1}=\sum_{i=1}^{n-2}A_{i}h_{i}, \quad  h_{n}=\sum_{i=1}^{n-2}B_{i}h_{i}.
    \label{eq:decomp}
\end{equation}
Consequently, eq. (\ref{eq:inv}) becomes
\begin{equation}
    \sum_{j=1}^{n-2}(y_{j}^{2}\partial_{j}I+y_{n-1}^{2}A_{j}\partial_{n-1}I+y_{n}^{2}B_{j}\partial_{n}I)h_{j}=0.
\end{equation}
Since by assumption the $n-2$ functions $h_{j}$ are linearly independent, 
each of their coefficients must vanish separately.
This means that any function of the variable
\begin{equation}
    V_{T}=\frac{(y_{1}^{2})^{A_{1}+B_{1}}(y_{2}^{2})^{A_{2}+B_{2}}\dots(y_{n-2}^{2})^{A_{n-2}+B_{n-2}}}{y_{n-1}^{2}y_{n}^{2}}.
    \label{eq:wvariable}
\end{equation}
is a flow invariant.

For example, in the case considered above
($X=1$, $Y=0$, $Z=0$, $W=1$)
we have the linear relations
\begin{equation}
h_{t}=h_{u}+h_{d}-h_{b}, \quad h_{c}=h_{u}+h_{d}-h_{s}.
\label{eq:depend}
\end{equation}
or $A_u=1$, $A_b=-1$, $A_s=0$, $A_d=1$, $B_u=1$, $B_b=0$, $B_s=-1$, $B-d=1$. Hence, $W$ coincides with $V_1^2$.
(Obviously any function of an invariant is an invariant).

Alternatively, we can also write
\begin{equation}
    h_{t}=h_{c}+h_{s}-h_{b}, \quad h_{u}=h_{c}+h_{s}-h_{d},
\end{equation}
and
\begin{equation}
    h_{u}=h_{t}+h_{b}-h_{d}, \quad h_{c}=h_{t}+h_{b}-h_{s}.
\end{equation}
which are obtained from the previous linear relation
by the permutations ($u\leftrightarrow c$, $d\leftrightarrow s$) and ($u\leftrightarrow t$, $d\leftrightarrow b$). 
These give rise to the invariants $V_2$ and $V_3$.

In summary, we see that the surfaces of FPs and the one-loop invariants
both originate from linear relations between the beta functions.
If we allow one coupling to be zero, the number of equations decreases and then only one linear relation remains. As a result, we obtain a line of fixed points for this case.

\section{Fixed points for two generations}
\label{app:2gen}

\subsection{Classification}

In this section of the appendix we list the fixed points for the Yukawa couplings
for the two-generation case, 
allowing for mixing and assuming that the gauge couplings are at their respective fixed point 
$g_{Y \ast}^2 = \frac{96}{41} f_g \pi^2$ and $g_{2 \ast}^2 = 0 = g_{3 \ast}^2$. 
The equations for the fixed points of the Yukawa couplings and the CKM parameter $W$ constitute a system of quadratic equations which possesses, following a na\"ive counting scheme, two lines of solutions and 20 further discrete solutions. These will be discussed in the following. 

Let us clarify that when we invoke phenomenological viability in the following, we refer to the IR values of the Yukawa couplings of the second and third generation only.

First of  all, we note that -- based on the discussion in the preceding section \ref{app:invariants} --
the two lines of solutions will collapse to a number of discrete
solutions upon the inclusion of two-loop terms. From the discussion below it will become clear that, based on permutation symmetries, one expects there to be four discrete solutions. We do not determine
these fixed points explicitly because the two-generation case serves only as an intermediate step to understand the SM case, {\it i.e.}, three generations. However, as two-loop terms will reduce the lines to points, we will list these (anticipated) 
solutions on the same footing as the other solutions which are discrete already on the one-loop level.

Second, the permutation symmetries $t \leftrightarrow c$ and   $b\leftrightarrow s$ lead to the appearance of all fixed points in two doublets. We exemplified
this behavior based on the solution discussed in Sec.~\ref{sec:2GenFPstructure}, see 
Table~\ref{tab:twogen} for the explicit values of the couplings at the respective fixed point.

First, regarding Table~\ref{tab:twogen} the relation between the solution labeled 1a to the one labeled 1b 
is due to the intergenerational permutation $b \leftrightarrow s$, changing the value 
$W_\ast =0$ changes to $W_\ast =1$.
Correspondingly, the solutions 1c and 1d can be obtained by obvious permutations. To identify the phenomenologically interest fixed points, we applied the following consideration: We start from the observation that 
$g_{Y \ast}^2 = \frac{96}{41} f_g \pi^2$ only makes sense if $f_g$ is positive. In order to avoid
negative values for the $y_{\mathrm{i} \ast}^2$, one needs to additionally require for this quartet that
$f_y \geq \frac{19}{82} f_g$. This implies that $f_y$ is positive. As there are poles in the beta 
functions for the CKM matrix element, a phenomenologically viable solution can only come from a fixed point with
$y_{t \ast}^2 > y_{c \ast}^2$ and 
$y_{b \ast}^2 > y_{s \ast}^2$.
The first condition excludes the solutions 1c and 1d, the second 1b and 1c,
leaving us with the case 1a as the only phenomenologically viable one.

Second, to describe the two lines of fixed points we note that these can be defined via the relations
\be
 y_{t \ast}^2 + y_{c \ast}^2 =  \left( \frac{140}{123} f_g + \frac 8 3 f_y \right) \pi^2  \, ,  
\quad 
y_{b \ast}^2 + y_{s \ast}^2 =  \left( - \frac{52}{123} f_g + \frac 8 3 f_y \right) \pi^2 \, ,
\ee
and
\be
 y_{c \ast}^2 - y_{b \ast}^2 = y_{t \ast}^2 - y_{s \ast}^2 
= \frac{1}{3} g_{Y \ast}^2 = \frac{32}{41} f_g \pi^2 \, ,  
\ee
for the line of solutions with $W_\ast =0$ and 
\be
y_{t \ast}^2 - y_{b \ast}^2 = y_{c \ast}^2 - y_{s \ast}^2 
= \frac{1}{3} g_{Y \ast}^2 = \frac{32}{41} f_g \pi^2 \, ,
\ee
for the line of solutions with $W_\ast =1$. Requiring $y_{i \ast}^2 \geq 0$
 $\; \forall \, i \in \left\lbrace t, c, b, s \right\rbrace$ 
and assuming $f_g > 0$,
the constraint
\be
f_y \ge \frac{13}{82} f_g, 
\ee
has to be fulfilled. Hence, the two lines will be of finite length, {\it i.e.}, they will have endpoints. 
As already stated, we will not resolve how two-loop terms will lift the degeneracy. To represent 
these FPs, one of the endpoints, namely the one with  $y_{s \ast}^2=0$,
will be shown in Table \ref{tab:2_gen_yukawa_FP} as case 2. 

\begin{table}[ht]
\begin{center}
\begin{tabular}{@{}|c|cccc|c|c|@{}} \hline
\#   & $y_{t \ast}^2/\pi^2$ & $y_{c \ast}^2/\pi^2$ & $y_{b \ast}^2/\pi^2$ & $y_{s \ast}^2/\pi^2$ & $W_\ast $ & Allowed \\[-5mm]
    & & & & & &  range \\ \hline
 1 & $\frac{16}{15} \left( f_g + 2 f_y \right)$ & $0$ 
    & $ \frac{16}{615} \left( - 19 f_g + 82 f_y \right)$ & $0$ &  $0$ & $f_y \geq \frac{19}{82} f_g$ \\
 2 & $\frac{4}{123} \left( 11 f_g + 82 f_y \right)$ & $\frac{32}{41} f_g $ 
    &  $\frac{4}{123} \left( -13 f_g + 82 f_y \right)$ & $0$ & $1$ & $f_y \geq \frac{13}{82} f_g$ \\
 3 & $ \frac{4}{123} \left( 23 f_g + 82 f_y \right) $ & $0$ 
    & $ \frac{4}{123} \left( - f_g + 82 f_y \right)$ & $0$ &  $1$ & $f_y \geq \frac{1}{82} f_g$ \\
 4 & $t_4$ & $0$ & $b_4$ & $s_4$ &  $W_{4 \ast}$ & $f_y = \frac{7}{82} f_g$ \\
 5 & $t_5$ & $c_5$ & $b_5$ & 0 & $W_{5 \ast}$ & $f_y = \frac{43}{82} f_g$  \\
 6 & $\frac{4}{123} \left( 35 f_g + 82 f_y \right)$ & 0 
    & $\frac{4}{123} \left( 11 f_g + 82 f_y \right)$ & $- \frac{32}{41} f_g$  & 1 & --- \\
\hline
\end{tabular}
\end{center}
\caption{Fixed point solutions in the Yukawa sector for the gauge coupling fixed points 
$g_{Y \ast}^2 = \frac{96}{41} f_g \pi^2$ and $g_{2 \ast}^2 = 0 = g_{3 \ast}^2$ such that 
$y_{t \ast}^2 > y_{c \ast}^2$ and 
$y_{b \ast}^2 > y_{s \ast}^2$.  The last column gives the allowed range 
of values for $f_y$, for which 
$y_{i \ast}^2 \geq 0$ $\; \forall \, i \in 
\left\lbrace t, c, b, s \right\rbrace$. 
For further definitions see the main text.}
\label{tab:2_gen_yukawa_FP}
\end{table}

From the six cases displayed in Table \ref{tab:2_gen_yukawa_FP},  one can generate the four endpoints of the two lines and the 20 discrete solutions by applying the permutations 
$t\leftrightarrow c$ and $b\leftrightarrow s$. From fixed point 1 in 
Table \ref{tab:2_gen_yukawa_FP} one can generate the cases 1a - 1d, and analogously for the 
other classes 2 - 6. Note that with an odd number of permutations of quarks, the value of $W_\ast$ changes. 
Each representative has been chosen such that the conditions 
$y_{t \ast}^2 > y_{c \ast}^2$ and 
$y_{b \ast}^2 > y_{s \ast}^2$
are fulfilled.

The \underline{fixed point 1} in Table \ref{tab:2_gen_yukawa_FP}, respectively, 1a in Table \ref{tab:twogen}
provides a phenomenologically viable fixed point, and therefore the flow 
corresponding to this fixed point is studied in Sec.~\ref{sec:2gen}. 
We note that the relation $y_{t \ast}^2 - y_{b \ast}^2 =
\frac {64}{41} f_g \pi^2 = \frac 2 3 g^2_{Y^\ast}$ is fulfilled at this fixed point.

The \underline{fixed point 2}, which is an endpoint of a line of solutions, fulfils the relations
$y_{t \ast}^2 - y_{b \ast}^2 = \frac 1 3 g^2_{Y^\ast}$ as well as
$y_{c \ast}^2 = \frac 1 3 g^2_{Y^\ast}$. Note that the value of $y_{c \ast}^2$ depends only 
on $f_g$ and not on $f_y$. Hence it is approximately determined by the value of the hypercharge 
coupling in the UV. This leads to a much too large Yukawa coupling for the charm, and therefore this 
fixed point is excluded by phenomenology.

The \underline{fixed point 3} corresponds to the extension of (\ref{eq:fptby}).
However, it enforces
a vanishing strange mass, see Sec.~\ref{sec:2gen} for a discussion of this point. 
Note that also in this case $y_{t \ast}^2 - y_{b \ast}^2 = \frac 1 3 g^2_{Y^\ast}$.

For the \underline{fixed point 4}, the values for the Yukawa couplings and the CKM matrix element are given by
\bea
y_{t \ast}^2 &=& t_4 = \frac{16}{1107} \left( 65 f_g + 82 f_y \right) \pi^2 \, ,
 \nonumber \\
y_{b \ast}^2 &=& b_4 = \frac{8}{1107} \left( - 21 f_g + 246 f_y - r \right) \pi^2 \, ,
\nonumber \\
y_{s \ast}^2 &=& s_4 = \frac{8}{1107} \left( - 21 f_g + 246 f_y + r \right) \pi^2 \, ,
\nonumber \\
W_{4 \ast} &=& \frac 1 2 + \frac{r}{2(65 f_g + 82 f_y)} ,
\eea
with $r = \sqrt{3 (7 f_g - 82 f_y) (65 f_g + 82 f_y)}$.
Requiring $y_{i \ast}^2 \geq 0$ $\; \forall \, i \in 
\left\lbrace t, c, b, s \right\rbrace$
as well as a real value for $r$ leads to
\be
f_y = \frac 7{82}  f_g \, ,
\ee
for $f_g>0$.
This implies, in addition to  $y_{c \ast}^2 =0$, 
\bea
y_{t \ast}^2 &=& t_4 = \frac{128}{123} f_g \pi^2  =  \frac 4 9 g^2_{Y^\ast} \, ,
 \nonumber \\
y_{b \ast}^2 &=& b_4 = 0 \, ,
\nonumber \\
y_{s \ast}^2 &=& s_4 = 0 \, ,
\nonumber \\
W_{4 \ast} &=& \frac 1 2 \, .
\eea
This does clearly not lead to a phenomenologically viable fixed point.

For the \underline{fixed point 5}, the values for the Yukawa couplings and the CKM matrix element are 
given by
\bea
y_{t \ast}^2 &=&  t_5=\frac{8}{1107} \left( 87 f_g + 246 f_y - s \right) \pi^2\, ,
\nonumber \\
y_{c \ast}^2 &=&  c_5=\frac{8}{1107} \left( 87 f_g + 246 f_y + s \right) \pi^2  \, ,
\nonumber \\
y_{b \ast}^2 &=&  b_5 = \frac{16}{1107} \left( - 43 f_g + 82 f_y \right) \pi^2 \, , 
\nonumber \\
W_{5 \ast} &=& \frac 1 2 - \frac{s}{2(43 f_g - 82 f_y)}, 
\eea
with 
$s=\sqrt{3 (43 f_g - 82 f_y) (29 f_g + 82 f_y)}$.
Requiring $y_{i \ast}^2 \geq 0$ $\; \forall \, i \in 
\left\lbrace t, c, b, s \right\rbrace$
as well as a real value for $s$ leads to
\be
f_y = \frac {43} {82}  f_g \, ,
\ee
for $f_g>0$.
In addition to  $y_{s \ast}^2 =0$, this also implies
\bea
y_{t \ast}^2 &=& t_5 = \frac{64}{41} f_g \pi^2  =  \frac 2 3 g^2_{Y^\ast} \, ,
 \nonumber \\
y_{c \ast}^2 &=& c_5 = \frac{64}{41} f_g \pi^2  =  \frac 2 3 g^2_{Y^\ast} \, ,
\nonumber \\
y_{b \ast}^2 &=& b_5 = 0 \, .
\eea
 Again, this is not a  phenomenologically viable fixed point.

The \underline{fixed point 6} is excluded right away because $y_{s \ast}^2$ is negative
for all $f_g > 0$.

\subsection{Scaling exponents}

The set of scaling exponents is the same for all four fixed points within each of the classes 1 --- 6. Here, we provide the eigenvalues of the stability 
matrix multiplied by an additional negative sign, i.e.,, the scaling exponents
\be
\theta_{(I)} = - {\mathrm {eig}}^{(I)} \left( \frac{\partial \beta_{g_i}}{\partial g_j} \right) \, ,
\ee
where $\vec{g}=(y_t, y_c,y_b, y_s, g_Y, W)$.

\begin{table}[H]
{\scriptsize{
\begin{tabular}{|c|c|c|c|}
\hline
 & 1 & 2 & 3 \\
\hline
$\theta_1$ & $-2 f_g$ &  $-2 f_g$ & $-2 f_g$ \\
$\theta_2$ & $\frac{2}{205}\left(11f_g+82 f_y \right)$ & 0 & $-\frac{3}{41}f_g$\\
$\theta_3$ & $\frac{1}{205}\left(11f_g+82 f_y \right)$ & $-\frac{11}{82}f_g-f_y$ & $\frac{3}{41}f_g$ \\
$\theta_4$ & $\frac{1}{205}\left(11f_g+82 f_y \right)$ &  & $-\frac{11}{82} f_g- f_y$  \\
$\theta_5$ & $\frac{1}{205} \Bigl(-3 (11f_g +82 f_y) $ & &
$\frac{1}{164} \Bigl(-3 (11f_g +82 f_y) $ \\ 
                  & $-2\sqrt{1246 f_g^2+ 2 (11f_g) \, (82 f_y) + (82 f_y)^2} \Bigr)$ & 
& $- \sqrt{1273 f_g^2 +2 (11f_g) \, (82 f_y) + (82 f_y)^2} \Bigr)$\\
$\theta_6$ &$\frac{1}{205}\Bigl(-3 (11f_g+82 f_y) $ &
&
$\frac{1}{164} \Bigl(-3 (11f_g+82 f_y) $\\
                  & $+2\sqrt{1246 f_g^2+ 2 (11f_g) \, (82 f_y) + (82 f_y)^2} \Bigr)$ &
& $+ \sqrt{1273 f_g^2 +2 (11f_g) \, (82 f_y) + (82 f_y)^2} \Bigr)$\\
\hline
\end{tabular}}}
\caption{\label{tab:2genCritExp}
Scaling exponents for the fixed points 1 - 3 given in Tab.~\ref{tab:2_gen_yukawa_FP}. As $f_g > 0$ and $f_y > 0$, most scaling exponents are strictly positive or negative. 
However, the sign of $\theta_6$ for fixed points 1 and 3 depends on the precise values of the parameters  $f_g$ and $f_y$. $\theta_6$ is negative 
for $f_y > \frac{19}{82} f_g$ and $f_y > \frac{1}{82} f_g$, respectively, which are just the allowed ranges for the fixed points 1 and 3 (cf.~Tab.~\ref{tab:2_gen_yukawa_FP}), excluding the lower bound, at which the scaling exponents are zero. The scaling exponents $\theta_4$, $\theta_5$ and $\theta_6$ for the fixed point 2 cannot be uniquely determined as these fixed points are the endpoints of 
a line.}
\end{table}

\section{Two loop beta functions}\label{app:twoloop}

Here, we present the two-loop expressions for the quark Yukawa beta functions in the absence of leptons and scalar quartic coupling. For the matrices $Y_{U}$ and $Y_{D}$
introduced in Section \ref{sec:betas}, we have the two-loop coefficients
{\small\begin{align}
    &\beta_{Y_{U}}^{\ (2)}=\left[ \tfrac{3}{2}(Y_{U}Y_{U}^{\dg})^{2}-\tfrac{1}{4}Y_{U}Y_{U}^{\dg}Y_{D}Y_{D}^{\dg}-Y_{D}Y_{D}^{\dg}Y_{U}Y_{U}^{\dg}\right. \nonumber \\
    &\left.\phantom{\tfrac{3}{2}} +\tfrac{11}{4}(Y_{D}Y_{D}^{\dg})^{2}+A_{UU}Y_{U}Y_{U}^{\dg}+A_{UD}Y_{D}Y_{D}^{\dg}+B_{U}\right]\frac{Y_{U}}{(4\pi)^{4}},\\
    &\beta_{Y_{D}}^{\ (2)}=\left[\tfrac{3}{2}(Y_{D}Y_{D}^{\dg})^{2}-\tfrac{1}{4}Y_{D}Y_{D}^{\dg}Y_{U}Y_{U}^{\dg}-Y_{U}Y_{U}^{\dg}Y_{D}Y_{D}^{\dg}\right. \nonumber \\
    &\left.\phantom{\tfrac{3}{2}}+\tfrac{11}{4}(Y_{U}Y_{U}^{\dg})^{2}+A_{DD}Y_{D}Y_{D}^{\dg}+A_{DU}Y_{U}Y_{U}^{\dg}+B_{D}\right]\frac{Y_{D}}{(4\pi)^{4}},
\end{align}}
where the superscript $(2)$ denotes two loop contribution. Here we have made use of the trace
\begin{equation}
    Y_{2}(S)=\Tr\left(3Y_{U}Y_{U}^{\dg}+3Y_{D}Y_{D}^{\dg}\right).
\end{equation}
This is useful in defining the following contributions
\begin{align}
    A_{UU}&=\left(\tfrac{223}{48}g_{1}^{2}+\tfrac{135}{16}g_{2}^{2}+16g_{3}^{2}\right)-\tfrac{9}{4}Y_{2},\\
    A_{UD}&=\tfrac{5}{4}Y_{2}(S)-\left(\tfrac{43}{48}g_{1}^{2}-\tfrac{9}{16}g_{2}^{2}+16g_{3}^{2}\right),\\
    A_{DD}&=\left(\tfrac{187}{48}g_{1}^{2}+\tfrac{135}{16}g_{2}^{2}+16g_{3}^{2}\right)-\tfrac{9}{4}Y_{2}(S),\\
    A_{DU}&=\tfrac{5}{4}Y_{2}(S)-\left(\tfrac{79}{48}g_{1}^{2}-\tfrac{9}{16}g_{2}^{2}+16g_{3}^{2}\right).
\end{align}
Similarly, we have the quartet contribution in the gauge couplings
\begin{align}
    B_{U}=-\chi_{4}(S)+&\left(\tfrac{1}{8}+\tfrac{145}{81}N_{sm}\right)g_{1}^{4}-\left(\tfrac{35}{4}-N_{sm}\right)g_{2}^{4}-\left(\tfrac{404}{3}-\tfrac{80}{9}N_{sm}\right)g_{3}^{4} \nonumber \\
    &-\tfrac{3}{4}g_{1}^{2}g_{2}^{2}+\tfrac{19}{9}g_{1}^{2}g_{2}^{3}+9g_{2}^{2}g_{3}^{2}+\tfrac{5}{2}Y_{4}(S),
\end{align}
\begin{align}
    B_{D}=-\chi_{4}(S)-&\left(\tfrac{29}{72}+\tfrac{5}{81}N_{sm}\right)g_{1}^{4}-\left(\tfrac{35}{4}-N_{sm}\right)g_{2}^{4}-\left(\tfrac{404}{3}-\tfrac{80}{9}N_{sm}\right)g_{3}^{4} \nonumber \\
    &-\tfrac{9}{4}g_{1}^{2}g_{2}^{2}+\tfrac{31}{9}g_{1}^{2}g_{2}^{3}+9g_{2}^{2}g_{3}^{2} +\tfrac{5}{2}Y_{4}(S) ,
\end{align}
where
\begin{equation}
    \chi_{4}(S)=\tfrac{9}{4}\Tr\left[3(Y_{U}Y_{U}^{\dg})^{2}+3(Y_{D}Y_{D}^{\dg})^{2}-\tfrac{2}{3}Y_{U}Y_{U}^{\dg}Y_{D}Y_{D}^{\dg}\right],
\end{equation}
\begin{equation}
    Y_{4}(S)=(\tfrac{17}{12}g_{1}^{2}+\tfrac{9}{4}g_{2}^{2}+8g_{3}^{2})\Tr(Y_{U}Y_{U}^{\dg})+(\tfrac{5}{12}g_{1}^{2}+\tfrac{9}{4}g_{2}^{2}+8g_{3}^{2})\Tr(Y_{D}Y_{D}^{\dg}).
\end{equation}
Herein, $N_{sm}$ is the number of families in the SM.\\

In the mass basis, we have
 \begin{align}
     &\beta_{y_{i}^{2}}^{\ (2)}=\left[3y_{i}^{4}-\frac{5}{2}y_{i}^{2}\sum_{\rho}y_{\rho}^{2}|V_{i\rho}|^{2}+\frac{11}{2}\sum_{\rho}y_{\rho}^{4}|V_{i\rho}|^{2} \right. \nonumber \\
     &\left.\phantom{\tfrac{3}{2}} +2A_{UU}y_{i}^{2}+2A_{UD}\sum_{\rho}y_{\rho}^{2}|V_{i\rho}|^{2}+2B_{U}\right]\frac{y_{i}^{2}}{(4\pi)^{4}},
\end{align}
\begin{align}
     &\beta_{y_{\rho}^{2}}^{\ (2)}=\left[3y_{\rho}^{4}-\frac{5}{2}y_{\rho}^{2}\sum_{i}y_{i}^{2}|V_{i\rho}|^{2} +\frac{11}{2}\sum_{i}y_{i}^{4}|V_{i\rho}|^{2}\right. \nonumber \\
     &\left.\phantom{\tfrac{3}{2}} +2A_{DD}y_{\rho}^{2}+2A_{DU}\sum_{i}y_{i}^{2}|V_{i\rho}|^{2}+2B_{D}\right]\frac{y_{\rho}^{2}}{(4\pi)^{4}},
\end{align}
where Latin indices stand for up-type quarks while Greek indices stand for down-type quarks. In this case, the $Y_{F}$-dependent quantities defined before become 
\begin{equation}
    Y_{2}(S)=3\sum_{i}y_{i}^{2}+3\sum_{\rho}y_{\rho}^{2},
\end{equation}
\begin{equation}
    \chi_{4}(S)=\tfrac{9}{4}\Big[3\sum_{i}y_{i}^{4}+3\sum_{\rho}y_{\rho}^{4}-\tfrac{2}{3}\sum_{i,\rho}|V_{i\rho}|^{2}y_{i}^{2}y_{\rho}^{2}\Big],
\end{equation}
\begin{equation}
    Y_{4}(S)=(\tfrac{17}{12}g_{1}^{2}+\tfrac{9}{4}g_{2}^{2}+8g_{3}^{2})\sum_{i}y_{i}^{2}+(\tfrac{5}{12}g_{1}^{2}+\tfrac{9}{4}g_{2}^{2}+8g_{3}^{2})\sum_{\rho}y_{\rho}^{2}. 
\end{equation}
For the magnitude square of the CKM elements, we have the two loop expression
\begin{align}
    &\beta_{|V_{i\rho}|^{2}}^{\ (2)}=\left[\sum_{\beta,j\neq i}\frac{y_{\beta}^{2}}{y_{i}^{2}-y_{j}^{2}}Q_{ij\beta}(V_{i\beta}V_{j\beta}^{*}V_{j\rho}V_{i\rho}^{*}+V_{i\beta}^{*}V_{j\beta}V_{j\rho}^{*}V_{i\rho})\right. \nonumber \\
    &\left.\phantom{\frac{3}{2}}+\sum_{j,\beta\neq\rho}\frac{y_{j}^{2}}{y_{\rho}^{2}-y_{\beta}^{2}}Q_{\rho\beta j}\left(V_{j\beta}^{*}V_{j\rho}V_{i\beta}V_{i\rho}^{*}+V_{j\beta}V_{j\rho}^{*}V_{i\beta}^{*}V_{i\rho}\right)\right]\frac{1}{(4\pi)^{4}}
   \label{betackm}
\end{align}
where we have introduced the factors
\begin{align}
 & Q_{ij\beta}=\tfrac{1}{2}y_{i}^{2}y_{j}^{2}+ (y_{i}^{4}+y_{j}^{4})-\tfrac{11}{4}(y_{i}^{2}+y_{j}^{2})y_{\beta}^{2}-A_{UD}(y_{i}^{2}+y_{j}^{2}),\\
 & Q_{\rho\beta j}=\tfrac{1}{2}y_{\rho}^{2}y_{\beta}^{2}+ (y_{\rho}^{4}+y_{\beta}^{4})-\tfrac{11}{4}(y_{\rho}^{2}+y_{\beta}^{2})y_{j}^{2}-A_{DU}(y_{\rho}^{2}+y_{\beta}^{2}).
\end{align}

\section{General linear terms in the Yukawa beta functions}\label{app:genbeta}

In this section, we explore the impact of general BSM contributions on the beta functions of the Yukawa matrices $Y_{U}$ and $Y_{D}$.
As in the main text, we consider linear modifications, that is, corrections that are proportional
to the matrices themselves. Yet, as the key difference, instead of taking a parameter $f_{y}$ multiplying each matrix ($Y_{U}$, $Y_{D}$), we study 
BSM contributions in matrix form. 
Our goal is to understand whether and how we can obtain BSM contributions to the Yukawa
beta functions that distinguish different generations of quarks. The main motivation lies in the fact that the flavor-blind contribution $f_y$ explored in the main text
does not appear to be sufficient structure to enable the prediction of hierarchies of quark masses, beyond the hierarchy between top and bottom quark. Giving up on the premise of flavor-blindness is a well-motivated way of exploring whether the inclusion of additional free parameters in the BSM sector can encode more of the flavor structure of the SM.
We, therefore, start with a generic structure of linear BSM modifications to the beta functions and then specialize to the case where the contributions to each 
$\beta_{y_i}$ are proportional to a different $f_i$.\\

We consider beta functions for $Y_{U}$ and $Y_{D}$ of the form
\begin{equation}
 \beta_{Y_{F}}=\beta_{Y_{F}}^{{\small SM}}+\beta_{Y_{F}}^{{\small BSM}}\equiv\beta_{Y_{F}}^{SM}-F_{y}^{F}Y_{F}G^{F}_{y},
 \label{eq:betagen}
\end{equation}
with $F=U,D$. For the remaining part of this discussion, we focus on the BSM part. The expressions for $\beta_{Y_{F}}^{{\small SM}}$ are given in (\ref{eq:YukawamatrixflowU}) and
(\ref{eq:YukawamatrixflowD}). We drop
the superscript BSM in order to simplify the notation. As shown in section \ref{sec:betas}, it is useful to go from $Y_{F}$ to the matrix $M_{F}=Y_{F}Y_{F}^{\dg}$ whose beta function is
\begin{equation}
 \beta_{M_{F}}=-F_{y}^{F}Y_{F}G_{y}^{F}Y_{F}^{\dg}-Y_{F}G_{y}^{F\dg}Y_{F}^{\dg}F_{y}^{F\dg}.
\end{equation}
Once again, we diagonalize $M_{F}$ with the unitary matrices $V_{L}^{F}$ such that $V_{L}^{U}M_{U}V_{L}^{U\dg}=\mathrm{diag}[y_{u}^2,y_{c}^2,y_{t}^2]$ and
$V_{L}^{D}M_{D}V_{L}^{D\dg}=\mathrm{diag}[y_{d}^2,y_{s}^2,y_{b}^2]$. All the important information on the running of the essential couplings is contained in 
\begin{equation}
 V_{L}^{F}\beta_{M_{F}}V_{L}^{F\dg}=-V_{L}^{F}F_{y}^{F}Y_{F}G_{y}^{F}Y_{F}^{\dg}V_{L}^{F\dg}-V_{L}^{F}Y_{F}G_{y}^{F\dg}Y_{F}^{\dg}F_{y}^{F\dg}V_{L}^{F\dg}.
\end{equation}
From the SM Lagrangian we also know that $Y_{F}$ are diagonalized by the unitary matrices ($V_{L}^{F}$, $V_{R}^{F}$) such that 
$V_{L}^{F}Y_{F}V_{R}^{F\dg}=D_{F}$. Introducing $V_{L}^{F}$ and $V_{R}^{F}$ in the previous equation, we find that
\begin{align}
 V_{L}^{F}\beta_{M_{F}}V_{L}^{F\dg}=&-V_{L}^{F}F_{y}^{F}V_{L}^{F\dg}(V_{L}^{F}Y_{F}V_{R}^{F\dg})V_{R}^{F}G_{y}^{F}V_{R}^{F\dg}(V_{R}^{F\dg}Y_{F}^{\dg}V_{L}^{F\dg}) \nonumber \\
 &-(V_{L}^{F}Y_{F}V_{R}^{F\dg})V_{R}^{F}G_{y}^{F\dg}V_{R}^{F\dg}(V_{R}^{F}Y_{F}^{\dg}V_{L}^{F\dg})V_{L}^{F}F_{y}^{F\dg}V_{L}^{F\dg} \nonumber \\
 =&-\widetilde{F}_{y}^{F}D_{F}\widetilde{G}_{y}^{F}D_{F}-D_{F}\widetilde{G}_{y}^{F\dg}D_{F}\widetilde{F}_{y}^{F\dg},
 \label{eq:nontrivial}
\end{align}
where we have defined the quantities $\widetilde{F}_{y}^{F}=V_{L}^{F}F_{y}V_{L}^{F\dg}$, $\widetilde{G}_{y}^{F}=V_{R}^{F}G_{y}V_{R}^{F\dg}$.\\

The beta functions for the Yukawa couplings are given by the diagonal elements of (\ref{eq:nontrivial}). Thus, for the up-type Yukawas we have
\begin{equation}
 \beta_{y_{i}^2}=-y_{i}\sum_{l}y_{l}\widetilde{F}^{U}_{y,il}\widetilde{G}^{U}_{y,li}-y_{i}\sum_{l}y_{l}\widetilde{F}_{y,il}^{U\ast}\widetilde{G}_{y,li}^{U\ast},
 \label{eq:genup}
\end{equation}
while for the down-type Yukawas we obtain
\begin{equation}
 \beta_{y_{\rho}^2}=-y_{\rho}\sum_{\alpha}y_{\alpha}\widetilde{F}^{D}_{y,\rho\alpha}\widetilde{G}^{D}_{y,\alpha\rho}
 -y_{\rho}\sum_{\alpha}y_{\alpha}\widetilde{F}_{y,\rho\alpha}^{D\ast}\widetilde{G}_{y,\rho\alpha}^{D\ast}.
 \label{eq:gendown}
\end{equation}
We observe that the beta function of $y_i^2$ ($y_\rho^2$) is not simply proportional to $y_i^2$ ($y_\rho^2$), which is a key difference to the case considered in 
the main text.

The beta functions for the CKM elements are derived from the off-diagonal terms in (\ref{eq:nontrivial}). In the present case, we have
\begin{align}
 &\frac{d|V_{i\rho}|^2}{dt}=\sum_{j\neq i}\frac{1}{y_{i}^{2}-y_{j}^{2}}\left[-y_{i}V_{j\rho}V^{\ast}_{i\rho}\sum_{l}y_{l}\widetilde{F}_{y,jl}^{U\ast}\widetilde{G}_{y,li}^{U\ast}
 -y_{j}V_{j\rho}V^{\ast}_{i\rho}\sum_{l}y_{l}\widetilde{F}_{y,il}^{U}\widetilde{G}_{y,lj}^{U}+c.c.\right] \nonumber \\
 &+\sum_{\beta\neq \rho}\frac{1}{y_{\rho}^{2}-y_{\beta}^{2}}\left[-y_{\beta}V_{i\beta}V^{\ast}_{i\rho}\sum_{\alpha}y_{\alpha}\widetilde{F}_{y,\rho\alpha}^{D\ast}\widetilde{G}_{y\alpha\beta}^{D\ast}
 -y_{\rho}V_{i\beta}V^{\ast}_{i\rho}\sum_{\alpha}y_{\alpha}\widetilde{F}_{y,\beta\alpha}^{D}\widetilde{G}_{y,\alpha\rho}^{D}+c.c.\right].
\end{align}
We see that the running of the CKM elements gets modified for general $F_{y}^{F}$ and $G_{y}^{F}$. The independence of the CKM beta functions from the new-physics 
contribution in the main text is therefore a consequence of the flavor-blindness that was a central premise of our main analysis.

The new type of contribution in the beta functions of the Yukawa couplings could also be worthwhile to explore, but we leave this to the future. Instead, we 
specialize to the case where
the BSM-induced running of each
Yukawa coupling is just proportional to itself ($\dot{y}_i^2\propto y_i^2$, $\dot{y}_\rho^2\propto y_\rho^2$). Moreover, we investigate the conditions under which
the proportionality factor is different for each generation of quarks ending up with a flavor-dependent type of BSM physics. 
In order to achieve $\beta_{{y}_i^2}\propto y_i^2$, $\beta_{{y}_\rho^2}\propto y_\rho^2$, we note that either $\widetilde{F}^{F}$ or
$\widetilde{G}^{F}$ has to be diagonal, so that we pick only one of the contributions in the sum of Eqs. (\ref{eq:genup})-(\ref{eq:gendown}).
When this happens, we obtain
\begin{equation}
 \beta_{y_{i}^2}=-2y_{i}^2\Re(\widetilde{F}^{U}_{y,ii}\widetilde{G}^{U}_{y,ii}),
 \label{eq:diagup}
\end{equation}
\begin{equation}
 \beta_{y_{\rho}^2}=-2y_{\rho}^2\Re(\widetilde{F}^{D}_{y,\rho\rho}\widetilde{G}^{D}_{y,\rho\rho}),
 \label{eq:diagdown}
\end{equation}
where $\Re(x)$ means the real part of $x$. In order to obtain the same contribution for each family, the quantities $\Re(\widetilde{F}^{U}_{y,ii}\widetilde{G}^{U}_{y,ii})$ and 
$\Re(\widetilde{F}^{D}_{y,\rho\rho}\widetilde{G}^{D}_{y,\rho\rho})$ should be equal within each generation. If we assume the matrices
$G_{y}^{F}$ to be diagonal, the running of the CKM elements becomes
\begin{align}
 \frac{d|V_{i\rho}|^2}{dt}=&\sum_{j\neq i}\frac{1}{y_{i}^{2}-y_{j}^{2}}\left[-y_{i}^2V_{j\rho}V^{\ast}_{i\rho}\widetilde{F}_{y,ji}^{U\ast}\widetilde{G}_{y,ii}^{U\ast}
 -y_{j}^2V_{j\rho}V^{\ast}_{i\rho}\widetilde{F}_{y,ij}^{U}\widetilde{G}_{y,jj}^{U}+c.c.\right] \nonumber \\
 &+\sum_{\beta\neq \rho}\frac{1}{y_{\rho}^{2}-y_{\beta}^{2}}\left[-y_{\beta}^2V_{i\beta}V^{\ast}_{i\rho}\widetilde{F}_{y,\rho\beta}^{D\ast}\widetilde{G}_{y,\beta\beta}^{D\ast}
 -y_{\rho}^2V_{i\beta}V^{\ast}_{i\rho}\widetilde{F}_{y,\beta\rho}^{D}\widetilde{G}_{y,\rho\rho}^{D}+c.c.\right].
\end{align}
There are two potential choices such that the BSM contribution to the CKM beta functions vanishes: i) in order to have a vanishing contribution to the CKM running at any scale, one possibility is to assume the matrices
$\widetilde{F}^{U}_{y}$ and $\widetilde{F}^{D}_{y}$ to diagonal, ii)
we can find particular $\widetilde{F}^{F}_{y}$ and $\widetilde{G}_{y}^{F}$ such that there are non-trivial cancellations among all the terms, which seems very unlikely.\\

In summary, we conclude that in order to obtain an unaltered CKM running and a BSM contribution to the Yukawa couplings distinguishing different generations of quarks, 
$\widetilde{F}_{y}^{F}$ and $\widetilde{G}_{y}^{F}$ should be diagonal. Note that these
conditions do not imply that $F_{y}^{F}$ and $G_{y}^{F}$ are diagonal, or that $F_{y}^{U}=F_{y}^{D}$, $G_{y}^{U}=G_{y}^{D}$. There are several ways we can obtain
an expression of the form given in (\ref{eq:betagen}). For instance, we can add fermions or scalars charged under the SM gauge group. Constructing interactions among
the new fields and the SM particles, we can modify the beta function of the quark Yukawa couplings. \\

In the following, instead of exploring which type of new physics might give rise to (\ref{eq:betagen}), we explore the particular case in which each generation of quarks gets
a different BSM contribution and the CKM sector remains unaltered. We can, in fact, show that the inclusion of new colored fermions leads to beta functions of the form
 \begin{equation}
     \frac{dy_{i}}{dt}=\left[(Y_{2}(S)-G_{U})+\frac{3}{2}y_{i}^{2}-\frac{3}{2}\sum_{\rho}y_{\rho}^{2}|V_{i\rho}|^{2}+f_{y,i}\right]\frac{y_{i}}{(4\pi)^{2}},
 \end{equation}
\begin{equation}
     \frac{dy_{\rho}}{dt}=\left[(Y_{2}(S)-G_{D})+\frac{3}{2}y_{\rho}^{2}-\frac{3}{2}\sum_{i}y_{i}^{2}|V_{i\rho}|^{2}-f_{y,\rho}\right]\frac{y_{\rho}}{(4\pi)^{2}},
\end{equation}
where, as before, $y_{i}=(y_{u},y_{c},y_{t})$, $y_{\rho}=(y_{d},y_{s},y_{b})$, $G_{U}=\frac{17}{12}g_{1}^{2}+\frac{9}{4}g_{2}^{2}+8g_{3}^{2}$,
$G_{D}=\frac{5}{12}g_{1}^{2}+\frac{9}{4}g_{2}^{2}+8g_{3}^{2}$, $Y_{2}(S)=3\sum_{i}y_{i}^{2}+3\sum_{\rho}y_{\rho}^{2}$, $V_{i\rho}$  is the CKM matrix, and
$f_{y,t}=f_{y,b}=f_{y,3}$, $f_{y,c}=f_{y,s}=f_{y,2}$, $f_{y,u}=f_{y,d}=f_{y,1}$, so that there are only three different $f_{y}$-parameters. 
The new-physics effect distinguishes the three generations explicitly.
Since the CKM beta functions are not modified in this particular configuration, we can use the fixed points given in Eq.~\eqref{eq:ckmconf}.
Each of these CKM configurations generates fixed points in the Yukawa couplings. We obtain isolated fixed points, as well
as lines parameterized by one of the Yukawa couplings. 
The lines of fixed points are expected to reduce to isolated points under the impact of higher-loop effects. A solution of the higher-order system of equations for fixed points is beyond the scope of this paper; thus we cannot determine which part of the fixed lines might potentially be of interest. For this reason, we focus on 
the isolated fixed points of the one-loop system.

For each of the cases in Eq.~\eqref{eq:ckmconf}, we find that the minimum number of vanishing couplings at the fixed point is equal to two. That is, only four couplings are non-trivial and correspond to irrelevant directions.
Therefore, only four couplings can be predicted. Since $f_{y,3}$ already generates a hierarchy between $y_{t}$ and $y_{b}$, we need only one extra 
parameter, $f_{y,1}$ or $f_{y,2}$, to set up the remaining hierarchy.
We highlight that if the two extra parameters $f_{y,1(2)}$ are used to generate the desired IR-values of the two additional non-zero couplings, there is no enhanced predictivity within our general study\footnote{A specific BSM model in which the $f_{y,i}$ are calculable would, of course, have an increased predictivity.}. Each additional calculable quark mass would be tuned with a corresponding free parameter. 
Before deciding which parameter to keep, we impose other important conditions on the solutions. 
The presence of poles in the beta functions precludes the possibility that the RG flows of up(down)-Yukawa couplings cross each other. Therefore, only $y_{d}$ and 
$y_{u}$ can be zero at the fixed point.
Consequently, we single out solutions
with $y_{d*}=y_{u*}=0$ only. This selection rule leaves us with only two cases: 
(1) $X_{\ast}=0$, $Y_{\ast}=1$, $Z_{\ast}=0$, $W_{\ast}=0$ with
\begin{align}
y_{t\, \ast}^2&=\frac{16\pi^2}{123}(6f_g+41f_{y,2}-123f_{y,3})\ ,\quad y_{c\, \ast}^2=\frac{4\pi^2}{123}(23f_g-451f_{y,2}+287f_{y,3}) \ , \nonumber \\
y_{b\, \ast}^2&=\frac{4\pi^2}{123}(-f_g-287f_{y,2}+451f_{y,3})\ , \quad y_{s\, \ast}^2=\frac{16\pi^2}{123}(-6f_g+123f_{y,2}-41f_{y,3})
\ ,
\label{eq:linesbone}
\end{align}
or (2) $X_{\ast}=0$, $Y_{\ast}=0$, $Z_{\ast}=1$, $W_{\ast}=0$ with
\begin{align}
y_{t\, \ast}^2&=\frac{4\pi^2}{123}(23f_g+287f_{y,2}-451f_{y,3})\ ,\quad y_{c\, \ast}^2=\frac{16\pi^2}{123}(6f_g-123f_{y,2}+41f_{y,3}) \ , \nonumber \\
y_{b\, \ast}^2&=\frac{16\pi^2}{123}(-6f_g-41f_{y,2}+123f_{y,3})\ , \quad y_{s\, \ast}^2=\frac{4\pi^2}{123}(-f_g+451f_{y,2}-287f_{y,3})
\ .
\label{eq:linesbtwo}
\end{align}
In both situations $f_{y,1}$ is absent, as it is expected since $y_{d*}=y_{u*}=0$. Working with $f_{y,3}$ and $f_{y,2}$ as free parameters, we now impose conditions
on the Yukawa couplings in order to narrow down the allowed region in the parameter space.

\begin{figure}[H]
\centering
\includegraphics[width=0.8\linewidth]{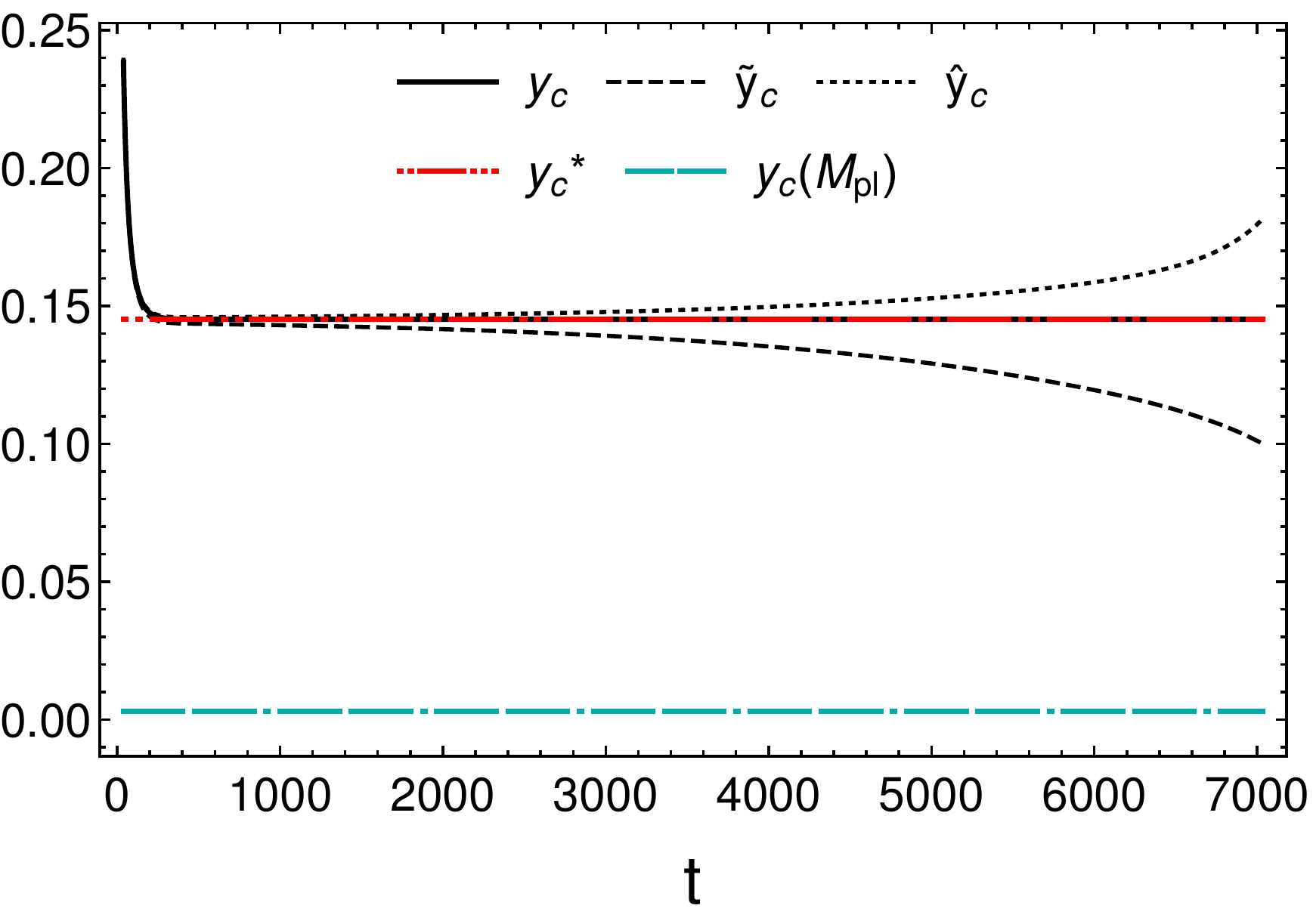}
\caption{We show the evolution of the coupling $y_{c}$ starting from the fixed point (\ref{eq:linesbone}), as well as other two trajectories starting a little 
above and below the fixed-point value (solid, dotted and dashed lines respectively). 
We also depict the fixed-point value and the corresponding value of the charm Yukawa coupling at the Planck scale.}
\label{fig:ycbreaking}
\end{figure}

We start by demanding $y_{t*}>y_{c*}>0$, $y_{b*}>y_{s*}>0$. These conditions translate into a constrained interval in the $f_{y,2}-f_{y,3}$ space. 
In the allowed parameter space there is always at least one coupling for which we have 
$y_{j\, \ast}\gg y_j(M_{\rm Planck})$.
For instance, in the case given by Eq.~\eqref{eq:linesbone}, we find that the minimum value of the charm Yukawa coupling we can obtain at a fixed point is
$y_{c\, \ast}\gtrsim0.136$, whereas at the Planck scale in the SM it holds that $y_{c}(M_{\rm Planck})=0.00293$.
This fact makes a matching between fixed-point trajectories and the observed IR value impossible because $y_{c}$ always increases towards low energies, as shown in Fig.~\ref{fig:ycbreaking}. Thus, the new contribution 
$f_{y,2}$ is not appropriate to generate the adequate hierarchy in the SM. 

The difficulty in the explanation of the full quark flavor structure arises from the two split patterns we have to generate. Firstly, there is the difference between the up- and down-type quarks
\emph{within} each generation. The type of new-physics contribution discussed here is in principle  able to explain this difference, at least for one generation.
Secondly, there is the quantitative difference \emph{among} the three SM generations. While our flows exhibit upper bounds on the quark masses of the lower 
generations, we have not discovered a mechanism that allows to predict the inter-generational hierarchy.
We stress that the inter- and the intragenerational hierarchy are two distinct aspects that might 
have different origin, and whose nature needs to be explored in future studies.

\end{document}